\documentclass[12pt]{article}
\usepackage{"conroy_latex3"}
\usepackage{"conroy_notations1"}

\usepackage{tocloft}

\usepackage{titletoc}
\usepackage{pgfplots}
\pgfplotsset{compat=1.18}

\title{\bf \Large \vspace{-25pt} Aggregating treatment effects across multiple outcomes }

\author{\normalsize Chun Pong Lau\thanks{\mbox{\footnotesize Darla Moore School of Business,
University of South Carolina, \href{mailto:cplau@sc.edu}{\texttt{cplau@sc.edu}}.} \\
This paper is a revised version of my job market paper.
I am grateful to Alex Torgovitsky, Stéphane Bonhomme, Azeem Shaikh, and Max Tabord-Meehan for their continued guidance, support, and encouragement throughout the project.
I also thank
Eric Auerbach,
Yuehao Bai,
Dylan Balla-Elliott,
Xinyue Bei,
Debopam Bhattacharya,
Chris Blattman,
Yong Cai,
Tim Christensen,
Joshua Dean,
Deniz Dutz,
Max Farrell,
Joseph Hardwick,
Eyo Herstad,
Thomas Hierons,
Samuel Higbee,
Damon Jones,
Omkar Katta,
Jacob Leshno,
Xinran Li,
Jonas Lieber,
José Luis Montiel Olea,
Ian Pitman,
Kirill Ponomarev,
Guillaume Pouliot,
Doron Ravid,
Frank Schorfheide,
Joshua Shea,
Myungkou Shin,
Suhas Vijaykumar,
Davide Viviano, and
Thomas Wiemann for helpful feedback and discussions, as well as participants in various seminars for valuable comments. All errors are my own.
}}
\date{\normalsize  \today{} }
\usepackage{amssymb}
\usepackage{makecell}

\usepackage[inline]{enumitem}
\usepackage[mathscr]{euscript}

\graphicspath{{./images/}}

\usepackage{centernot}
\usepackage{tcolorbox}
\usepackage{marginnote}
\usepackage{multicol}
\usepackage{lscape}
\usepackage{accents}
\usepackage{bbding}
\usepackage{mathtools}

\newcommand{\dx}{d_x}
\newcommand{\dy}{q}

\usepackage{textgreek}
\newcommand{\cUpsilon}{\text{\fontfamily{cmr}\selectfont\textUpsilon}}

\usepackage{ragged2e}

\allowdisplaybreaks

\usepackage[sc]{mathpazo}

\makeatletter
\renewcommand\paragraph{\@startsection{paragraph}{4}{\z@}%
            {-2.5ex\@plus -1ex \@minus -.25ex}%
            {1.25ex \@plus .25ex}%
            {\normalfont\normalsize\bfseries}}
\makeatother
\newcommand{\pca}{Principal component analysis (PCA)}
\newcommand{\kling}{Equally weighted standardized averaging (SA)}
\newcommand{\anderson}{Inverse variance matrix weighting (IVM)}

\newcommand{\pres}{Precision}
\newcommand{\inter}{Interpretability}

\newcommand{\mynote}[1]{ \begin{minipage}[t]{1\columnwidth}
	{\small {\it Notes:}#1}
\end{minipage}}

\begin{document}
\fontfamily{ppl}\selectfont

\maketitle

\begin{abstract}
Researchers often aggregate outcomes to estimate treatment effects on abstract concepts, such as ``crime'' or ``wealth.''
First, I document that principal component analysis (PCA), inverse variance matrix (IVM) weighting, and standardized averaging (SA) are commonly used but have unexpected shortcomings.
PCA can assign negative weights, need not maximize precision, depends on an arbitrary sign normalization, and can have nonstandard asymptotics.
Unlike SA, IVM can assign negative weights, but neither necessarily maximizes precision.
Second, I develop a statistical decision-theoretic approach that uses mean-squared error to choose nonnegative weights while accounting for outcome quality.
I revisit a study on social distancing.
\end{abstract}

\newpage

\startcontents[main]
{
}

\newpage

\section{Introduction}
Researchers commonly observe multiple related outcomes that measure an underlying abstract variable.
For instance, \citet{bruhnetal2018jpe} measure entrepreneurial confidence and goal setting using entrepreneurs' responses to questions about attitudes and beliefs.
\citet{jonesetal2019qje} measure productivity using different employment and workplace-related outcomes.
\citet{bau2022jpe} measures wealth using the ownership of various assets.
\citet{bhattetal2023qje} measure crime involvement using different types of offenses and arrests.
To learn the treatment effect on the abstract variable, to summarize the effect across outcomes, or to improve interpretability, researchers often aggregate related outcomes into a single index.

Some commonly used approaches to aggregating outcomes are the principal component analysis (PCA) approach popularized by \citet{filmerpritchett2001d}, the equally weighted standardized averaging (SA) approach by \citet{klingetal2007ecta}, and the inverse variance matrix (IVM) weighting approach by \citet{anderson2008jasa}.
To illustrate the popularity of these methods, Table \ref{tab:count} counts the number of articles published in ``top-five'' economics journals between 2020 and 2024 that applied these methods.

In this paper, I analyze the properties and shortcomings of these approaches and propose a new method for optimally aggregating outcomes.
In the first part of the paper, I show how using PCA, SA, or IVM affects precision and interpretability.
Precision is measured by the asymptotic variance of the treatment effect on the aggregated outcome or, more generally, by its mean-squared error.
Interpretability requires that the weights on the outcomes be nonnegative and sum to one.
This is because negative weights can cause sign reversal in the aggregated treatment effect.
The negative weights issue here is similar to, but distinct from, the negative weights issue for subpopulations, such as in difference-in-differences (e.g., \citealp{dechaisemartindHaultfŒuille2020aer}; \citealp{callawaysantanna2021joe}; \citealp{goodmanbacon2021joe}; \citealp{sunabram2021joe}; \citealp{borusyaketal2024restud}) and local average treatment effects (e.g., \citealp{blandholetal2026restud}; \citealp{słoczyński2026wp}).

I show that PCA-aggregated treatment effects need not maximize precision, can be difficult to interpret because of negative weights, are sensitive to an arbitrary sign normalization, and can lead to a nonstandard asymptotic distribution.
These properties contrast with how PCA is sometimes motivated as a way to increase power or improve interpretability.
IVM can also have negative weights and does not always maximize precision.
SA assigns equal and positive weights to the outcomes, so it does not suffer from interpretability issues caused by negative weights.
However, SA does not use the correlation structure among the outcomes, which leaves room for precision gains.

In the second part of the paper, I use statistical decision theory to develop a new approach to aggregating treatment effects.
The approach takes precision into account by using the mean-squared error of the aggregated treatment effect as the objective, and it imposes nonnegative weights to avoid interpretability issues.
Since the abstract variable of interest is unobserved, I consider two optimality criteria.
The first is the minimax criterion, which minimizes the worst-case risk.
The second is the recently proposed adaptation regret criterion \citep{armstrongetal2025ecta}, which measures the cost of deviating from the optimal weight relative to the smallest attainable worst-case risk.
The second criterion has the advantage of not requiring the researcher to commit to a level of misspecification, which the minimax criterion requires.
I show that, under either criterion, the weights can be computed using convex optimization.
I demonstrate how to conduct inference using fixed-length confidence intervals that account for potential bias under the restriction of nonnegative weights.

My decision-theoretic approach can incorporate various economic restrictions.
It allows for shape restrictions, such as some treatment effects being more important than others.
I show that the decision-theoretic framework can also be motivated by a communication model in which the reader has subjective weights on the treatment effects.

I illustrate the tools proposed in this paper by revisiting \citet{karlanetal2026jde}, who study how a randomized cash transfer program affected social distancing during the COVID-19 pandemic.
To measure social distancing, the authors construct an index from six outcomes using PCA.
I focus on their analysis in which the program's contemporaneous effect on the social distancing index is positive and significant at the 1\% level, although none of the effects on the individual outcomes is significant at that level.
I find that this effect is partly driven by the outcome with the largest negative treatment effect receiving the most negative PCA weight.
I then apply my decision-theoretic approach to choose the weights, without restricting the relative quality of the outcomes.
The estimated effect is positive but smaller than the PCA estimate, and it is not significant at the 10\% level, even when the weights are chosen to minimize the asymptotic variance of the aggregated treatment effect.
Next, I show how to incorporate domain knowledge that two of the outcomes (such as the number of days with visitors, as noted by the authors) are less informative about the effect on social distancing.
My approach can impose this domain knowledge as shape restrictions and can recover a positive effect on social distancing that is significant at the 5\% level under the minimax criterion when the treatment effects on these two outcomes are sufficiently downweighted and assumed to be more biased than those on the other four.
Hence, my approach makes precise how strong the assumptions must be for that conclusion to hold.
In contrast, PCA, SA, and IVM cannot use such knowledge and have the shortcomings documented in Section~\ref{sec:problems}.

\subsection{Survey of the use of PCA, SA, and IVM}
Table~\ref{tab:count} documents how often PCA, SA, and IVM are used to aggregate outcomes.
The table shows that these methods are prominent in applied work.
I restrict the search to articles published in ``top-five'' economics journals between 2020 and 2024.
The counts for PCA are based on searching Google Scholar and journal websites for articles that contain the phrase ``principal component.''
The counts for SA and IVM are based on articles that cite \citet{klingetal2007ecta} and \citet{anderson2008jasa}, respectively, on Google Scholar.
The counts in Table \ref{tab:count} include only articles that use these methods to aggregate outcomes.

\begin{table}[!t]
	\centering
	\caption{Usage of PCA, SA, and IVM in ``top-five'' economics journals in 2020--2024.}
	\label{tab:count}
	\setlength{\tabcolsep}{20pt}
	\begin{tabular}{lccc}
		\toprule
		\bf Journal $\bm{\backslash}$ Method &\bf  PCA &\bf  SA &\bf IVM \\
		\midrule
		American Economic Review
		&10	&13	&14	\\
		Econometrica
		&4	&2	&2 	\\
		Journal of Political Economy
		&4	&6	&0 	\\
		Quarterly Journal of Economics
		&7	&7	&5 	\\
		Review of Economic Studies
		&5	&5	&2 	\\
		\midrule
		\bf Total
		&30	&33	& 23	\\
		\bottomrule
	\end{tabular}\par \smallskip\smallskip
	\mynote{
	PCA refers to principal component analysis,
	SA refers to equally weighted standardized averaging \citep{klingetal2007ecta},
	and IVM refers to inverse variance matrix weighting \citep{anderson2008jasa}.
	The table focuses on articles using PCA, SA, and IVM to aggregate outcomes. }
\end{table}

\subsection{Related literature}
This paper contributes to two literatures in econometrics.

First, this paper is connected to the literature on aggregating outcomes.
\citet{obrien1984biom} is an early work in biostatistics that studies the use of generalized least squares (GLS) to aggregate outcomes.
Recently, \citet{andersonmagruder2023wp} show that SA is the most powerful index when outcomes are equicorrelated and have equal treatment effects; I give necessary and sufficient conditions in Appendix~\ref{app:t-stat}.
\citet{gomez2024wp} studies the power properties of PCA, SA, and IVM by simulation but neither discusses negative weights nor derives theoretical properties of precision or power; I address the former and derive the latter in my analysis.
\citet{alleeetal2022pca} review the use of PCA and factor analysis in accounting research.
Consistent with my survey of economics articles in Table~\ref{tab:count}, the authors find that 219 articles used PCA or factor analysis in ten major accounting journals during 2015--2019.
They do not discuss asymptotic variance, power, negative weights, sensitivity to normalization, or nonstandard asymptotic distributions for PCA and factor analysis.
They also do not discuss SA or IVM in their review.

In addition to PCA, SA, and IVM, other approaches have recently been proposed to aggregate outcomes.
These methods involve different objectives or require additional assumptions or data; examples include \citet{andersonmagruder2023wp}, \citet{huetal2024wp}, \citet{fugreen2025wp}, and \citet{stoetzeretal2025ajps}.
I review these approaches in Appendix~\ref{sec:other-methods}.

Second, this paper is related to the literature on statistical decision theory and optimal estimation and inference by \citet{donohoetal1990aos}, \citet{donoho1994aos}, \citet{cailow2004aos}, \citet{armstrongkolesar2018ecta, armstrongkolesar2021ecta, armstrongkolesar2021qe}, and others.
The idea of using adaptation regret to adapt over a range of misspecification was recently proposed by \citet{armstrongetal2025ecta}, which is related to \citet{bickel1984aos}, who adapts over granular sets.
Although I use the adaptation regret concept in \citet{armstrongetal2025ecta}, my setting differs in that I allow all estimators to be biased, whereas they assume that an unbiased estimator is available.
I also restrict attention to convex combinations of the estimators to ensure interpretability of the treatment effect on the aggregated outcome.
This paper is also related to the use of statistical decision theory in site selection and evidence aggregation problems, as in the recent work by \citet{gechteretal2024wp}, \citet{ishiharakitagawa2024wp}, and \citet{montieloleaetal2025wp}.
These papers have a different goal from reporting an aggregated treatment effect.
See also \citet{wald1950bk}, \citet{savage1951jasa}, and \citet{manski2004ecta} for statistical decision theory.

In concurrent and independent work, \citet{fedchenko2026wp} studies treatment effect estimation with summary indices, focusing on SA and IVM.
Both my paper and \citet{fedchenko2026wp} point out that negative weights in IVM can cause interpretability issues, show how to conduct valid inference under data-dependent weights, and note that SA or IVM need not improve power.
Despite the similarities, there are several notable differences between my paper and the work of \citet{fedchenko2026wp}.
First, I develop a new statistical decision-theoretic approach to aggregating outcomes and study properties of existing approaches, whereas \citet{fedchenko2026wp} primarily studies SA and IVM, and recommends avoiding indices when they generate negative weights.
Second, I consider the treatment effect on the latent outcome as the target parameter, in addition to the treatment effect on the summary index, whereas  \citet{fedchenko2026wp} considers only the latter.
Third, I study and point out issues with PCA, which my survey in Table~\ref{tab:count} shows is commonly used, while \citet{fedchenko2026wp} mentions PCA only briefly, noting that its results can extend to PCA indices under additional regularity conditions.
I also show that PCA is sensitive to an arbitrary sign normalization and can have a nonstandard asymptotic distribution.

\subsection{Organization of the paper}

Section~\ref{sec:setup} introduces the setup and describes the criteria for aggregating treatment effects.
Section~\ref{sec:problems} studies commonly used methods and their shortcomings.
Section~\ref{sec:var-min-weight} presents an approach to choosing interpretable weights that minimize variance.
Section~\ref{sec:decision} presents the decision-theoretic approach.
Section~\ref{sec:sdeg} revisits the analysis of how cash transfers affect social distancing and applies the framework of this paper to aggregate treatment effects.
Section~\ref{sec:concl} concludes.
All proofs are collected in the appendix.

\section{Framework} \label{sec:setup}
In this section, I introduce the setup and discuss some natural criteria for aggregation.

\subsection{Setup and notation} \label{sec:notations}
Consider a researcher who observes multiple outcomes $\bmY_i \equiv (Y_{i,1}, \ldots, Y_{i,\dy})' \in \bR^{\dy}$ to measure the effect of a treatment $D_i \in \bR$, where $\dy \geq 2$ is the number of outcomes and $i = 1, \ldots, n$ indexes units.
Let $\bmbe \equiv (\be_1, \ldots, \be_{\dy})' \in \bR^{\dy}$ be the vector of treatment effects on these outcomes (possibly estimated using additional covariates).
Researchers are often interested in learning the treatment effect on a latent outcome of interest $L_i \in \bR$, in addition to (or in lieu of) the treatment effect on each of the outcomes.
Let $\theta \in \bR$ be the target parameter that measures the treatment effect, such as the average treatment effect of $D_i$ on $L_i$.
Combining several asset outcomes into a single wealth measure is common in empirical work \citep[e.g.,][]{blattmanetal2020aeri,banerjeeetal2021aeri,bau2022jpe}.
Hence, I use this setting as a running example for exposition.

\begin{eg}[Running example of wealth index]
\label{eg:run-wealth}
A researcher is interested in learning the effect of a cash transfer program on wealth.
The researcher collects multiple asset-related outcomes $\bmY_i$, such as the ownership of goats, cows, cars, televisions, and cooling devices.
Let $L_i$ be wealth and $D_i \in \{0, 1\}$ be the indicator that equals 1 if treatment is received and equals 0 otherwise.
Here, $\theta$ is the treatment effect of $D_i$ on $L_i$.
\end{eg}

\noindent Section \ref{sec:setup-empeg} provides additional empirical examples illustrating the interpretation of $\theta$.

The outcomes are required to satisfy two conditions.
First, they are oriented in the same direction.
In the running example, the orientation requirement is satisfied when each $Y_{i,j}$ is increasing in the number of assets.
Second, the outcomes are standardized in line with how the researcher wants to interpret the effect (e.g., using the full sample or the control sample).
These two requirements are summarized below and are maintained throughout the paper.
\begin{assu}
\label{assu:outcome}
The outcomes $\bmY_i$ are oriented in the same direction as $L_i$, so that higher values indicate better outcomes, and they are suitably standardized.
\end{assu}

Suppose the researcher uses the following weighted average of treatment effects to estimate $\theta$:
\begin{equation}
	\label{eq:tau-1}
	\hta
	\equiv \bmw'\hbmbe
	= w_1 \hbe_1 + \cdots + w_{\dy} \hbe_{\dy}
\end{equation}
for an estimator $\hbmbe \equiv (\hbe_1, \ldots, \hbe_{\dy})'$ of $\bmbe$ and a given weight vector $\bmw \equiv (w_1, \ldots, w_{\dy})' \in \bR^{\dy}$.
I adopt the representation in \eqref{eq:tau-1} because this is the weighted average of treatment effects produced by a linear model when the left-hand side variable is $\bmw'\bmY_i$ (see Lemma~\ref{lem:linear-out} in the appendix for details).
This representation nests PCA, SA, and IVM because they aggregate outcomes linearly.
Throughout, $\hta$ denotes the weighted average of estimated treatment effects, whose estimand $\bmw'\bmbe$ may differ from $\theta$.
In the running example (Example \ref{eg:run-wealth}), the treatment effect of the cash transfer program on the aggregated outcome $\bmw'\bmY_i$ may not be equal to $\theta$ (i.e., the treatment effect on wealth) without additional assumptions.

The above setup assumes that the treatment effects on different outcomes are related to the same $\theta$.
This single-domain assumption is not necessary when one believes there is one latent outcome for each domain.
Researchers can create a summary index for each domain of outcomes (as suggested by, e.g., \citealp{anderson2008jasa}).
In the running example on wealth, researchers could create separate indices for livestock and durable assets.
I focus on a single domain for clarity.

\subsection{Additional empirical examples} \label{sec:setup-empeg}
Beyond the running example on wealth, I discuss additional empirical examples to explain the setup.

\begin{eg}[Public good provision]
\label{eg:cd}
\citet{campantedo2014aer} study the impact of isolated state capital cities on corruption, accountability, and public good provision.

In this example, let $\theta$ be the effect of state capital isolation on public good provision.
The authors observe three outcome variables $\bmY_i$ related to public good provision, namely, the percentage of the population with health insurance, the log number of hospital beds, and a measure of inputs and outcomes in education.
Here, $\hbmbe$ is the regression estimator of how isolation affects each of the three outcomes.
They aggregate the three variables using PCA to create a public good provision index, so $\bmw$ is the corresponding vector of PCA weights.
\end{eg}

\begin{eg}[Entrepreneurial spirit index]
\label{eg:bks}
\citet{bruhnetal2018jpe} study how management consulting services affect the productivity, return on assets, and ``entrepreneurial spirit'' of small and medium-sized enterprises in Mexico.
Entrepreneurial spirit measures the confidence of entrepreneurs and their goal setting.

Let $\theta$ be the impact of consulting on entrepreneurial spirit.
The authors create an entrepreneurial spirit index using responses to eight survey questions $\bmY_i$.
Each outcome is the entrepreneur's response (coded as 1 to 5) to a survey question, such as ``I have professional goals.''
Here, $\hbmbe$ is the vector of estimated treatment effects of consulting on the responses.
They create indices using both PCA and SA, so $\bmw$ is the corresponding vector of PCA or SA weights.
\end{eg}

\begin{eg}[Violence involvement]
\label{eg:bhkbb}
\citet{bhattetal2023qje} study the impact of community programs on involvement in serious violence $\theta$.
To measure involvement in serious violence, the authors use related outcomes $\bmY_i$, namely, shooting and homicide victimizations, arrests, and other serious violent-crime arrests.
Here, $\hbmbe$ is the vector of estimated treatment effects on each outcome.
They create a standardized index that averages the three outcomes.
They also create an index using PCA.
Hence, $\bmw$ depends on the approach used.
\end{eg}

\begin{eg}[Social distancing]
\label{eg:sd}
\citet{karlanetal2026jde} study how cash transfers affect social distancing, food expenditure, income, and psychological well-being, using a randomized controlled trial in Ghana during the COVID-19 pandemic.

Let $\theta$ be the impact of the cash transfer program on social distancing.
Social distancing is an abstract concept that cannot be observed, but the authors collect six related outcomes, such as the number of days on which the respondent stays at home all day, whether the respondent tries to keep a distance from nonfamily members, and whether the respondent wears a mask.
The authors aggregate the six outcomes using PCA to create a social distancing index.
\end{eg}

\subsection{Criteria for choosing the weights} \label{sec:criteria}
Even if one focuses on using a linear combination in \eqref{eq:tau-1}, there are many choices for the weights.
In this subsection, I discuss two criteria for choosing the weights $\bmw \in \bR^{\dy}$.

\subsubsection{Interpretability} \label{sec:criteria-inter}
The interpretability criterion requires that the weights be nonnegative and sum to one.
Negative weights can cause the aggregated treatment effect to be negative even though the treatment effect on each outcome is positive.
In the running example on wealth (Example~\ref{eg:run-wealth}), suppose the cash transfer program has a positive treatment effect on each asset.
With negative weights, the treatment effect on the aggregated outcome (the wealth index) can be negative.
Moreover, negative weights can cause a reversed ordering of the aggregated outcome.
For instance, suppose the weight on cows is negative in the running example.
This means that, holding other assets fixed, an individual with more cows receives a lower value of the wealth index.
Hence, negative weights on the outcomes lead to interpretability issues.

The sign reversal issue described above is related to, but different from, the negative weights issues pointed out in the recent econometrics literature, such as \citet{blandholetal2026restud} and \citet{słoczyński2026wp} on instrumental variables, and \citet{dechaisemartindHaultfŒuille2020aer}, \citet{callawaysantanna2021joe}, \citet{goodmanbacon2021joe}, \citet{sunabram2021joe}, and \citet{borusyaketal2024restud} on difference-in-differences.
In the aforementioned papers, the concern is with negative weights on the treatment effects for subpopulations.
In the context of aggregating outcomes, negative weights are placed on the treatment effects for different outcomes within the same population.
Regarding the issue of reversed ordering of the aggregated outcome, \citet{vyaskumaranayake2006hpp} and \citet{kolenikovangeles2009riw} point out that PCA socioeconomic status indices can have this problem because some variables can receive negative weights.
\citet{andersonmagruder2023wp} also mention that GLS-weighted indices can have negative weights.

Motivated by these considerations, I focus on the class of convex weights that are nonnegative and sum to one:
\begin{align}
	\label{eq:convex-w}
	\cWcvx
	\equiv
	\{ \bmw \in \bR^{\dy} : \bmw'\oneq = 1 , \bmw \geq \zeroq
	\},
\end{align}
where $\oneq$ is a vector of $\dy$ ones and $\zeroq$ is a vector of $\dy$ zeros.
Requiring the weights to sum to one has two purposes.
First, together with nonnegativity, this pins down the scale of the weights to avoid arbitrarily large weights.
Second, the sum-to-one requirement ensures that $\hta$ recovers the homogeneous treatment effects case, i.e., if $\hbe_{j} = \obe$ for all $j = 1, \ldots, \dy$, then $\hta = \obe$ for any $\bmw \in \cWcvx$.

\begin{re}
\label{re:postprocess}
After computing the weights, researchers sometimes ``post-process'' the aggregated outcome $\bmw'\bmY_i$ by standardizing it (e.g., \citealp{parkervogl2023ej}) or rescaling it to $[0, 1]$ (e.g., \citealp{ajzenman2021aeja}).
I discuss the implications of this post-processing step in Appendix~\ref{supp:results-weight}.
\end{re}

\subsubsection{Precision} \label{sec:criteria-pres}
A natural criterion for $\hta$ to be precise is to choose the weights to minimize the mean-squared error (MSE).
This amounts to using the squared loss function $(\hta - \theta)^2$.
The MSE can be written as a function of $\bmw$ as follows:
\begin{align}
	\label{eq:mse}
	\bE[(\hta - \theta)^2]
	=
	\var[\bmw'\hbmbe] + \left(\bE[\bmw'\hbmbe] - \theta \right)^2.
\end{align}
The MSE criterion evaluates the quality of the estimator in terms of variance and bias.
In the running example, the bias component measures how well the treatment effect on the weighted average of assets captures the treatment effect on wealth.
The variance component evaluates the noisiness of the aggregated treatment effect.
This criterion nests the special case in which the treatment effect on each asset $\beta_j$ equals the treatment effect on wealth $\theta$, i.e., $\beta_j = \theta$ for each $j = 1, \ldots, \dy$.
In this special case that $\hbe_j$ is unbiased for $\theta$ for each $j = 1, \ldots, \dy$, the MSE reduces to the variance for any $\bmw \in \cWcvx$.
Section~\ref{sec:var-min-weight} shows how to choose the weights in this special case.
My decision-theoretic approach, introduced in Section~\ref{sec:decision}, shows how to handle the bias formally without assuming that $\be_j = \theta$ for all $j$.

\section{Common aggregation methods and their shortcomings} \label{sec:problems}
In this section, I show the shortcomings of common methods for aggregating outcomes.
Each method is mainly evaluated using the precision and interpretability criteria described in Section \ref{sec:criteria}.
I also study how these methods affect the asymptotic variance of the aggregated treatment effect, the $t$-statistic, and power.
I show that the PCA approach is even subject to additional issues: sensitivity to an arbitrary sign normalization and a nonstandard asymptotic distribution.

\subsection{\pca} \label{sec:pca-main}
PCA is a dimension-reduction tool developed more than a century ago \citep{pearson1901, hotelling1933}.
PCA transforms a set of correlated variables into a new set of uncorrelated variables called principal components.
Table \ref{tab:count} shows that PCA is a popular approach to aggregating outcomes.

PCA is often performed on the correlation matrix of the outcomes, denoted by $\bmSY$.
Using the correlation matrix is preferred to using the covariance matrix because the outcomes have different units, so PCA on the covariance matrix is not scale-invariant \citep{jolliffe2002bk}.
Under Assumption~\ref{assu:outcome}, $\bmY_i$ represents the standardized outcomes and $\bmbe$ is the vector of treatment effects on such standardized outcomes.

The PCA approach aggregates outcomes using the first principal component (PC1) of $\bmSY$, i.e., it finds the vector $\bmw \equiv (w_1, \ldots, w_{\dy})' \in \bR^{\dy}$ that maximizes the variance of the linear combination of the standardized outcomes $\var[\bmw'\bmY_i] = \bmw'\bmSY\bmw$.
For textbook expositions of PCA, see, for instance, \citet{jolliffe2002bk} and \citet{jackson2005bk}.

Section \ref{sec:pca-outline} reviews the relevant theory of PCA.
Sections \ref{sec:pca-interpret}  and \ref{sec:pca-precise} evaluate PCA using the interpretability and precision criteria, respectively.
Section \ref{sec:pca-sign} explains that PCA-aggregated treatment effects are sensitive to an arbitrary sign normalization.
Section \ref{sec:pca-dist} shows that the PCA-aggregated treatment effect can have a nonstandard asymptotic distribution.

\subsubsection{The PCA problem} \label{sec:pca-outline}

The problem of finding the PC1 of $\bmSY$ can be written as
\begin{align}
	\label{eq:pca-problem}
	\wpca = \argmax_{\bmw \in \cWunit} \ \ \bmw'\bmSY\bmw,
\end{align}
where the class of weights is
\begin{align}
	\label{eq:weight-pca}
	\cWunit
	\equiv
	\{ \bmw \in \bR^{\dy} : \bmw'\bmw = 1, \bmc'\bmw \geq 0
	\}
\end{align}
for a given $\bmc \in \bR^{\dy} \backslash \{\zeroq\}$.
I explain each piece of \eqref{eq:pca-problem}, including $\bmc$ in \eqref{eq:weight-pca}, below.

First, the PC1 is the leading eigenvector of $\bmSY$.
The PC1 of $\bmSY$ is unique up to a nonzero scalar multiple when the largest eigenvalue is unique.
To formalize this, let $\{(\bvec_j, \lambda_j)\}^{\dy}_{j=1}$ be the eigenpairs of $\bmSY$, where $\{\lambda_j\}^{\dy}_{j=1}$ are the eigenvalues of the matrix $\bmSY$ and $\bvec_j$ is a unit-length eigenvector corresponding to $\lambda_j$ for each $j = 1, \ldots, \dy$.
I assume the eigenvalues are ordered as $\lambda_1 \geq \cdots \geq \lambda_{\dy}$.
The problem of having repeated leading eigenvalues is discussed in Section~\ref{sec:pca-dist}.
\begin{assu}[Unique leading eigenvalue of $\bmSY$]
\label{assu:eval}
$\lambda_1 > \lambda_2$.
\end{assu}

Second, a normalization on the length of $\bmw$ is required because the variance $\var[\bmw'\bmY_i]$ is unbounded without any restrictions on $\bmw$.
The unit-length restriction
\begin{equation}
	\label{eq:unit-length}
	\bmw'\bmw = w_1^2 + \cdots + w_{\dy}^2 = 1
\end{equation}
in \eqref{eq:weight-pca} is commonly used.
Other length normalizations are possible, although the results would change with the choice of normalization, and some normalizations could lead to a more difficult optimization problem (see, for instance, \citealp{jolliffe2002bk}).

\begin{figure}[!t]
\centering
\caption{Sign normalization for the PCA problem.}
\label{fig:pca-sign-1}
\begin{tikzpicture}[scale=1]
\begin{axis}[
    axis equal,
    xlabel={$w_1$},
    ylabel={$w_2$},
    xmin=-1.35, xmax=1.35,
    ymin=-1.35, ymax=1.35,
    axis lines=middle, 
    xtick={0}, 
    ytick={0}, 
    xlabel style={anchor=west}, 
    ylabel style={anchor=south}, 
]

  \draw[ccivy, thick, dashed, rotate around={-45:(0,0)}, line width = 2pt] (0,0) ellipse ({sqrt(2)} and {sqrt(1)});
  \draw[ccivy, thick, rotate around={-45:(0,0)}, line width = 2pt] (0,0) ellipse ({sqrt(2)} and {sqrt(1)});

  \draw[cclake, thick, line width = 3.5pt] (0,0) circle (2.1cm);

\node[circle, fill=ccgrey, inner sep=0pt, minimum size=10pt] at (axis cs:.707,.707) {};
\node[circle, fill=ucmaroon, inner sep=0pt, minimum size=10pt] at (axis cs:.707,.707) {};
\node[circle, fill=ccgrey, inner sep=0pt, minimum size=10pt] at (axis cs:-.707,-.707) {};

\end{axis}
\end{tikzpicture}
   	\mynote{
	The green ellipse represents the objective of the PCA problem.
	The blue unit circle represents the unit-length constraint.
	Both the gray and red dots are on the largest ellipse that is tangent to the unit circle.
	The red dot is identified by the $w_1 + w_2 \geq 0$ constraint.}
\end{figure}
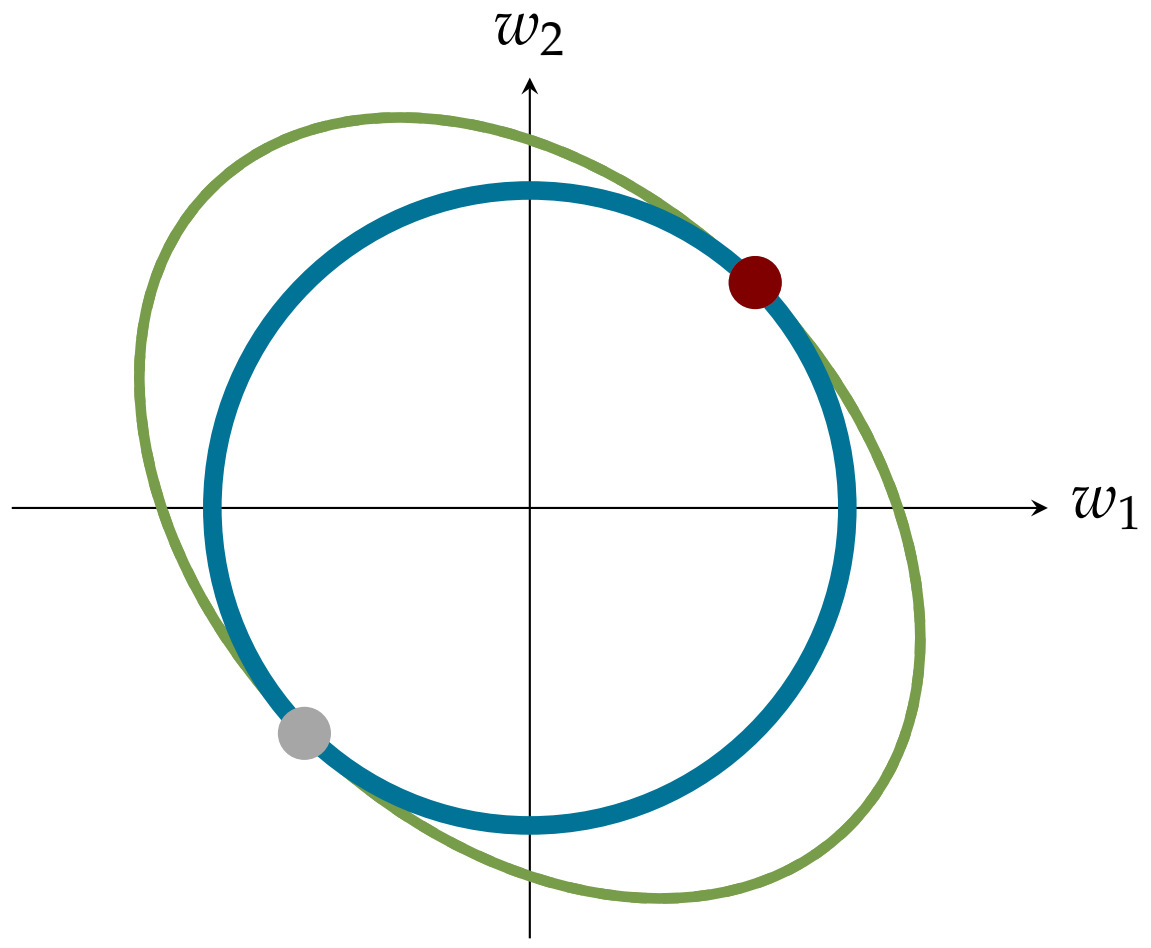

Third, the inequality $\bmc'\bmw \geq 0$ in \eqref{eq:weight-pca} is an identification condition used to obtain a unique solution.
This identification condition is needed because the leading eigenvector is unique up to a sign change even when \eqref{eq:unit-length} is imposed (i.e., both $\bvec_1$ and $-\bvec_1$ are leading eigenvectors of $\bmSY$).
Figure \ref{fig:pca-sign-1} explains this via the geometry of the PCA problem, where the PC1 problem finds the largest ellipse (i.e., the objective $\bmw'\bmSY\bmw$ in \eqref{eq:pca-problem}) constrained to the unit circle (i.e., the unit-length constraint \eqref{eq:unit-length}).
The unit circle alone is insufficient to determine a unique solution because both the gray and the red dots are valid solutions.
The inequality constraint $w_1 + w_2 \geq 0$ can be used to rule out the gray dot, yielding a unique solution.
Table \ref{tab:sign-pca} shows how several common programming languages implement the additional sign constraints to identify the unique solution to the PCA problem.

\begin{table}[!t]
	\centering
	\caption{Sign constraints for PCA in common implementations.}
	\label{tab:sign-pca}
	\begin{tabular}{p{0.29\textwidth} p{0.61\textwidth}}
		\toprule
		\bf Programming language
		& \bf Sign constraints	\\
		\midrule
		Stata
		& The \texttt{pca} command chooses the sign of the loadings such that the sum of the entries of each principal component is positive \citep{statapca}.	\\
		\hline
		Matlab
		& The \texttt{pca} function chooses the sign of the loadings such that the entry with the largest absolute value in each principal component is positive \citep{matlabpca}. \\
		\hline
		R
		& The \texttt{princomp} function sets the signs of the loadings such that the first entry of each principal component is nonnegative by default \citep{rpca}.	\\
		\bottomrule
	\end{tabular}
\end{table}
As shown in Table \ref{tab:sign-pca}, the constraint $\bmc'\bmw > 0$ covers the Stata implementation by setting $\bmc$ to a vector of ones (note that Stata imposes a strict inequality per the manual).
The same constraint also covers the R implementation by setting $\bmc$ to a vector with the first entry equal to 1 and the remaining entries equal to 0.
This normalization means the ordering of the variables matters.
The normalization used by Matlab cannot be written as one linear constraint, but it can be analyzed in a similar way.
These implementations differ, and the data alone do not pin down the sign normalization.
Section~\ref{sec:pca-sign} shows the consequences.

\subsubsection{\inter} \label{sec:pca-interpret}
The class of weights \eqref{eq:weight-pca} used by PCA does not restrict the weights to be nonnegative.
Whether all weights can be positive depends on the correlation of the outcomes.
By the Perron--Frobenius theorem (see, for instance, \citealp[Exercise 8.7.1]{mardiaetal1979bk}), if $\bmSY$ has only positive entries, then all the entries in the leading eigenvector of $\bmSY$ have the same sign.
Without this condition, the PC1 of $\bmSY$ can have entries of opposite signs.
This occurs in the empirical application in Section~\ref{sec:sdeg}, where two of the six outcomes receive negative PCA weights (Figure~\ref{fig:jde-te}).

Outcomes can be negatively correlated with each other even if they are all positively affected by the treatment.
I discuss two scenarios in which negative correlations can occur.

The first possible scenario arises when the set of outcomes includes goods that are substitutes.
In the running example on wealth, the wealth index might include outcomes that are substitutes for one another, such as air conditioners, air coolers, and fans, all of which appear in the household asset data of \citet{bau2022jpe}.
To understand how negative correlation arises with substitutes, I consider a stylized two-good example below.

\begin{eg}[Negative PCA weights with substitutes]
\label{eg:consumer-optim-2}
Suppose the researcher creates a summary index of wealth as in Example \ref{eg:run-wealth} using two goods that are substitutes (e.g., cows and goats).
In this case, it is reasonable to assume that more cows or goats lead to greater wealth.
In addition, the two goods can be negatively correlated because they are substitutes.
A farmer with a given budget who keeps more cows has less left over for goats.

However, if PCA is used to create a summary index for these two goods, there can be a negative weight when they are negatively correlated.
I formalize this via a consumer optimization framework in Appendix \ref{app:consumer-optim-2}.
\end{eg}

\begin{eg}[Negative PCA weights from mutually exclusive binary indicators]
Negative correlation in outcomes arises when a variable is discretized into mutually exclusive binary indicators.
To see this in the running example on wealth (Example~\ref{eg:run-wealth}), let $Y_{i}$ be a categorical variable with support $\{y_0, y_1, \ldots, y_M\}$ that indicates the type of cooling device owned by a household.
If the following binary indicators are created to represent ownership of each type of cooling device $Z_{i,l} = \ind[Y_i = y_l]$ for $l = 1, \ldots, M$, then $\cov[Z_{i,l}, Z_{i,k}] = - \bE[Z_{i,l}]\bE[Z_{i,k}] \leq 0$ for $l \neq k$. See \citet{kolenikovangeles2009riw} for more discussion.
\end{eg}

Finally, the class of weights $\cWunit$ in \eqref{eq:weight-pca} requires the weights to be of unit length.
Beyond possibly being negative, the PCA weights typically do not sum to one.
Hence, even if $\be_j = \obe$ for all $j = 1, \ldots, \dy$, the aggregated treatment effect $\bmw'\bmbe$ can be different from $\obe$.

\subsubsection{\pres} \label{sec:pca-precise}
PCA is often motivated on the grounds that it improves precision or that it is a tool for dimension reduction.
Although these statements about the outcomes follow from the definition of PCA, I show that they do not necessarily translate into a precise aggregated treatment effect.

The analysis below shows that running PCA on the correlation matrix of the outcomes does not necessarily yield an aggregated treatment effect with low variance.
For simplicity, I assume a binary treatment and let the variance matrices be known.
In practice, the variance matrices have to be estimated, and the weights are computed by running PCA on the estimated correlation matrix.
Computing the correct standard errors of the aggregated treatment effect requires accounting for these estimated weights; see Appendix \ref{app:gen-out} for details.

\begin{figure}[!t]
    \centering
    \caption{A two-outcome numerical example that evaluates the precision of PCA.}
    \label{fig:pca-1}
		\includegraphics[scale=1]{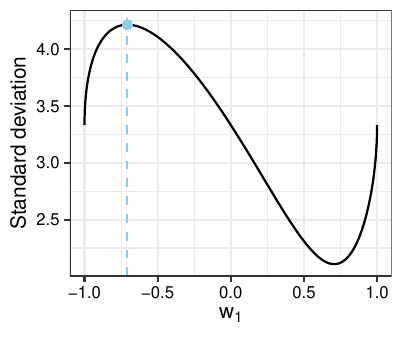}
  	\mynote{
	The figure plots the asymptotic standard deviation of the aggregated treatment effect against $w_1$.
	See Example \ref{eg:binary-te-1} for the data-generating process (DGP).
	The vertical dashed line is the choice made by PCA.}
\end{figure}

\begin{eg}
\label{eg:binary-te-1}
In the setup of Example~\ref{eg:run-wealth}, let $\bmY_i$ be the asset outcomes, standardized as in Assumption~\ref{assu:outcome} so that $\bmSY$ is their correlation matrix.
Let $Y_{i,j} = \xi_j + \be_j D_i + U_{i,j}$ with $\bE[U_{i,j} | D_i] = 0$, where $\xi_j$ is an intercept and $U_{i,j}$ an error term, and let $\hbmbn$ be an estimator of $\bmbe$ with asymptotic variance matrix $\bmSb$ (the closed-form expression is given in \eqref{eq:pre-gen-4} in the appendix).
For a fixed $\bmw \in \cWunit$, the aggregated estimator $\htan \equiv \bmw'\hbmbn$ has asymptotic variance $\si_\tau^2(\bmw) \equiv \bmw'\bmSb\bmw$ by Proposition~\ref{prop:pre-gen-1}.

PCA chooses $\bmw$ to maximize the variance of the aggregated outcome $\bmw'\bmSY\bmw$, but precision requires minimizing the variance of the aggregated treatment effect $\si_\tau^2(\bmw)$.
In Proposition~\ref{eg:prop:gen-precision}\ref{prop:pca-eq-1}, I show that these coincide only under a strong condition: the direction along which the outcomes vary the most must also be the one along which the estimated treatment effects vary the least.
If the errors are homoskedastic and $\bmbe = \zeroq$, then $\bmSb = \frac{1}{p_D(1 - p_D)}\bmSY$ by \eqref{eq:pre-gen-8}, where $p_D \equiv \bP[D_i = 1]$.
In this case, PCA even selects the weights that minimize precision.

Figure~\ref{fig:pca-1} quantifies this case when $\dy = 2$ with $\corr[Y_{i,1}, Y_{i,2}] = -0.6$ and $p_D = 0.1$.
The figure plots the asymptotic standard deviation of the aggregated treatment effect against $w_1$, with $w_2 = \sqrt{1 - w_1^2} \geq 0$.
The leading eigenvector of $\bmSY$ is $(-\frac{1}{\sqrt{2}}, \frac{1}{\sqrt{2}})'$ and is at the maximum of the curve.
The asymptotic standard deviation there is twice the smallest value attainable in $\cWunit$.
\end{eg}

Next, consider the MSE in \eqref{eq:mse}.
By an argument analogous to that in Example \ref{eg:binary-te-1}, PCA does not minimize the MSE in general since the MSE depends on $\theta$ through the bias term.

Apart from variance, PCA also requires strong conditions to maximize the $t$-statistic.
This contrasts with how PCA is sometimes motivated as a way to improve power (e.g., \citealp{huangzhang2021aejae}).
The analysis of the $t$-statistic differs from that of the asymptotic variance because the $t$-statistic also involves the point estimate.
I sketch the main idea here and provide the details, including more general cases, in Appendix \ref{app:t-stat} because the analysis requires additional notation.
In the homoskedastic setting of Example~\ref{eg:binary-te-1}, the square of the $t$-statistic has a one-to-one relationship with the $R$-squared.
Thus, maximizing the $R$-squared requires choosing the weights to ``separate'' the means of $\bmw'\bmY_i$ for the treated and control groups as much as possible, relative to the variance of $\bmw'\bmY_i$.
This means $\bmw'\bmbe$ is also important, but PCA on $\bmSY$ does not achieve this goal without additional conditions because the PC1 depends only on $\bmSY$ and ignores $\bmbe$.

\subsubsection{Sensitivity to the sign (identification) constraint}  \label{sec:pca-sign}
The sign identification condition in~\eqref{eq:weight-pca} is arbitrary.
As Table~\ref{tab:sign-pca} shows, common implementations impose different constraints to achieve identification.
This matters because the signs of the weights directly determine the sign of the aggregated treatment effect (see \eqref{eq:tau-1}).
Changing the constraint can therefore flip the sign of the estimated effect, as the following example shows.

\begin{figure}[!t]
    \centering
    \caption{The PCA-aggregated treatment effects replicated using R.}
    \label{fig:pca-issue-new-1}
    \includegraphics[scale=1]{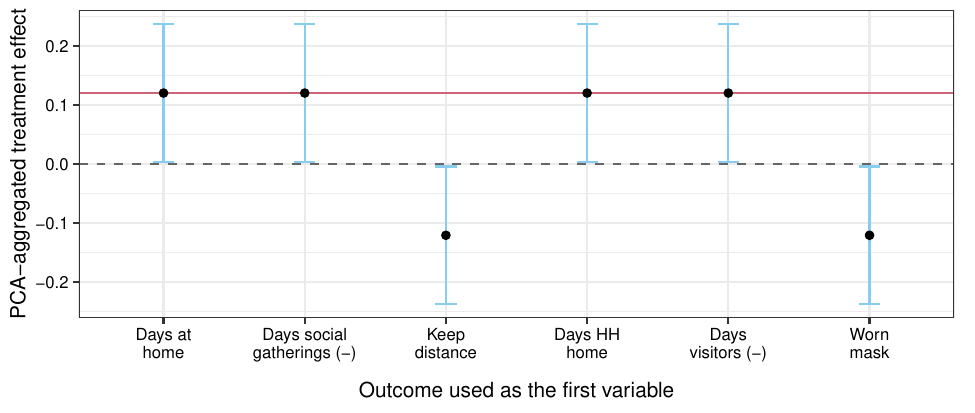}
    \mynote{
		The $x$-axis shows which variable is the first variable passed to the function \texttt{princomp} in R.
		Outcomes marked with ($-$) are reverse-coded, so a higher value means more social distancing.
		``HH'' stands for household members.
		The points show the PCA-aggregated treatment effects.
		The blue bars show the 99\% confidence intervals with standard errors clustered at the household level as in the original Stata replication code.
		The red horizontal line shows the PCA-aggregated treatment effect obtained from Stata.}
\end{figure}

\revisiteg{\ref{eg:sd}}{%
\citet{karlanetal2026jde} generate a social distancing index using PCA.
In their literature review, the authors note that there is mixed evidence on whether cash transfers have a positive or negative effect on social distancing.

I reanalyze the part of their analysis in which they find a positive contemporaneous impact of the cash transfer program on the social distancing index that is significant at the 1\% level.
I replicate the analysis in R to generate a PCA index and then regress it on the treatment variable and controls; see Section~\ref{sec:sdeg} for details.
As reviewed in Table \ref{tab:sign-pca}, the sign chosen by the \texttt{princomp} function depends on what the ``first variable'' is.
Figure~\ref{fig:pca-issue-new-1} reports the estimated effect on the PCA social distancing index when each of the six outcomes is used as the first variable via the \texttt{princomp} function.
It shows that positive or negative effects are possible when different outcomes are used as the first variable.
Each effect is significant at the 1\% level when standard errors are clustered at the household level as in the authors' analysis.
}

When all treatment effects and PCA weights have the same sign, it might be clear what the ``correct'' choice of the sign identification constraint is.
However, PCA should not be used if the sign of the treatment effect is a priori ambiguous and the PCA weights have mixed signs, which is exactly the situation in the above example.

\subsubsection{Nonstandard asymptotic distribution} \label{sec:pca-dist}
The PCA approach is further subject to a weak identification problem.
This can make conventional standard errors for the aggregated treatment effect invalid.
The source is the sign identification constraint in~\eqref{eq:weight-pca}.
If $\bmc'\wpca$ is close to zero, this creates a weak identification issue.
Intuitively, this is because it becomes more difficult to uniquely identify the solution than in the strong identification case, where $\bmc'\wpca$ is ``far away'' from zero.

The proposition below summarizes this observation and its implications for inference, using a drifting sequence to model the case in which $\bmc'\wpca$ is close to zero, as \citet{staigerstock1997ecta} do for weak instruments.
To analyze the large-sample properties, I require that the estimators of $\bmbe$ and $\bmSY$ be consistent, that a suitable central limit theorem apply, and that a degeneracy condition be ruled out.
These conditions are stated as Assumptions~\ref{assu:limit-dist} and~\ref{assu:pca:nondegen} in the appendix.

\begin{prop}
\label{prop:pca-main-weak}
Let Assumptions~\ref{assu:outcome} and~\ref{assu:eval} hold.
For each $\delta > 0$, let $\{\bm{\bmSi}_{Y,n}\}$ be a sequence of correlation
matrices with $\bm{\bmSi}_{Y,n} \longrightarrow \bmSY$, let $\wpcan$
solve~\eqref{eq:pca-problem} with $\bmSY$ replaced by $\bm{\bmSi}_{Y,n}$, and
suppose that $\bmc'\wpcan = \frac{\delta}{\sqrt{n}}$ and $\bvec_1'\bmbe \neq 0$.
Let Assumptions~\ref{assu:limit-dist} and~\ref{assu:pca:nondegen} hold with $\bmOm = \bmSi_{Y,n}$ and $\hbmOn = \widehat{\bmSi}_{Y,n}$.
Let $\hwpca$ solve~\eqref{eq:pca-problem} with $\bmSY$ replaced by $\widehat{\bmSi}_{Y,n}$,
and define $\tau_{\mathrm{pca},n} \equiv \wpcan'\bmbe$ and
$\htpca \equiv \hwpca'\hbmbn$.
Consider $\cC = [t_{\mathrm{lb}}, t_{\mathrm{ub}}]$, where
$-\infty < t_{\mathrm{lb}} < t_{\mathrm{ub}} < \infty$.
Then,
\[
	\limsup_{\delta \downarrow 0} \
	\limsup_{n\to\infty} \
	\bP
		[ \sqrt{n}(\htpca - \tau_{\mathrm{pca},n}) \in \cC ]
	\leq
	0.5.
\]
\end{prop}

The proof uses eigenvector perturbation (see \citealp{stewartsun1990bk}, for a comprehensive reference) because the PCA problem is not convex and eigenvectors do not generally have a closed-form expression except in special cases.
With strong identification, it is still possible to obtain asymptotic normality.
I show the details for the strong identification case in Appendix~\ref{app:gen-out}.
I illustrate the weak identification issue in Appendix~\ref{app:weak-sim} with a calibrated simulation in which the sampling distribution of $\htpca$ becomes bimodal as $\bmc'\wpca$ approaches zero, which suggests standard methods lead to incorrect inference.

Having repeated leading eigenvalues (i.e., relaxing Assumption \ref{assu:eval}) can lead to a similar identification issue because the leading eigenvector is no longer unique.

\subsection{\anderson} \label{sec:anderson-main}
\citet{anderson2008jasa} creates a summary index as a weighted average of standardized outcomes using the inverse of the covariance matrix of the outcomes.
Following \citet[pages 1485 and 1494]{anderson2008jasa}, I impose Assumption~\ref{assu:outcome} specialized to ``control group standardization,'' i.e., each outcome is demeaned and divided by its control group standard deviation.
The weights are given by
\begin{align}
	\label{eq:anderson-w}
	\wivm \equiv \frac{\bmSY^{-1}\oneq}{\oneq'\bmSY^{-1}\oneq},
\end{align}
where $\bmSY$ now denotes the covariance matrix of the outcomes standardized using the control group outcomes as described above.
The weights in~\eqref{eq:anderson-w} can be obtained by solving
\begin{align}
	\label{eq:framework-simp-ivm}
	\min_{\bmw \in \cWsto} \ \bmw'\bmSY\bmw,
\end{align}
where the class of weights whose entries sum to one (``sto'') is
\begin{align}
	\label{eq:weight-ivm}
	\cWsto
	\equiv
	\{ \bmw \in \bR^{\dy} : \bmw'\oneq = 1\},
\end{align}
but the weights are not required to be nonnegative.

\citet[page 1485]{anderson2008jasa} notes that using the covariance matrix in weighting ensures that highly correlated outcomes receive less weight and that the resulting weight is the efficient GLS estimator \citep{obrien1984biom}.
The connection with GLS can be found in the theorem on page 1082 of \citet{obrien1984biom}, which states that if $\bmY$ is a vector of $\dy$ unbiased estimators of a scalar parameter and the vector has covariance matrix $\bmSY$, then the corresponding best linear unbiased estimator is given by $\wivm'\bmY$.

\subsubsection{\inter} \label{sec:anderson-interpret}
The class of weights~\eqref{eq:weight-ivm} used by IVM leaves the signs of the individual weights unrestricted.
Although the denominator of \eqref{eq:anderson-w} is positive ($\bmSY$ is positive definite), the numerator sums the rows of $\bmSY^{-1}$, and its entries can be negative.
Hence, the weights can be negative.

In Appendix \ref{app:ivm-pos}, I derive a sufficient condition on $\bmSY$ that ensures the components of the weights $\wivm$ are positive.
Intuitively, the condition states that the weights are positive provided that no outcome is too ``strongly'' correlated with the others.

\subsubsection{\pres} \label{sec:anderson-precise}
Minimizing the objective in~\eqref{eq:framework-simp-ivm} is not necessarily the same as minimizing the asymptotic variance of the aggregated treatment effect because that objective focuses on the variance of the outcomes.
It is also not equivalent to minimizing the MSE in~\eqref{eq:mse} without additional assumptions on $\theta$.

I revisit the binary treatment example.
As in Section~\ref{sec:pca-precise}, I assume for simplicity that the variance matrices are known.
In practice, such matrices have to be estimated, and \eqref{eq:anderson-w} is computed using the estimate of $\bmSY$.
I defer the discussion of estimated weights and of how to correctly compute standard errors to Appendix~\ref{app:gen-out}.

\begin{eg}
\label{eg:binary-te-2}
Consider the binary treatment setup of Example \ref{eg:binary-te-1}, but with each outcome divided by its control group standard deviation as in \citet{anderson2008jasa}.
The derivation of the asymptotic variance carries over with this different standardization.

Minimizing $\bmw'\bmSY\bmw$ and minimizing $\bmw'\bmSb\bmw$ over $\bmw \in \cWsto$ do not necessarily lead to the same solution.
They coincide only under restrictive conditions, such as when the errors are homoskedastic and $\bmbe = \zeroq$.
In this case, $\bmSb$ is a scalar multiple of $\bmSY$ as in Example~\ref{eg:binary-te-1}.
However, this may not be true under heteroskedasticity or in a more general setup.
I show the details in Appendix~\ref{app:pre-gen}.

Next, consider the MSE in \eqref{eq:mse}.
Even if the variance of $\hbmbn$ is known, the MSE still depends on $\theta$, so IVM does not minimize the MSE without additional assumptions.
\end{eg}

As in the PCA analysis, $\wivm$ does not necessarily maximize the $t$-statistic or improve power.
The details on the $t$-statistic and IVM can be found in Appendix~\ref{app:t-stat}.

\subsection{\kling} \label{sec:kling-main}
\citet{klingetal2007ecta} create a summary index by taking an equally weighted average of the outcomes after each outcome is demeaned using the control group's mean and divided by the standard deviation of the corresponding outcome in the control group.
Hence, the weight is
\[
	\wsa
	\equiv
	\left( \frac{1}{\dy}, \ldots, \frac{1}{\dy} \right)'
	=
	\frac{1}{\dy} \oneq.
\]
Since $\wsa$ is strictly positive, SA avoids the interpretability issues that negative weights cause in PCA and IVM.

\citet{klingliebman2004wp} and the appendix of \citet{klingetal2007ecta} discuss the rationale behind dividing by the control group standard deviations.
\citet[footnote 13]{klingliebman2004wp} explain that this is to compare effect sizes relative to the control group, which is related to Glass's delta (see, for instance, \citealp{glassetal1981bk}).
They compute the sample variance of the weighted average of treatment effects using seemingly unrelated regression to account for the correlation in $\hbmbn$.

\subsubsection{\pres} \label{sec:kling-precise}
Equal weighting minimizes the asymptotic variance of $\htan$ if the asymptotic variance matrix of $\hbmbn$ is an equicorrelation matrix.
This standard result is summarized below.

\begin{prop}
\label{prop:kling-var-1}
Let $\obmSi$ be an equicorrelation matrix, i.e., $\obmSi$ has diagonal entries equal to 1 and off-diagonal entries all equal to $\overline\rho \in (-\frac{1}{\dy - 1}, 1)$.
Then, $\wsa = \argmin_{\bmw \in \cWsto} \bmw'\obmSi\bmw$.
\end{prop}
Requiring the entries of $\bmw$ to sum to one in Proposition~\ref{prop:kling-var-1} is a natural minimal normalization assumption ensuring that $\wsa \in \cWsto$ as in \eqref{eq:weight-ivm}.
The correlation $\overline\rho$ is restricted to keep $\obmSi$ positive definite (see, for instance, \citealp[page 241]{abadirmagnus2005matrix}).

Although these conditions show when equal weighting is optimal, using the actual covariance matrix to weight the outcomes can increase precision.
I demonstrate this in Appendix \ref{app:equal-v-eg} with a stylized example in which duplicated variables serve as a limiting case of highly correlated outcomes.

As in the previous analysis, $\wsa$ does not necessarily maximize the $t$-statistic or improve power.
The details on the connection between the $t$-statistic and SA can be found in Appendix \ref{app:t-stat}.

\section{Choosing weights to minimize the variance of the aggregated treatment effect}  \label{sec:var-min-weight}

In this section, I describe a simple method for choosing weights that satisfy the interpretability and precision criteria in Section~\ref{sec:criteria} when the treatment effects on different outcomes are unbiased, i.e., $\be_j = \theta$ for all $j = 1, \ldots, \dy$.
In the running example on wealth (Example~\ref{eg:run-wealth}), this is reasonable when the treatment effect on each asset captures $\theta$ exactly.
In this case, the bias term in the MSE criterion~\eqref{eq:mse} is zero for every $\bmw \in \cWcvx$, so being precise means minimizing the variance of the aggregated treatment effect.

Let $\hbmbn$ be an estimator of $\bmbe$.
Let $\bmSi$ be the asymptotic variance matrix of the treatment effects estimator, where $\bmSi$ is positive definite.
Let $\hbmSn$ be a consistent estimator of $\bmSi$.
The weights $\hwvmc \in \cWcvx$ that minimize the estimated variance of the aggregated treatment effect over $\cWcvx$, i.e., the class of nonnegative weights summing to one defined in~\eqref{eq:convex-w}, can be obtained by solving the following problem:
\begin{align}
	\label{eq:vbmin-w}
	\min_{\bmw \in \cWcvx} \ \bmw'\hbmSn\bmw,
\end{align}
where ``vmc'' is the abbreviation for variance minimization subject to convex weights.
Let $\wvmc$ be the solution to~\eqref{eq:vbmin-w} when $\hbmSn$ is replaced by $\bmSi$.
The following proposition summarizes the large-sample properties of the aggregated treatment effect estimator.

\begin{prop}
\label{prop:vmc-b0-d}
Let Assumption~\ref{assu:outcome} hold.
Suppose $\be_j = \theta$ for all $j = 1, \ldots, \dy$,
$\sn(\hbmbn - \bmbe) \Dt \scN(\zeroq, \bmSi)$, $\hbmSn \Pt \bmSi$, and $\bmSi$ is positive definite.
Then, $\hwvmc \Pt \wvmc$ and
$\sn( \hwvmc'\hbmbn - \wvmc'\bmbe ) \Dt \scN(0, \wvmc'\bmSi\wvmc)$.
\end{prop}

The proposition suggests that, under the given assumptions, a $100(1 - \alpha)\%$ confidence interval for the aggregated treatment effect $\wvmc'\bmbe$ can be constructed as $[\hwvmc'\hbmbn \pm z_{1-\frac{\alpha}{2}}\sqrt{\frac{1}{n}\hwvmc'\hbmSn\hwvmc}]$, where $z_{1 - \frac{\alpha}{2}}$ is the $(1-\frac{\alpha}{2})$ quantile of the standard normal distribution.
Although the weights are estimated, the confidence interval does not require an additional term to adjust for the estimated weights in conducting inference on the aggregated treatment effect because $\be_j = \theta$ for all $j = 1, \ldots, \dy$ is assumed and the weights sum to one.
Hence, $\sn( \hwvmc'\hbmbn - \wvmc'\bmbe ) = \sn( \hwvmc'\hbmbn - \theta) = \hwvmc' \sn(\hbmbn - \bmbe)$.

Although the limiting distribution of the aggregated treatment effect in Proposition~\ref{prop:vmc-b0-d} is normal, the estimated weights need not be, because the weights are constrained to be nonnegative.
I show the details in Appendix~\ref{sec:inference-b0}.

Choosing the weights by~\eqref{eq:vbmin-w} avoids a drawback shared by PCA, SA, and IVM, namely, that they do not target the precision of the aggregated treatment effect.
In addition, the weights in~\eqref{eq:vbmin-w} are nonnegative by construction, which avoids the interpretability issues of PCA and IVM.

However, the approach in~\eqref{eq:vbmin-w} assumes $\be_j = \theta$ for all $j = 1, \ldots, \dy$, which can be a strong assumption in practice.
This is because the outcomes have varying quality in capturing the underlying variable of interest.
Hence, the bias term in the MSE criterion~\eqref{eq:mse} is no longer zero and depends on the choice of the weights, so minimizing the variance is not the right objective.
In Section~\ref{sec:decision}, I take a statistical decision-theoretic approach and show how to choose the weights by trading off bias against variance.

\section{A statistical decision approach}		\label{sec:decision}
In this section, I develop a statistical decision framework to aggregate treatment effects across multiple outcomes.
The framework provides a principled approach that uses the mean-squared error introduced in Section~\ref{sec:criteria} as its objective, while ensuring interpretable weights.
It also allows researchers to incorporate prior information about the relative quality of the outcomes.

To develop the theory, I introduce the decision problem in Section \ref{sec:decision-problem}.
Section \ref{sec:param-space} discusses the parameter space.
Sections~\ref{sec:minimax} and \ref{sec:adapt} then develop two ways to choose the weights, the minimax and adaptive approaches, respectively.
Section \ref{sec:inference} shows how to conduct inference.
Section \ref{sec:implementation} presents an implementation algorithm and summarizes the procedure.

\subsection{The decision problem} \label{sec:decision-problem}
This subsection introduces the researcher's problem and defines the risk function.
Assume the researcher observes a vector of estimated treatment effects $\hbmbe$ that follows
\begin{equation}
	\label{eq:decision-1}
	\hbmbe
	\sim \scN(\bmbe, \bmSi),
\end{equation}
where $\bmbe \equiv (\be_1, \ldots, \be_{\dy})' \in \bR^{\dy}$ is the vector of population means and the covariance matrix $\bmSi$ is known.
Following Assumption \ref{assu:outcome}, I assume that the vector of estimated treatment effects $\hbmbe$ is obtained from the suitably standardized outcomes.
The normal distribution in \eqref{eq:decision-1} can be justified by a large-sample approximation.
I also assume that the researcher reports $\hta = \bmw'\hbmbe$, a weighted average of treatment effects on the observed outcomes, as in \eqref{eq:tau-1} with weights $\bmw \in \cWcvx$ that are nonnegative and sum to one as in~\eqref{eq:convex-w}.
Focusing on the class of weights $\cWcvx$ avoids the interpretability issues discussed in Section \ref{sec:criteria-inter}.

The following assumption ensures that the covariance matrix $\bmSi$ in \eqref{eq:decision-1} is positive definite.
\begin{assu}
\label{assu:sig-pd}
$\bmSi$ has real-valued eigenvalues bounded below by $\la_{\mathrm{lb}} > 0$ and above by $\la_{\mathrm{ub}} < \infty$, where $\la_{\mathrm{ub}} \geq \la_{\mathrm{lb}}$.
\end{assu}

To measure the quality of $\bmbe$ in capturing $\theta$, define
\begin{equation}
	\label{eq:decision-bias-1}
	\bmb \equiv \bmbe - \theta \oneq \in \bR^{\dy}.
\end{equation}
In the running example (Example \ref{eg:run-wealth}), the $j$-th component of $\bmb$ can be interpreted as how well the treatment effect on asset $j$ measures the treatment effect on wealth.
For example, if the first outcome is the ownership of goats, then $b_1$ measures how well the treatment effect on the ownership of goats captures that of wealth.

I use the squared loss function to model the performance of the estimator $\hta$ relative to $\theta$ as in Section \ref{sec:criteria-pres}.
Since $\hta$ is a linear combination of $\hbmbe$ in \eqref{eq:tau-1}, I can write the loss function as follows:
\begin{equation}
	\label{eq:decision-case1-2}
	L(\bmw, \hbmbe, \theta)
	\equiv
	(\hta - \theta)^2
	=
	(\bmw'\hbmbe - \theta)^2.
\end{equation}
The loss function is written as a function of $\bmw$ because the researcher's action is to choose the weights $\bmw$ when forming the estimator $\hta$.

The risk function of the estimator $\hta$ is the expectation of the loss function \eqref{eq:decision-case1-2} with respect to the distribution of $\hbmbe$ in \eqref{eq:decision-1}.
For any $\bmb \in \bR^{\dy}$ and $\bmw \in \cWcvx$, the risk function is defined as follows:
\begin{align}
	R(\bmw, \bmb)
	& \equiv
	\bE[ L(\bmw, \hbmbe, \theta)]
	= \bE[(\bmw'\hbmbe - \bmw'\bmbe)^2 + (\bmw'\bmbe - \theta)^2]
	= \bmw'\bmSi\bmw + (\bmw'\bmb)^2,
	\label{eq:decision-case1-3}
\end{align}
where the equalities follow from \eqref{eq:decision-1}, the definition of $\bmb$ in \eqref{eq:decision-bias-1}, and the fact that the entries of $\bmw$ sum to one.

\subsection{Parameter space and maximum risk} \label{sec:param-space}
In this subsection, I introduce several ways to model and motivate the parameter space of $\bmb$.
In the running example on wealth, this parameter space is needed because how well each asset's treatment effect captures the effect on wealth is unknown, although the covariance matrix $\bmSi$ in \eqref{eq:decision-case1-3} is known (or is consistently estimable).
Researchers may also have prior information about the relative quality of treatment effects across assets.

Let $B \geq 0$ be a maximum level of misspecification chosen by the researcher, and let $\cS(B)$ be a nonempty and compact parameter space for $\bmb$.
I describe three examples showing that $\cS(B)$ flexibly incorporates the researcher's restrictions or prior information on treatment effects.
More details can be found in Appendix \ref{app:param}.

\begin{eg}[Bounds on the overall magnitude]
\label{eg:param-b}
I start with the case in which the researcher places an upper bound on $\bmb$ in $\cS(B)$ using the $\ell_p$-norm of $\bmb$ for $p \geq 1$ (as in \citealp{armstrongkolesar2021qe}).
In this case, the parameter space can be written as
\begin{equation}
	\label{eq:param-b-1}
	\cS(B) = \{ \bmb \in \bR^{\dy} : \|\bmb\|_p \leq B\}.
\end{equation}

In the running example on wealth, this restriction places an upper bound $B$ on the quality of the treatment effect across different assets.
If $B = 0$, then $\be_j = \theta$ for each $j = 1, \ldots, \dy$ (i.e., the treatment effect on each asset equals the treatment effect on wealth as in Section~\ref{sec:var-min-weight}).
If $B > 0$, choosing $p = \infty$ can be interpreted as setting the same bound on each $b_j$.
Choosing $p \in [1, \infty)$  bounds the misspecification jointly, so it allows some treatment effects to be more important or to capture $\theta$ better than others.
\end{eg}

\begin{eg}[Shape restrictions]
\label{eg:param-shape}
Researchers can impose shape restrictions when they believe the treatment effects on some outcomes are more important than those on others.

In the entrepreneurial spirit index from Example \ref{eg:bks}, \citet{bruhnetal2018jpe} note that two of the outcomes might not be directly affected by the consulting service but might instead be driven by business improvements.
Thus, they construct another entrepreneurial spirit index that excludes these two outcomes.

In this empirical example, shape restrictions that make these two outcomes less important can be an informative alternative to dropping them directly.
Let $\cJ_0$ index the two outcomes they exclude, and $\cJ_1$ index the remaining outcomes.
A natural shape restriction here is $\kappa | b_{j_1} | \leq |b_{j_0}|$ for $j_1 \in \cJ_1$ and $j_0 \in \cJ_0$, where $\kappa \geq 0$ controls the relative importance of the outcomes in $\cJ_0$ and $\cJ_1$.
Setting $\kappa > 1$ can be interpreted as the outcomes in $\cJ_1$ being more ``important.''

Together with a bound on $\bmb$ as in Example \ref{eg:param-b}, the above can be described as
\[
	\cS(B) = \{ \bmb \in \bR^{\dy} : \|\bmb\|_p \leq B, \bmQ|\bmb| \leq \zero_l\},
\]
where $\bmQ \in \bR^{l\times \dy}$ and $|\bmb|$ denotes the componentwise absolute value of $\bmb$.
Other types of constraints can be used, such as $\bmQ\bmb \leq \zero_l$.
\end{eg}

\begin{eg}[Subjective weights and communication]
\label{eg:param-comm}
Researchers may also worry about how the weighted average of treatment effects $\hta$ will be communicated to readers.
In the running example (Example \ref{eg:run-wealth}), readers may think that the treatment effect of the cash transfer program on wealth $\theta$ equals the treatment effect on the weighted average of assets, but have subjective views on what the weights should be for each asset.

This communication problem can be described in terms of a parameter space for $\bmb$.
Suppose $\theta = \bmga'\bmbe$ for some known $\bmga \in \cWcvx$ and $\|\bmb\|_p \leq B$ as in Example~\ref{eg:param-b}.
Then, the quality of treatment effects can be written as $\bmb = \bmbe - (\bmga'\bmbe)\oneq$.
In Appendix~\ref{sec:param-comm}, I show that this can be studied via a shape restriction in the parameter space.
In addition to a known $\bmga$, I show how to allow for an unknown $\bmga$ or ambiguity around a given $\bmga$.
\end{eg}

Given the parameter space $\cS(B)$, the maximum risk can be evaluated as the maximum of \eqref{eq:decision-case1-3} over $\bmb \in \cS(B)$, i.e.,
\begin{equation}
	\label{eq:decision-case1-4}
	\Rmax(B, \bmw)
	\equiv
	\max_{\bmb \in \cS(B)}
	R(\bmw, \bmb)
	=
	\bmw'\bmSi\bmw + \max_{\bmb \in \cS(B)} (\bmw'\bmb)^2,
\end{equation}
where the equality holds because the variance component is independent of $\bmb$.
By Assumption \ref{assu:sig-pd}, the variance component $V(\bmw) \equiv \bmw'\bmSi\bmw$ is strictly convex in $\bmw$ because $\bmSi$ is positive definite.
Since $\cS(B)$ is nonempty, the maximum squared bias $\oM(B, \bmw) \equiv \max_{\bmb \in \cS(B)} (\bmw'\bmb)^2$ is convex in $\bmw$ for a given $B\geq0$ because it is a maximum of convex functions \citep[page 81]{boydbandenberghe2004bk}.

\subsection{Choosing weights using the minimax approach} \label{sec:minimax}
With a given parameter space, a natural approach is to choose the weights that minimize the worst-case risk over the parameter space.
This can be done using the minimax criterion.
In the running example on wealth, using the minimax approach means choosing the weight for each asset by minimizing the worst-case MSE over all possible levels of treatment effect quality $\bmb \in \cS(B)$.

The weights can be obtained by solving the following minimax problem:
\begin{equation}
	\label{eq:minimax-case1-1}
	\Rmm(B)
	\equiv
	\min_{\bmw \in \cWcvx} \ \Rmax(B, \bmw)
	=
	\min_{\bmw \in \cWcvx} \ \left[ V(\bmw) + \oM(B, \bmw)\right],
\end{equation}
where the maximum risk $\Rmax(B,\bmw)$ is defined in~\eqref{eq:decision-case1-4}.
The quantity $\Rmm(B)$ is referred to as the minimax risk.
Suppose $\cS(B)$ is bounded and nonempty.
As discussed in Section \ref{sec:param-space}, $V(\bmw)$ is strictly convex in $\bmw$ and $\oM(B, \bmw)$ is convex in $\bmw$.
Hence, the objective of \eqref{eq:minimax-case1-1} is strictly convex.
It follows that \eqref{eq:minimax-case1-1} has a unique optimal solution.
I denote the optimal solution to the minimax problem \eqref{eq:minimax-case1-1} as follows:
\begin{equation}
	\label{eq:minimax-case1-1soln}
	\bmwsb = \argmin_{\bmw \in \cWcvx} \ \Rmax(B, \bmw).
\end{equation}
The expression for $\oM(B, \bmw)$ depends on the parameter space.
The optimal solution can be computed via convex optimization; details are provided in Appendix \ref{sec:comp-minimax}.

I discuss some special cases of the minimax problem \eqref{eq:minimax-case1-1} and an example with two outcomes to build intuition.

\begin{re}[Variance minimization when $B = 0$]
\label{re:minimax-b0}
Consider $B = 0$ so that $\be_j = \theta$ for each $j = 1, \ldots, \dy$.
This leads to $\oM(0, \bmw) = 0$, and the objective of the minimax problem~\eqref{eq:minimax-case1-1} reduces to $V(\bmw)$.
Hence, the optimal weight is the variance-minimizing weight that is nonnegative and sums to one, as in Section~\ref{sec:var-min-weight}.
\end{re}

\begin{re}[Connection with matrix regularization]
\label{re:mat-reg}
Consider using the $\ell_2$-norm for Example~\ref{eg:param-b}, i.e., $\cS(B) = \{ \bmb \in \bR^{\dy} : \|\bmb\|_2 \leq B\}$.
Then, $\oM(B, \bmw) = B^2 \|\bmw\|^2_2$.
The objective of \eqref{eq:minimax-case1-1} becomes $\Rmax(B, \bmw) = \bmw'(\bmSi + B^2\bmI_{\dy}) \bmw$, where $\bmI_{\dy}$ is the identity matrix of size $\dy \times \dy$.
Thus, the optimal weight for the minimax problem can be viewed as the minimizer of the quadratic form in the regularized matrix $\bmSi + B^2\bmI_{\dy}$ subject to $\bmw \in \cWcvx$.
\end{re}

\begin{re}[Connection with standardized averaging]
\label{re:minimax-binf}
Suppose $B \longrightarrow \infty$ in Example~\ref{eg:param-b} with $p \in (1, \infty)$.
Then, the solution satisfies (see Corollary~\ref{cor:limit-minimax} in the appendix)
\[
	\lim_{B \to \infty} \bmwsb = \frac{1}{\dy}\oneq.
\]
Thus, the limiting weights are similar to those of SA, which puts equal weight on each treatment effect (although SA divides outcomes by the standard deviation from the control sample as in Section \ref{sec:kling-main}).
This provides a statistical decision-theoretic justification for using equal weights as the optimal solution when $B \longrightarrow \infty$.
\end{re}

\begin{eg} \label{eg:run-1}
Suppose the researcher observes two assets as in Example \ref{eg:run-wealth}, so that $\hbmbe = (\hbe_{1}, \hbe_2)'$ and $\bmbe = (\be_1, \be_2)'$.
Assume that~\eqref{eq:decision-1} has the following form, where $\rho \in (-1, 1)$ and $\si_2 > 0$:
\[
	\begin{pmatrix}
		\hbe_{1}		\\	\hbe_{2}
	\end{pmatrix}
	\sim
	\scN
	\left(
		\begin{pmatrix}
			\be_1	\\	\be_2
		\end{pmatrix},
		\begin{pmatrix}
			1 	& 	\rho \si_2	\\
			\rho\si_2  & \si_2^2
		\end{pmatrix}
	\right).
\]
Suppose the Euclidean norm is used to bound $\bmb \equiv (b_1, b_2)'$ in $\cS(B)$ as in Example \ref{eg:param-b}.
Since $w_1 + w_2 = 1$, I focus on characterizing the optimal first weight $w_1^\star(B)$.
The optimal solution $w_1^\star(B)$ for the minimax problem with two outcomes is
\begin{equation}
	\label{eq:run-1-2}
	w_1^\star(B)
	=
	\begin{cases}
		0
		& \text{ if $\si_2^2 - \rho \si_2 \leq - B^2$,}	\\
		1
		& \text{ if $1 - \rho \si_2 \leq - B^2$,}	\\
		\frac{\si_2^2 - \rho \si_2 + B^2}{1 + \si_2^2 - 2\rho\si_2 + 2B^2}
		& \text{ otherwise.}
	\end{cases}
\end{equation}

The optimal weight in \eqref{eq:run-1-2} depends on $B$, $\si_2^2$, and $\rho$.
Figure~\ref{fig:egminimax-1} plots the optimal weight on the first treatment effect as a function of $B$ using $\si_2 = 0.2$ and $\rho = 0.3$.
\begin{figure}[!t]
	\centering
	\caption{Illustration of the optimal weight $w_1^\star(B)$ in \eqref{eq:run-1-2}.}
	\label{fig:egminimax-1}
	\includegraphics[scale=1]{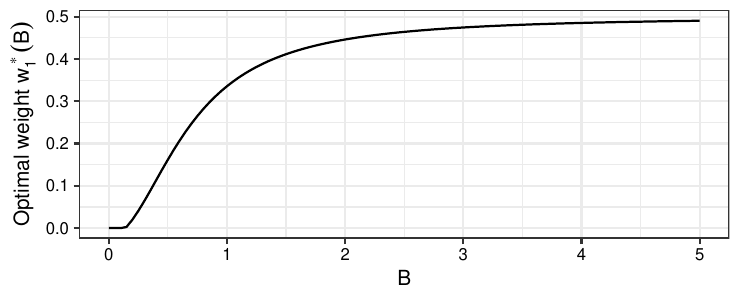}
    \mynote{
    The figure plots $w_1^\star(B)$ from \eqref{eq:run-1-2} against $B$.
    The DGP uses $\si_2 = 0.2$ and $\rho = 0.3$.}
\end{figure}
A few observations are as follows.
First, consider $B = 0$.
The optimal weight is chosen to minimize the variance of the aggregated treatment effect, as explained in Remark \ref{re:minimax-b0}.
If the second treatment effect is sufficiently precise (i.e., $\si_2 \leq \rho < 1$, which requires $\rho > 0$), then it is optimal to set $w_1^\star(0) = 0$.
This is because both treatment effects are unbiased relative to the treatment effect on wealth when $B = 0$, so only the variance matters.
The numerical example in Figure~\ref{fig:egminimax-1} shows that indeed $w_1^\star(0) = 0$.

Second, as $B$ increases, \eqref{eq:run-1-2} suggests that it may not be optimal to put all the weight on the second treatment effect even if $\si_2 \leq \rho$.
This behavior depends on the correlation between the two treatment effects and also reflects the possibility that the second treatment effect may capture the treatment effect on wealth less well when $B > 0$.

Third, as $B \longrightarrow \infty$, $w_1^\star(B) \longrightarrow \frac{1}{2}$.
This reflects that researchers want to be agnostic and put equal weights on both treatment effects as in Remark \ref{re:minimax-binf}.
\end{eg}

\subsection{Choosing weights using the adaptive approach} \label{sec:adapt}

Solving the minimax problem in Section \ref{sec:minimax} requires the researcher to choose a value of $B \geq 0$.
However, researchers may not want to commit to a specific value of $B$, but may be willing to specify a set $\cB \subseteq \bR$ such that $B \in \cB$.
In the running example on wealth, researchers may want to choose asset weights that perform well across different bounds on the quality of treatment effects.
In particular, researchers might want to find the weights that are a middle ground between variance minimization ($B = 0$) and a large $B$.
To avoid committing to weights that are optimal for a specific value of $B$, I consider a recent adaptive framework by \citet{armstrongetal2025ecta}.

Before going into the details, I discuss two differences between my setup and that of \citet{armstrongetal2025ecta}.
First, I assume that all treatment effects $\hbmbe$ in \eqref{eq:decision-1} can be misspecified relative to $\theta$, whereas \citet{armstrongetal2025ecta} assume that an unbiased estimator is available.
Allowing all components in $\hbmbe$ to be potentially misspecified relative to $\theta$ is important in aggregating outcomes.
For instance, in the running example (Example \ref{eg:run-wealth}), it is unlikely that the treatment effect on any given asset captures the treatment effect on wealth perfectly.
Second, I focus on the linear estimator \eqref{eq:tau-1}, in which the weights $\bmw$ are restricted to the convex class of weights as in \eqref{eq:convex-w}.
\citet{armstrongetal2025ecta} do not restrict their estimators to this class, and their solution approach is different.
I impose this restriction to prevent interpretability issues in the resulting aggregated treatment effects, as explained in Section~\ref{sec:criteria-inter}.

\subsubsection{Definitions} \label{sec:adapt-review}
I briefly review the relevant definitions from the methodology proposed by \citet{armstrongetal2025ecta}, adapted to my setup, in which all estimators can be biased and $\bmw \in \cWcvx$.
For a given $\bmw \in \cWcvx$ and $B \in \cB$, the adaptation regret is defined as the following ratio:
\begin{equation}
	\label{eq:areg-1}
	A(B, \bmw)
	\equiv
	\frac{\Rmax(B, \bmw)}{\Rmm(B)}
	= \frac{\Rmax(B, \bmw)}{\min_{\tbmw \in \cWcvx} \ \Rmax(B, \tbmw)},
\end{equation}
where the maximum risk $\Rmax(B, \bmw)$ and the minimax risk $\Rmm(B)$ are defined in \eqref{eq:decision-case1-4} and \eqref{eq:minimax-case1-1}, respectively.
For a given $\bmw \in \cWcvx$, the adaptation regret can be interpreted as how much worse $\bmw$ performs relative to an oracle that knows the minimax optimal weight.
In particular, $100[A(B, \bmw) - 1]\%$ equals the percentage increase in the worst-case MSE.

They define $\Amax(\cB, \bmw) \equiv
	\sup_{B \in \cB}
	A(B, \bmw)$
as the worst-case adaptation regret, and
\begin{equation}
	\label{eq:areg-3}
	\Amm(\cB)
	\equiv
	\inf_{\bmw \in \cWcvx}
	\sup_{B \in \cB}
	A(B, \bmw).
\end{equation}
A choice of weights $\bmwa$ is optimally adaptive if
\begin{equation}
	\label{eq:areg-3b}
	\Amm(\cB) = \Amax(\cB, \bmwa).
\end{equation}

\subsubsection{Solving for the optimally adaptive weight} \label{sec:adapt-results}

Solving \eqref{eq:areg-3} directly for the optimally adaptive weight involves two steps.
First, the inner problem finds the worst-case adaptation regret over $B \in \cB$ for each $\bmw \in \cWcvx$.
Second, the outer problem chooses the weight that minimizes the worst-case adaptation regret.

In this subsection, I derive new properties of the adaptation regret for the setup in Section \ref{sec:decision-problem}.
These properties are new because of the differences between my setup and that of \citet{armstrongetal2025ecta}, as discussed at the beginning of Section~\ref{sec:adapt}.

\begin{assu}
\label{assu:convex-m2}
There exists a nonnegative function $m:\bR^{\dy} \longrightarrow \bR$, convex on $\cWcvx$, such that $\oM(B, \bmw) = B^2 m(\bmw)$ for all $B \geq 0$ and $\bmw \in \cWcvx$.
\end{assu}

For instance, in Example \ref{eg:param-b}, $\oM(B, \bmw) = B^2 \|\bmw\|_{\ps}^2$, where the $\ell_{\ps}$-norm is the dual norm of the $\ell_p$-norm and $\frac{1}{p} + \frac{1}{\ps} = 1$ (see, e.g., \citealp[Appendix A.1.6]{boydbandenberghe2004bk}).
In this case, $m(\bmw) = \|\bmw\|^2_{\ps}$.
I show the other cases in Appendix \ref{app:param}.
The following proposition shows a useful property of the adaptation regret.

\begin{prop}
\label{prop:areg-main}
Let Assumptions \ref{assu:outcome}, \ref{assu:sig-pd}, and \ref{assu:convex-m2} hold.
Let $\cS(B)$ be a parameter space that has a nonempty interior for every $B > 0$, as defined in Section \ref{sec:param-space}.
Consider the minimax and adaptive problems defined in \eqref{eq:minimax-case1-1} and \eqref{eq:areg-1}, respectively.
Let $\cB = [\uB, \oB]$ with $\oB \geq \uB \geq 0$, and let $\bmw \in \cWcvx$.
\begin{enumerate}[label=(\alph*)]
\item \label{prop:areg-main-a} $A(B, \bmw)$ has one of the following shapes:
\begin{enumerate}[label=(\roman*)]
	\item $A(B, \bmw)$ is monotone in $B \in \cB$.
	\item There exists $B_0 \in (\uB, \oB)$ such that $A(B, \bmw)$ is nonincreasing in $B$ on $[\uB, B_0)$ and nondecreasing in $B$ on $[B_0, \oB]$.
\end{enumerate}
\item \label{prop:areg-main-b}  The worst-case adaptation regret is given by
\[
	\Amax(\cB, \bmw)
	=
	\sup_{B \in \cB}
	A(B, \bmw)
	=
	\max\{ A(\uB, \bmw), A(\oB, \bmw)\}.
\]
If $\oB = \infty$, then $A(\oB, \bmw)$ refers to $\lim_{B \to \infty} A(B, \bmw)$.
\end{enumerate}
\end{prop}

Proposition \ref{prop:areg-main}\ref{prop:areg-main-a} describes the shape of the adaptation regret over $B \in \cB$ for each $\bmw \in \cWcvx$.
It implies that, under Assumption \ref{assu:convex-m2}, the supremum over $B \in \cB$ is achieved at one of the endpoints.
This simplifies the computation of the inner problem in \eqref{eq:areg-3}.

The optimally adaptive weight can be computed via convex optimization.
I provide the details and explain how to reformulate the problem as a single optimization problem in Appendix \ref{sec:comp-adaptive}.

Consider $\cB = [0, \infty]$ and the $\ell_p$-norm bound of Example~\ref{eg:param-b} with $p \in (1, \infty)$.
Then, the optimally adaptive weight can be interpreted as a middle ground between the variance-minimizing weight (for $\uB = 0$ in Section~\ref{sec:var-min-weight}) and equal weighting (for $\oB = \infty$ in Remark \ref{re:minimax-binf}).

The following proposition states another useful property of the optimally adaptive weight.

\begin{prop}
\label{prop:same-opt-adpt}
Consider the same assumptions and definitions as in Proposition \ref{prop:areg-main} with $\cB = [\uB, \oB]$ and $\uB < \oB$.
Suppose that $A(\uB, \bmw)$ and $A(\oB, \bmw)$ are strictly convex and continuous in $\bmw \in \cWcvx$.
Let $\bmwa \in \cWcvx$ be the optimally adaptive weight as defined in~\eqref{eq:areg-3b}.
Then,
\[
	A(\uB, \bmwa) = A(\oB, \bmwa).
\]
\end{prop}

The parameter spaces of Section \ref{sec:param-space} satisfy the strict convexity and continuity hypotheses of Proposition~\ref{prop:same-opt-adpt} whenever $\oB$ is finite and $\cS(B)$ is compact.
This follows from the definition $A(B, \bmw) = \frac{\Rmax(B, \bmw)}{\Rmm(B)}$ in \eqref{eq:areg-1} with $\Rmax(B,\bmw) = V(\bmw) + \oM(B, \bmw)$, the strict convexity (Assumption \ref{assu:sig-pd}) and continuity of
$V(\bmw)$, the convexity and continuity of $\oM(B, \bmw)$ (the latter by Berge's maximum theorem; see \citealp[Theorem 17.31]{aliprantisborder2006bk}), and the positivity of $\Rmm(B)$ and its independence of $\bmw$.

\begin{re}
\label{re:penalty-1}
When $\lim_{B \to \infty} A(B, \bmw)$ is convex in $\bmw \in \cWcvx$ instead of strictly convex, the worst-case adaptation regret is not necessarily strictly convex in $\bmw$.
Nevertheless, strict convexity can be restored by adding a small penalty term to the worst-case adaptation regret, i.e., by replacing $\Amax([0, \infty], \bmw)$ with $\Amax([0, \infty], \bmw) + \varrho \|\bmw\|^2_2$ for a small $\varrho > 0$.
This is because $\Amax([0, \infty], \bmw)$ is convex in $\bmw$ and $\varrho \|\bmw\|^2_2$ is strictly convex in $\bmw$ for $\varrho > 0$.
\citet{gafarov2025joe} also adds a strictly convex penalty term for moment inequality models.
\end{re}

\conteg{\ref{eg:run-1}}{Assume that $\cB = [0, \infty]$ and $\rho\si_2 < 1$.
The optimally adaptive weight on $\hbe_1$ is given by
\begin{equation}
	\label{eq:run-1-adapt-optim}
		w_{\mathrm{A},1}(\mu_1) \equiv
		\frac{\mu_1 [ \Rmm(0)^{-1}(\si_2^2 - \rho\si_2) -2] + 2}{\mu_1[\Rmm(0)^{-1}(\si_2^2 - 2\rho\si_2 + 1) - 4] + 4},
\end{equation}
where $\mu_1 \in (0,1)$ is a Lagrange multiplier that satisfies the optimality condition given in Appendix \ref{sec:proof-claim-1-app}, and the minimax risk $\Rmm(0)$ is obtained by evaluating \eqref{eq:run-1-3} at $B = 0$.
The proof of~\eqref{eq:run-1-adapt-optim} is in Appendix~\ref{app:eg:run-1}.

\begin{figure}[!t]
	\centering
	\caption{Illustration of $A(0, w_1)$ and $\lim_{B \to \infty} A(B, w_1)$ for the two-outcome example.}
	\label{fig:claim4-1}
		\includegraphics[scale=1]{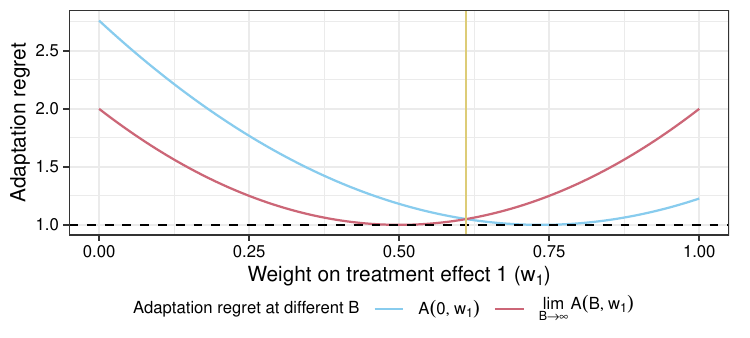}
    \mynote{
    The figure shows the adaptation regret for Example~\ref{eg:run-1} with $\si_2 = 1.5$ and $\rho = 0.2$.
    The yellow vertical line marks the optimally adaptive weight and the dashed horizontal line marks the smallest attainable adaptation regret.}
\end{figure}

For illustrative purposes, I present a numerical example with $\si_2 = 1.5$ and $\rho = 0.2$ in Figure \ref{fig:claim4-1}.
The blue curve is the adaptation regret for $B = 0$.
It is the regret relative to the variance-minimizing weight (see Remark~\ref{re:minimax-b0}).
The red curve is the adaptation regret for $B = \infty$.
It is the regret relative to equal weighting, which minimizes the worst-case squared bias (see Remark~\ref{re:minimax-binf}).
The optimally adaptive weight $w_{\mathrm{A},1}(\mu_1)$ is located at the point where the two curves intersect, as described in Proposition \ref{prop:same-opt-adpt}.
This is also reflected in \eqref{eq:run-1-adapt-optim}, where $w_{\mathrm{A},1}(\mu_1)$ is a weighted average of the variance-minimizing and bias-minimizing weights.

I provide some intuition for why $w_{\mathrm{A},1}(\mu_1)$ is not on the boundary in this example.
A boundary weight puts all the weight on one outcome, so the aggregated treatment effect inherits the full bias of the treatment effect on that outcome instead of averaging it with the bias of the other outcome.
Shifting a small amount of weight away from the boundary lowers the adaptation regret for $B = \infty$.
Unless the boundary weight minimizes variance, the shift also reduces the adaptation regret for $B = 0$.
In either case, the higher of the two regrets falls, so the optimally adaptive weight is in the interior.
}

\subsection{Inference} \label{sec:inference} \label{sec:flci}
This subsection describes how to perform valid inference based on the minimax or adaptive procedures.
When $B > 0$, standard Wald confidence intervals for $\theta$ are not valid because the aggregated treatment effect can be biased relative to $\theta$.
As a result, I construct fixed-length confidence intervals (FLCI) that take the bias into account \citep{donoho1994aos, armstrongkolesar2018ecta, armstrongkolesar2021ecta, armstrongkolesar2021qe}.

\subsubsection{Minimax approach} \label{sec:flci-minimax}
Let $\bmwsb$ be the solution to the minimax problem in~\eqref{eq:minimax-case1-1soln}.
By~\eqref{eq:decision-1},
\begin{equation}
	\label{eq:minimax-flci-1}
	\frac{\bmwsb'\hbmbe - \theta}{\sqrt{\bmwsb'\bmSi\bmwsb}} \sim
	\scN \left( \frac{\bmwsb'\bmb}{\sqrt{\bmwsb'\bmSi\bmwsb}}, 1 \right).
\end{equation}
The distribution in~\eqref{eq:minimax-flci-1} depends on $\bmb \in \cS(B)$ only through its mean.

In practice, $\bmSi$ is unknown, and the minimax problem is solved using an estimator $\hbmSn$ of the asymptotic variance matrix of $\sn(\hbmbn - \bmbe)$, so that $\cS(B)$ now bounds $\sn(\bmbe - \theta \oneq)$ (as in Appendix~\ref{app:flci-valid}).
Since $\hbmbn$ is computed from outcomes standardized by estimated standard deviations, $\hbmSn$ should account for this estimation step; see Appendix~\ref{app:gen-out}.
Let $\hbmwn$ be the weight from the minimax problem that uses $\hbmSn$ instead of $\bmSi$.
A $100(1 - \alpha)\%$ fixed-length confidence interval that takes the bias into account can be formed as
\begin{equation}
	\label{eq:minimax-flci-2}
	\cI_n \equiv \left[\hbmwn'\hbmbn \pm \frac{\hcvn \widehat s_n}\sn \right],
\end{equation}
where $\widehat s_n \equiv \sqrt{\hbmwn'\hbmSn\hbmwn}$ and the critical value $\hcvn \in \bR$ is the smallest value such that $\max_{\bmb \in \cS(B)} \bP[|\scN(\frac{\hbmwn'\bmb}{\widehat s_n}, 1)| > \hcvn] \leq \alpha$ following \eqref{eq:minimax-flci-1}.
In Appendix~\ref{app:flci-valid}, I establish the asymptotic validity of the FLCI; see also Appendix~\ref{app:decision:aux}.

\subsubsection{Adaptive approach} \label{sec:flci-adaptive}
By the same argument as in Section \ref{sec:flci-minimax}, one can construct a $100(1 - \alpha)\%$ confidence interval for $\theta$ based on the adaptive approach of Section~\ref{sec:adapt}.

Suppose the researcher adapts over $\cB = [\uB, \oB]$.
One choice, suggested by~\citet{armstrongetal2025ecta}, is to construct FLCIs as in~\eqref{eq:minimax-flci-2} but with $\hbmwn$ replaced by the optimally adaptive weight and $\cS(B)$ replaced by $\cS(\uB)$ and $\cS(\oB)$ to summarize the range of critical values that guarantee coverage under different choices of $\cS(B)$ for $B \in \cB$.
The asymptotic validity of~\eqref{eq:minimax-flci-2} with the optimally adaptive weight is established in Appendix~\ref{app:flci-valid}.

\subsection{Implementation} \label{sec:implementation}
This subsection summarizes the statistical decision-theoretic procedures.
\begin{description}[leftmargin=!,labelwidth=3.3em]
	\item[Step 1.] Estimate the treatment effects $\hbmbn$ and their asymptotic variance matrix $\hbmSn$.
	\item[Step 2.] Specify the parameter space (see Section \ref{sec:param-space}).
	\item[Step 3.] Estimate the weights (with $\bmSi$ replaced by~$\hbmSn$).
	\begin{description}[leftmargin=0pt, labelsep=.5em]
		\item[\normalfont (a)] For the minimax approach, choose $B \geq 0$ and solve \eqref{eq:minimax-case1-1soln}.
		\item[\normalfont (b)] For the adaptive approach, choose $\cB = [\uB, \oB]$ and solve \eqref{eq:areg-3}.
	\end{description}
	\item[Step 4.] Conduct inference as in Section~\ref{sec:flci}.
\end{description}

\section{Empirical application} \label{sec:sdeg}
In this section, I illustrate the methodology proposed in this paper by revisiting \citet{karlanetal2026jde}, which is introduced in Example~\ref{eg:sd}.
The authors evaluate a randomized cash transfer program for low-income households in Ghana during the COVID-19 pandemic.
In each of the urban and rural samples, they selected 1,550 households with the lowest food expenditure and randomized these households equally into treatment and control groups, stratified by rural/urban status and food expenditure cells.
Among these 3,100 households, 1,508 enrolled.
For these enrolled households, the treatment group (771 households) received eight mobile money transfers of 90 Ghanaian cedis each between 2020 and 2021, totaling around one month's income.
The control group received only the first transfer.

\citet{karlanetal2026jde} study how cash transfers affect social distancing, food expenditure, income, and psychological well-being.
They note that other COVID-19 cash transfer experiments have reported conflicting effects on social distancing.
I revisit their analysis of how the cash transfer program affects the social distancing index, aggregated from six outcomes using PCA.
In particular, I reanalyze the part of their analysis in which they find a significant and positive effect of the cash transfer program on the social distancing index using the second and fourth follow-up surveys.

\subsection{Treatment effects on individual outcomes and on the indices from three common methods of aggregation}
\begin{figure}[!t]
	\centering
	\caption{Treatment effects and PCA weights for the cash transfer application.}
	\label{fig:jde-te}
	\includegraphics[scale=1]{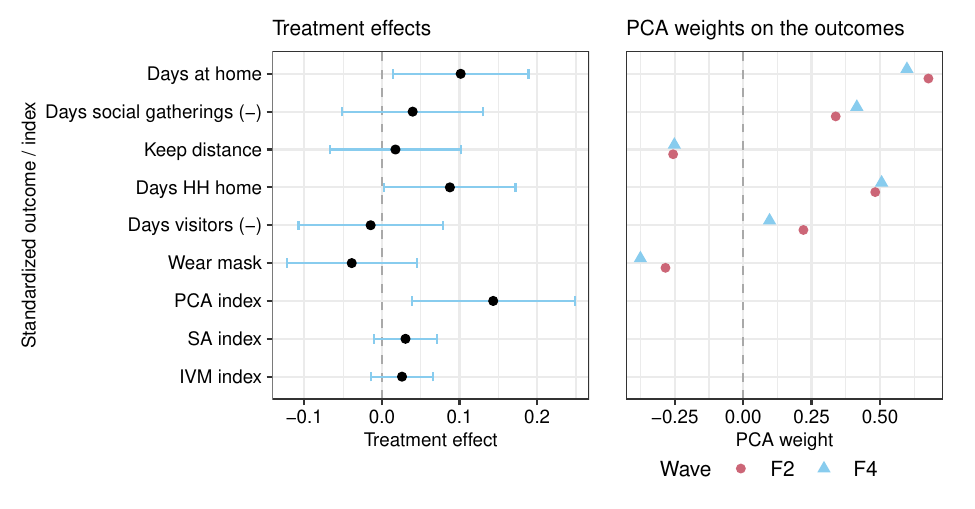}
    \mynote{
    The left panel shows the treatment effects on the individual outcomes and the PCA, SA, and IVM indices, as well as the 95\% confidence intervals.
    The right panel shows the PCA weights on each outcome.
		Outcomes marked with ($-$) are reverse-coded, so a higher value means more social distancing.
		``HH'' stands for household members.
		F2 and F4 are the two survey waves.}
\end{figure}

Figure \ref{fig:jde-te} presents the treatment effects and PCA weights.
The left panel shows the regression results on the individual outcomes and the indices.
I standardize the individual outcomes in Figure~\ref{fig:jde-te} (although the authors report them in raw units in their table), so that the treatment effects are comparable.
In addition, I do not restandardize the indices after they are created, so that each index remains a weighted sum of the standardized outcomes, although the authors did restandardize their PCA index; see also Remark~\ref{re:postprocess} and Lemma~\ref{lem:linear-post}.
Without restandardization, the estimated effect on the PCA index has the same sign as that on the authors' index and is also significant at the 1\% level.

The authors use PCA to aggregate the outcomes.
For comparison, I also apply SA and IVM, two other popular methods identified in my survey (Table \ref{tab:count}).
The authors create a separate PCA index for each survey wave and estimate the treatment effect on the index by pooling the two waves.
Following them, I construct all three indices separately for each wave and standardize their components using the full sample of that wave; standardizing by the control group instead for SA and IVM gives similar results.
The blue bars show $95\%$ confidence intervals computed as in the authors' original code.
As in their implementation, these confidence intervals do not adjust for the standardization step or the data-dependent weights; adjusting them as in Appendix~\ref{app:gen-out} yields similar results.
For the individual outcomes, dividing by the estimated standard deviation does not change the $t$-statistic when one tests for a zero effect.
The right panel shows the PCA weights on the six outcomes.
The weights are similar across the two waves.
In both waves, the same two outcomes receive negative PCA weights.

Some observations from Figure~\ref{fig:jde-te} are as follows.
First, as noted above, two outcomes receive negative PCA weights.
This creates the interpretability issue discussed in Section \ref{sec:criteria-inter}.
Second, the treatment effect on the PCA index falls outside the convex hull of the individual treatment effects.
This is partly because the outcome with the largest negative treatment effect (``wear mask'') also receives the most negative PCA weight.
The effect on the PCA index is significant at the 1\% level, whereas none of the individual effects is significant at that level (the effects on ``days at home'' and ``days household members at home'' are significant at the 5\% level).
Third, the treatment effects on the SA and IVM indices are smaller than that on the PCA index, and they are not significant at the 10\% level.
Hence, whether the aggregated effect is significant depends on which of the three methods is used.

\subsection{Applying the decision-theoretic approach}
The treatment effects in Figure \ref{fig:jde-te} raise two questions.
First, with PCA, SA, and IVM yielding indices that give treatment effects with different magnitudes and significance levels, what can be learned about the effect of cash transfers?

Second, how can one incorporate domain knowledge about the underlying abstract variable, which none of PCA, SA, and IVM can exploit?
Such domain knowledge is available in the present application.
For example, whether people wear masks may capture social distancing less well than the other outcomes because mask wearing is a precaution taken during contact.
In addition, the authors expect the variable ``days visitors'' to be less controllable by the household receiving the transfers \citep[page 12]{karlanetal2026jde}, presumably because whether nonfamily members visit depends more on the visitors themselves.

\begin{figure}[!t]
    \centering
    \caption{Results for the statistical decision-theoretic approach (without shape restrictions) for the cash transfer application.}
    \label{fig:jde-orig-d1}
    \includegraphics[scale=1]{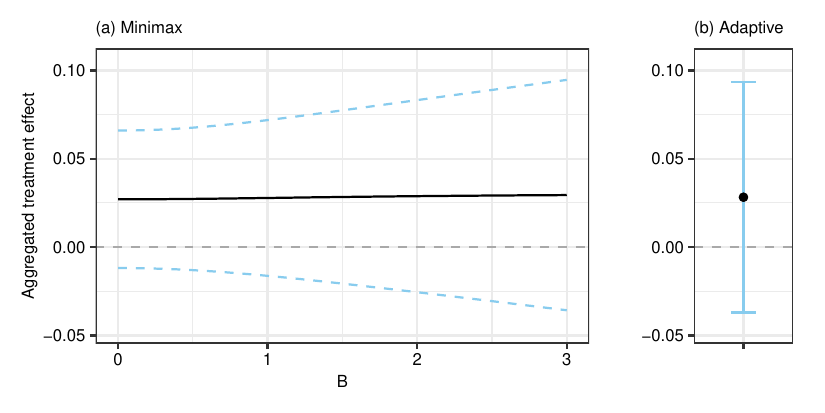}
    \mynote{
    In panel (a), the solid line shows the treatment effect using the minimax approach for different values of $B$, and the dashed lines show the 95\% FLCI.
		In panel (b), the dot represents the treatment effect when adapting over $[0, 3]$, and the bar represents the 95\% FLCI for $\cS(3)$.}
\end{figure}

To answer these questions, I apply the statistical decision-theoretic approach developed in Section~\ref{sec:decision}.
I first do so without imposing any domain knowledge about the relative quality of the outcomes.
Figure~\ref{fig:jde-orig-d1} reports the estimated treatment effects from the minimax approach in panel~(a) and the adaptive approach in panel~(b).
I use the $\ell_2$-norm to bound the vector $\bmb$, i.e., I set $p = 2$ in \eqref{eq:param-b-1}, so that $B$ can be interpreted as the upper bound on the overall magnitude of the bias on the scale of Section~\ref{sec:flci-minimax}.
At $B = 0$, the minimax weights minimize the variance of the aggregated treatment effect.
Hence, the approach minimizes the width of the confidence interval, while ensuring that the weights are nonnegative for interpretability.
Panel (b) shows that the adaptive approach with $\cB = [0, 3]$, the range of $B$ in panel (a), gives a similar effect.
Under both approaches, the estimated effect is positive but is not even significant at the 10\% level.
It is also smaller in magnitude than under PCA.
This is consistent with the earlier observation that PCA's larger effect is driven partly by the negative weight it places on the variable ``wear mask'' that has the most negative treatment effect.

\begin{figure}[!t]
    \centering
    \caption{Results for the statistical decision-theoretic approach (with shape restrictions) for the cash transfer application.}
    \label{fig:jde-shape-d1}
    \includegraphics[scale=1]{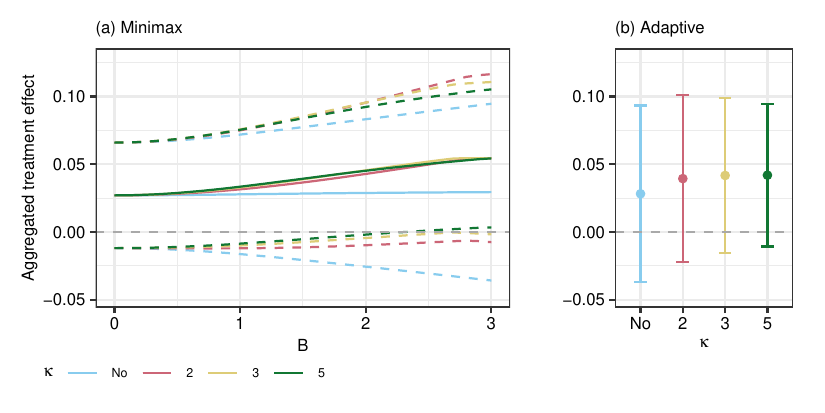}
    \mynote{
    Each color corresponds to a specific value of $\kappa$ in the shape restriction; ``No'' means no shape restriction, as in Figure~\ref{fig:jde-orig-d1}.
     In panel (a), the solid line shows the treatment effect using the minimax approach for different values of $B$, and the dashed lines show the 95\% FLCI.
		 In panel (b), the dot represents the treatment effect when adapting over $[0, 3]$, and the bar represents the 95\% FLCI for $\cS(3)$.}
\end{figure}

I then incorporate the domain knowledge to answer the second question.
Following Example~\ref{eg:param-shape}, I assume that ``wear mask'' captures social distancing less well than the other outcomes, and that ``days visitors'' is less controllable by the household.
I implement these two pieces of knowledge through shape restrictions, which downweight the two outcomes.
In the shape restriction, the parameter $\kappa$ controls how much less important the two outcomes are when $\kappa > 1$.
A larger $\kappa$ requires the biases of the other four outcomes to be smaller relative to those of the two outcomes.
Figure~\ref{fig:jde-shape-d1} reports the results.
For each $\kappa \in \{2, 3, 5\}$ and $B > 0$ considered, the aggregated treatment effect exceeds the effect without shape restrictions.
The minimax treatment effects become significant at the 5\% level only when the two outcomes are heavily downweighted (at large $B$) and the other four outcomes are assumed to be less biased (e.g., when $\kappa = 5$).
However, under the adaptive approach in panel~(b), which does not require committing to a specific value of $B$, the FLCI for $\cS(3)$ contains 0 for the specifications considered in the figure.

The exercise sheds light on what assumptions are needed to conclude that the cash transfer program has a positive contemporaneous effect on social distancing.
One has to put less emphasis on the two outcomes ``wear mask'' and ``days visitors'' and assume that the other four outcomes are less biased in order to support a positive and significant effect under the minimax criterion.
The decision-theoretic approach makes these assumptions explicit, without relying on ad hoc approaches such as PCA, SA, and IVM.

\section{Conclusion} \label{sec:concl}
Researchers commonly observe multiple outcomes that are related to an underlying abstract concept, such as crime or wealth.
They often aggregate the outcomes into an index to evaluate the treatment effect on such abstract concepts.
In this paper, I first examine the properties and issues of the three widely used approaches, namely, PCA, SA, and IVM.
I show that PCA has several unattractive properties.
PCA can have negative weights, does not necessarily maximize precision, is sensitive to arbitrary normalization choices, and can even lead to a nonstandard asymptotic distribution.
IVM does not suffer from the last two issues, but it can have negative weights and also need not maximize precision.
Although PCA and IVM incorporate correlation information when computing weights, neither uses the covariance matrix of the treatment effects.
SA avoids having negative weights, but it does not use the correlation structure of the outcomes.

I propose a statistical decision-theoretic approach to aggregating outcomes that uses the mean-squared error of the aggregated treatment effect as its risk function while ensuring interpretable weights.
The weights can be chosen by the minimax criterion or the adaptation regret criterion.
The adaptation regret criterion has the advantage of not requiring the researcher to commit to a specific level of misspecification.
I show that convex optimization can be used to compute the weights.
I illustrate my approach by revisiting the analysis in \citet{karlanetal2026jde} of the effect of a cash transfer program on social distancing.
I show that using PCA to aggregate outcomes can lead to negative weights and need not maximize precision.
I show how my approach chooses weights that do not suffer from interpretability issues, how it can flexibly trade off bias and variance, and how it can incorporate domain knowledge about outcome quality through shape restrictions.

\addcontentsline{toc}{section}{References}
{	\singlespacing{
	\bibliographystyle{ecta}
	\bibliography{refs}
	}
}

\appendix
\numberwithin{equation}{section}
\counterwithin{figure}{section}

\newpage
\begin{center}

{\bf \Large Supplemental appendix to ``Aggregating treatment effects across multiple outcomes''}
\end{center}

\stopcontents[main]

\startcontents[appendix]
{
  \hypersetup{linkcolor=black}
	\printcontents[appendix]{}{1}{\singlespacing\setcounter{tocdepth}{2}}
}

\bigskip  \bigskip	\bigskip

\newpage

\section{Supplemental results for Sections~\ref{sec:setup} and~\ref{sec:problems}} \label{app:main-supp-problems}

\subsection{Connecting the standardized and unstandardized outcomes} \label{sec:std-unstd-y}
This subsection presents the setup for the standardized and unstandardized outcomes. \par

\noindent \textbf{Linear model.}
Let $\tbmY_i \equiv (\tY_{i,1}, \ldots, \tY_{i,\dy})' \in \bR^{\dy}$ be the unstandardized outcomes and $\bmY_i \equiv (Y_{i,1}, \ldots, Y_{i,\dy})' \in \bR^{\dy}$ be the standardized outcomes.
The linear model for $\tbmY_i$ is
\begin{align}
	\label{eq:pre-gen-5}
	\tbmY_i
	& =
	\tbmxi
	+
	\tbmbe D_i
	+
	\tbmU_i,
\end{align}
where $\tbmbe \equiv (\tbe_1, \ldots, \tbe_{\dy})'$, $\tbmxi \equiv (\txi_1, \ldots, \txi_{\dy})'$, and $\tbmU_i \equiv (\tU_{i,1}, \ldots, \tU_{i, \dy})'$.

\noindent \textbf{Standardization.}
Outcomes are standardized by $\hbSn \equiv (\hS_{n,1}, \ldots, \hS_{n,\dy})'$ as $\bmY_i = \diag[\hbSn]^{-1} \tbmY_i$.
Let $\bms \equiv (s_1, \ldots, s_{\dy})' \in \bR^{\dy}$ be the population analog of $\hbSn$.
Depending on the method, $\hS_{n,j}$ can be the sample standard deviation of outcome $j$ using the full sample or the control sample.
I consider the following linear model for $\bmY_i$:
\begin{align}
	\label{eq:sep-reg-2}
	\bmY_i
	& = \bmxi + \bmbe D_i + \bmU_i,
\end{align}
where $\bmbe \equiv (\beta_1, \ldots, \beta_{\dy})'$, $\bmxi \equiv (\xi_1, \ldots, \xi_{\dy})'$, and $\bmU_i \equiv (U_{i,1}, \ldots, U_{i, \dy})'$.
Let $\htbmbn$ and $\hbmbn$ be the estimators of $\tbmbe$ and $\bmbe$ in \eqref{eq:pre-gen-5} and \eqref{eq:sep-reg-2}, respectively.
By Lemma \ref{lem:linear-out},
\begin{equation}
	\label{eq:pre-gen-2}
	\hbmbn = \diag[\hbSn]^{-1} \htbmbn
	\quad
	\text{ and }
	\quad
	\bmbe = \diag[\bms]^{-1} \tbmbe.
\end{equation}
I maintain the following assumption on~\eqref{eq:pre-gen-5}, where part~\ref{assu:main-reg-ols-1} requires random sampling and $D_i$ to have variation,
part~\ref{assu:main-reg-ols-2} imposes the standard independence assumption, and
part~\ref{assu:main-reg-ols-4} requires the variance matrices to be positive definite.
\begin{assu} \label{assu:main-reg-ols}
\begin{enumerate*}[label=(\alph*)]
	\item \label{assu:main-reg-ols-1} $\{(D_i, \tbmY_i')'\}_{i=1}^n$ is i.i.d., $D_i \in \{0, 1\}$, and $p_D \equiv \bP[D_i = 1] \in (0, 1)$.
	\item \label{assu:main-reg-ols-2} $\bE[\tbmU_i | D_i] = \zeroq$.
	\item \label{assu:main-reg-ols-4} $\var[\tbmY_i]$ and $\var[\tbmU_i]$ are symmetric positive definite matrices with bounded eigenvalues.
\end{enumerate*}
\end{assu}

To study the large-sample properties, I impose the following high-level assumption:
\begin{assu}
\label{assu:pre-gen-1}
\begin{enumerate*}[label=(\alph*)]
	\item $\bms > \zeroq$.
	\item
	$
	\sqrt{n} \left( \begin{smallmatrix}
		\htbmbn - \tbmbe \\
		\hbSn - \bms
	\end{smallmatrix} \right)
	\Dt
	\left( \begin{smallmatrix}
		\bmZtb \\ \bmZs
	\end{smallmatrix} \right)
	\sim
	\scN\!
	\left(
	\zero_{2\dy},
	\left(\begin{smallmatrix}
	\psbb & \pssb' \\
	\pssb & \psss
	\end{smallmatrix}\right)
	\right)$.
\end{enumerate*}
\end{assu}

\begin{prop}
\label{prop:pre-gen-1}
Consider~\eqref{eq:sep-reg-2} and the standardization as in \eqref{eq:pre-gen-2}.
Let Assumption~\ref{assu:pre-gen-1} hold.
Define $\bmH \equiv \diag[\bms]^{-2}\diag[\tbmbe]$.
Then, $\hbmbn \Pt \bmbe$ and  $\sn (\hbmbn - \bmbe) \Dt \scN(\zeroq, \bmSb)$, where
\begin{equation}
	\bmSb
	\equiv
	\diag[\bms]^{-1}\psbb \diag[\bms]^{-1}
	-
	\diag[\bms]^{-1}\pssb' \bmH'
	-
	\bmH\pssb \diag[\bms]^{-1}
	+
	\bmH \psss\bmH'.
	\label{eq:pre-gen-4}
\end{equation}
\end{prop}

\subsection{How aggregation affects precision} \label{app:pre-gen}
I provide details for the examples in Section~\ref{sec:problems}, and then present some general results.

\begin{eg}[PCA and precision]
\label{eg:app:pca-precision}
Assume $\bms = \oneq$ and $\tbmbe = \zeroq$.
For $d \in \{0, 1\}$, define $\bmSi_{\widetilde U, d} \equiv \var[\tbmU_i | D_i = d]$.
Under~\eqref{eq:pre-gen-5} and Assumption~\ref{assu:main-reg-ols}, $\var[\tbmY_i] = \tbmbe\tbmbe' p_D(1 - p_D) + p_D\bmStUi + (1-p_D)\bmStUo$.
The asymptotic variance of $\htbmbn$ is (see Appendix \ref{app:proof:app:a} for a proof)
\begin{equation}
	\smash{\psbb
	=
	\frac{1}{p_D^2}\var[D_i \tbmU_i] + \frac{1}{(1 - p_D)^2} \var[(1 - D_i)\tbmU_i]
	= \frac{1}{p_D}\bmStUi + \frac{1}{1 - p_D}\bmStUo},
	\label{eq:pre-gen-7b}
\end{equation}
where I use $\var[D_i \tbmU_i] = p_D\bmStUi$
and
$\var[(1 - D_i) \tbmU_i] = (1 - p_D)\bmStUo$ since $D_i$ is binary.
Because $\bms = \oneq$ and $\tbmbe = \zeroq$, \eqref{eq:pre-gen-4} becomes $\bmSb = \psbb$.
I consider two cases.

First, consider the homoskedastic case.
Write $\bmStUo = \bmStUi \equiv \bmStU$.
Using~\eqref{eq:pre-gen-7b}, I have
\begin{align}
	\smash{\bmSb
	=
	\psbb
	=
	\frac{1}{p_D} \bmStU + \frac{1}{1 - p_D} \bmStU
	=
	\frac{1}{p_D(1 - p_D)} \var[\tbmY_i]
	=
	\frac{1}{p_D(1 - p_D)} \bmSY}.
	\label{eq:pre-gen-8}
\end{align}
This shows that PCA maximizes, rather than minimizes, the asymptotic variance of the aggregated treatment effect because $\bmSb$ is proportional to $\bmSY$.

Next, consider the heteroskedastic case.
Using the law of total variance, $\bmSY = \tbmbe\tbmbe' p_D(1 - p_D) + p_D\bmStUi + (1 - p_D)\bmStUo$, $\tbmbe = \zeroq$, $\bms = \oneq$, and~\eqref{eq:pre-gen-7b},
\begin{align}
	p_D(1 - p_D) \bmw' \bmSb \bmw -  \bmw'\bmSY \bmw
	& =
	(1 - 2p_D) (\bmw' \bmStUi \bmw - \bmw' \bmStUo \bmw).
	\label{eq:pre-gen-12}
\end{align}
Thus, PCA still needs strong conditions to maximize precision.
\end{eg}

\begin{eg}[IVM and precision]
\label{eg:app:ivm-precision}
Consider the setup as in Example \ref{eg:app:pca-precision}.
Note that $\bms$ now represents the control group's standard deviations in IVM \citep{anderson2008jasa}.

Under homoskedasticity, $\argmin_{\bmw \in \cWsto} \bmw'\bmSY\bmw = \argmin_{\bmw \in \cWsto}  \bmw'\bmSb\bmw$ by~\eqref{eq:pre-gen-8}, but $\wivm$ can have negative weights.
Under heteroskedasticity, \eqref{eq:pre-gen-12} shows that minimizing $\bmw'\bmSY\bmw$ and minimizing $\bmw'\bmSb\bmw$ can lead to different solutions, unless $p_D = \frac{1}{2}$.
\end{eg}

\noindent Finally, I show that PCA, SA, and IVM maximize precision only under strong conditions.
\begin{prop}
\label{eg:prop:gen-precision}
Let Assumptions~\ref{assu:eval} and~\ref{assu:pre-gen-1} hold.
Let $\bmSY$ and $\bmSb$, as defined in~\eqref{eq:pre-gen-4}, be positive definite.
Write $\wpca$ as in \eqref{eq:pca-problem}, $\wivm$ as in \eqref{eq:framework-simp-ivm}, and $\wsa$ as in Section \ref{sec:kling-main}.
\begin{enumerate}[label=(\alph*)]
	\itemsep-1pt
	\item \label{prop:pca-eq-1} Let the smallest eigenvalue of $\bmSb$ be unique.
	Then, $\wpca = \smash{\argmin_{\bmw \in \cWunit} \bmw'\bmSb\bmw}$ if and only if $\bvec_1$ is parallel to the PC1 of $\smash{\bmSb^{-1}}$.
	\item \label{prop:anderson-eq-hetero-2} $\wivm = \argmin_{\bmw \in \cWsto} \bmw'\bmSb\bmw$ if and only if $\bmSY^{-1}\oneq = a \bmSb^{-1}\oneq$ for some $a \neq 0$.
	\item \label{prop:sa-eq-1} $\wsa = \smash{\argmin_{\bmw \in \cWsto} \bmw'\bmSb\bmw}$ if and only if $\oneq = \smash{a \bmSb^{-1}\oneq}$ for some $a \neq 0$.
\end{enumerate}
\end{prop}

\subsection{How aggregation affects the \emph{t}-statistic and power} \label{app:t-stat}
To understand how aggregation affects the $t$-statistic and power, I consider the linear model using the aggregated outcome, so that I can state the corresponding hypothesis.

\noindent \textbf{Linear model.}
Consider the binary treatment setup as in Example~\ref{eg:binary-te-1}:
\begin{equation}
	\label{eq:genout-reg-1}
	\bmw'\bmY_i = \overline\xi + \tau D_i + \oU_i.
\end{equation}
\eqref{eq:genout-reg-1} is related to \eqref{eq:sep-reg-2} by writing $\overline\xi = \bmw'\bmxi$, $\tau = \bmw'\bmbe$, and $\oU_i = \bmw'\bmU_i$.

\noindent \textbf{Hypothesis and test statistic.} Consider the following hypothesis:
\begin{equation}
	\label{eq:sep-hypo-1}
	H_0: \tau = 0
	\quad
	\text{vs.}
	\quad
	H_1: \tau \neq 0.
\end{equation}
Let $\Tnw$ be the conventional sample $t$-statistic for testing \eqref{eq:sep-hypo-1} and $\hRRw$ be the sample analog of the population $R$-squared.
Then, I have \citep[equation (4.41)]{wooldridge2016bk}
\begin{equation}
	\label{eq:r2-t-1}
	[\Tnw]^2
	=
	(n-2)\frac{\hRRw}{1 - \hRRw}.
\end{equation}
The $t$-statistic and $R$-squared are functions of $\bmw$ because the outcome in \eqref{eq:genout-reg-1} depends on $\bmw$.
Since \eqref{eq:r2-t-1} has a factor of $(n - 2)$, I consider a ratio of $t$-statistics for a fixed $\bmw_0$ such that $R^2(\bmw_0) \in (0, 1)$, so that the ratio has a well-defined probability limit.

\begin{prop}
\label{prop:t-final-all}
Let Assumptions \ref{assu:eval}, \ref{assu:main-reg-ols}, and \ref{assu:pre-gen-1} hold, and suppose $\bmbe \neq \zeroq$.
Consider \eqref{eq:genout-reg-1} and testing \eqref{eq:sep-hypo-1}.
Let ${\emph{\textrm{\cUpsilon}}}(\bmw; \bmw_0) \equiv
\plim_{n\to\infty} \  \frac{[\Tnw]^2}{[\widehat{T}_n(\bmw_0)]^2}
= \left.\frac{\RRw}{1 - \RRw} \right/ \frac{R^2(\bmw_0)}{1 - R^2(\bmw_0)}$
for any $\bmw_0 \in \bR^{\dy}$ such that $R^2(\bmw_0) \in (0, 1)$.
\begin{enumerate}[label=(\alph*)]
	\item \label{prop:t-final} The PC1 of $\bmSY$ solves $\max_{\bmw : \|\bmw\|_2 = 1} \text{\emph{\textrm{\cUpsilon}}}(\bmw; \bmw_0)$
if and only if it is parallel to $\bmbe$.
	\item \label{prop:t-ivm} $\wivm$ solves $\max_{\bmw \in \cWsto} \text{\emph{\textrm{\cUpsilon}}}(\bmw; \bmw_0)$ if and only if $\bmbe$ is parallel to $\oneq$.
	\item \label{prop:t-sa} $\wsa$ solves $\max_{\bmw \in \cWsto} \text{\emph{\textrm{\cUpsilon}}}(\bmw; \bmw_0)$ if and only if it is parallel to $\bmSY^{-1}\bmbe$.
\end{enumerate}
\end{prop}

Proposition~\ref{prop:t-final-all} implies that the weights that maximize the square of the ratio of limiting conventional $t$-statistics depend on both $\bmbe$ and $\bmSY$.
Hence, PCA, SA, and IVM do not maximize power without strong conditions.
The ``if'' direction of Proposition~\ref{prop:t-final-all}\ref{prop:t-ivm} on IVM is related to the GLS test of \citet{obrien1984biom}, which assumes that the treatment effects on the standardized outcomes are equal; the proposition shows that this assumption is also necessary for IVM to maximize $\cUpsilon(\bmw; \bmw_0)$ over $\cWsto$.

This can be seen from the one-to-one relationship between $[\Tnw]^2$ and $\hRRw$ in \eqref{eq:r2-t-1}.
In the binary treatment case, the $R$-squared is maximized if the means of $\bmw'\bmY_i$ for the $D_i = 1$ and $D_i = 0$ populations are as far apart as possible, relative to the variance of $\bmw'\bmY_i$.
Thus, $\bmbe$ matters because a larger $(\bmw'\bmbe)^2$ increases separation.

\begin{re}
Maximizing $\RRw$ (or the $t$-statistic squared in the homoskedastic case) is related to linear discriminant analysis (LDA), which classifies observations into different groups using a linear combination of the features (see, e.g., \citealp{hastieetal2009bk} or \citealp{mardiaetal2024bk}).
If one views $\bmY_i$ as ``features'' and $D_i$ as ``labels,'' then $\RRw$ is related to Fisher's linear discriminant, although the covariance matrices differ.
The connection between LDA and $\RRw$ highlights the importance of using $D_i$ if one wants to maximize the $t$-statistic (i.e., to improve power).
Here, running PCA on $\bmY_i$ is analogous to unsupervised learning, and maximizing the $t$-statistic is analogous to supervised learning.
\end{re}

\begin{figure}[!t]
    \centering
    \caption{A two-outcome numerical example that evaluates the $R$-squared of PCA.}
    \label{fig:pca-2}
    \begin{subfigure}[t]{0.5\textwidth}
        \centering
        \caption{DGP 1.}
        \label{fig:pca-1a}
			  \includegraphics[scale=1]{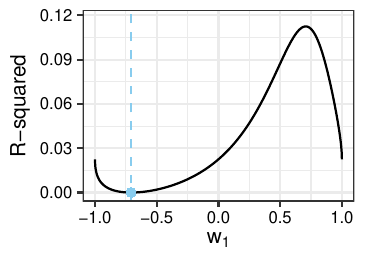}
    \end{subfigure}%
    ~
    \begin{subfigure}[t]{0.5\textwidth}
        \centering
        \caption{DGP 2.}
        \label{fig:pca-1b}
			  \includegraphics[scale=1]{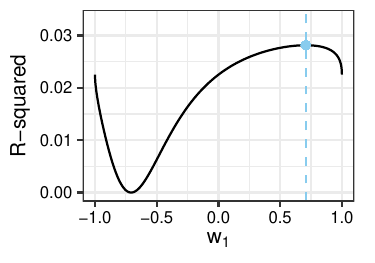}
    \end{subfigure}
    \mynote{
	The panels plot, against $w_1$, the population $R$-squared of the regression of the aggregated outcome on $D_i$.
	See Example \ref{eg:run-eg-3b} for the DGPs.
	The dashed lines show the choice made by running PCA on the outcomes.}
\end{figure}

\begin{eg}
\label{eg:run-eg-3b}
This numerical example demonstrates Proposition~\ref{prop:t-final-all}\ref{prop:t-final} and shows that PCA can maximize or minimize the $t$-statistic squared (or equivalently, the $R$-squared).

Let DGP 1 be the DGP in Example \ref{eg:binary-te-1} but with $\bmbe = (0.5, 0.5)'$, and DGP 2 be the same as DGP 1, except that $\corr[Y_{i,1}, Y_{i,2}] = 0.6$.
Figure~\ref{fig:pca-2} shows the (population) $R$-squared when $w_1Y_{i,1} + w_2 Y_{i,2}$ is used as the outcome, which suffices because there is a one-to-one relationship between the $t$-statistic squared and the $R$-squared (see \eqref{eq:r2-t-1}).
PCA minimizes the $R$-squared under DGP 1 but maximizes it under DGP 2, so PCA does not necessarily improve power.
\end{eg}

\subsubsection{Extensions}
I discuss various extensions to the earlier analysis, with appropriate adjustments.

First, when covariates are included, a similar analysis can be performed using the Frisch--Waugh--Lovell (FWL) theorem by replacing the outcome and $D_i$ with the residualized variables and adjusting for the degrees of freedom in the $t$-statistic to account for the number of covariates.

Second, one can test \eqref{eq:sep-hypo-1} using $\Tnw = \frac{\sqrt{n}(\bmw'\hbmbn)}{\widehat{s}_n(\bmw)}$, where $\widehat{s}_n(\bmw) \equiv \sqrt{\bmw'\hbmSb\bmw}$ and $\hbmSb \Pt \bmSb$.
An analysis similar to that of Proposition \ref{prop:t-final-all} applies using the Rayleigh quotient $\frac{\bmw'(\bmbe\bmbe')\bmw}{\bmw'\bmSb\bmw}$.
The weights that maximize the quotient are parallel to $\bmSb^{-1}\bmbe$.

Finally, the analysis can be used to study local power.
Let $\mathrm{cv} > 0$ be a given critical value that depends on the significance level $\alpha \in (0, 1)$.
Then, $\bP[|\Tnw| > \mathrm{cv}]  = \bP[| \frac{\sqrt{n}(\bmw'\hbmbn - \bmw'\bmbe)}{\widehat{s}_n(\bmw)} + \frac{\sqrt{n}(\bmw'\bmbe)}{\widehat{s}_n(\bmw)} | > \mathrm{cv}]$.
Consider a local alternative where $\bmbe_n =  \ns \bmd$ and $\bmd \in \bR^{\dy}$.
If Assumption~\ref{assu:pre-gen-1} and $\hbmSb \Pt \bmSb$ hold along the sequence of local alternatives, Slutsky's theorem implies that this converges to $\bP[ | Z + \frac{\bmw'\bmd}{\sqrt{\bmw'\bmSb\bmw}} | > \mathrm{cv}]$,
where $Z \sim \scN(0, 1)$ and $\bmSb$ is~\eqref{eq:pre-gen-4} evaluated at $\tbmbe = \zeroq$.
This is the same as evaluating the probability of a folded normal distribution $|\scN( \frac{\bmw'\bmd}{\sqrt{\bmw'\bmSb\bmw}}, 1) | $ and is increasing in the Rayleigh quotient $\frac{\bmw'\bmd\bmd'\bmw}{\bmw'\bmSb\bmw}$.
Hence, the earlier analysis again applies.

\subsection{Inference on generated outcomes} \label{app:gen-out}
Standardizing and aggregating the outcomes using data-dependent weights require one to account for them in inference.
This subsection shows how to adjust for them.
This generated outcome issue is related to generated regressors, studied since \citet{pagan1984ier} and \citet{murphytopel1985jbes}, although I focus on the dependent variable and the issue arises through the data-dependent weights; see \citet{battagliaetal2024wp} on inference with regressors generated from unstructured data.

I impose the following high-level assumption for the large-sample analysis, so that the treatment effect and variance estimators are consistent and a suitable central limit theorem applies.
Let $\bmOm$ be a suitably defined variance matrix depending on the aggregation (e.g., $\bmOm$ is the correlation matrix for PCA), and $\hbmOn$ be a consistent estimator of $\bmOm$.
Define the operator $\vech[\cdot]$ that stacks the columns of the lower triangular portion of a $(\dy \times \dy)$ matrix into one vector of length $\ell \equiv \frac{\dy(\dy+1)}{2}$.
For example, $\vech[(\begin{smallmatrix} x_{1} & x_{2} \\ x_{2} & x_{3} \end{smallmatrix})] = (x_1, x_2, x_3)'$.

\begin{assu}
\label{assu:limit-dist}
\begin{enumerate*}[label=(\alph*)]
	\item \label{assu:limit-dist-1} $\hbmbn \Pt \bmbe$, $\hbmOn \Pt \bmOm$, and $\bmOm$ is positive definite and symmetric.
	\item $\msbD$ is the duplication matrix such that $\msbD\vech[\bmOm] = \vecc[\bmOm]$. \\
	\item \label{assu:limit-dist-2} $\sqrt{n} \left(\begin{smallmatrix}
		\hbmbn - \bmbe \\
		\vech[\hbmOn - \bmOm]
	\end{smallmatrix}\right) \Dt
	\left(\begin{smallmatrix}
		\bmZb \\ \bmZO
	\end{smallmatrix}\right)
	 \sim
	\scN \left(
		\left(\begin{smallmatrix} \zeroq \\ \zerol \end{smallmatrix}\right),
		\left(\begin{smallmatrix} \bmSi & \vssb' \\  \vssb & \vsss \end{smallmatrix}\right)
	\right)$.
\end{enumerate*}
\end{assu}

Because $\hbmbn$ already incorporates the standardization step, its asymptotic variance matrix in Assumption~\ref{assu:limit-dist} does as well.
The following proposition characterizes the limiting distribution.
The term $\bmw_0'\bmSi\bmw_0$ is the variance for fixed weights.
The remaining two terms are the corrections from estimating $\bmw_0$ through $\hbmOn$.
Both terms vanish, e.g., when $\bmM'\bmbe = \zerol$, in particular when $\bmbe = \zeroq$, but in general not when only $\tau = 0$.
This highlights that, in general, one cannot ignore the fact that the weights are estimated, rather than fixed, and that the standard errors should account for this.

\begin{prop}
\label{prop:asym-tau-general}
Let Assumption \ref{assu:limit-dist} hold. Suppose $\hbmwn \Pt \bmw_0$ and
$\sqrt{n}(\hbmwn - \bmw_0) = \bmM \sqrt{n}\vech[\hbmOn - \bmOm] + o_{\bP}(1)$ for some
$\bmM \in \bR^{\dy \times \ell}$. Write $\htan \equiv \hbmwn'\hbmbn$ and $\tau \equiv \bmw_0'\bmbe$. Then,
\[
	 \smash{
		\sqrt{n}(\htan - \tau) \Dt
		\scN (0,  \bmw_0'\bmSi\bmw_0 + 2\bmw_0'\vssb'\bmM'\bmbe + \bmbe'\bmM\vsss\bmM'\bmbe)
	 }.
\]
\end{prop}

In the following, I apply Proposition \ref{prop:asym-tau-general} to PCA and IVM.

\begin{lem}
\label{lem:asym-tau-standard-c}
Let Assumption~\ref{assu:eval} hold, and let Assumption~\ref{assu:limit-dist} hold with $\bmOm = \bmSY$.
Let $\hwpca = \argmax_{\bmw \in \cWunit} \bmw'\hbmOn\bmw$, where $\cWunit$ is defined in \eqref{eq:weight-pca}.
Suppose $\bmc'\wpca > 0$.
Then, $\hwpca \Pt \wpca$.
\end{lem}

\begin{cor}
\label{prop:asym-tau-standard}
Let the assumptions in Lemma \ref{lem:asym-tau-standard-c} hold.
Write $\htpca \equiv \hwpca'\hbmbn$ and $\tpca \equiv \wpca'\bmbe$.
Assume $\bmc'\bvec_1 > 0$ and let $\bmBv$ be as defined in Assumption~\ref{assu:pca:nondegen}.
Then,
\[
	 \smash{
		\sn (\htpca - \tpca)
		\Dt
		\scN(0,
		\wpca'\bmSi\wpca
		+
		2 \wpca' \vssb' \bmBv' \bmbe
		+
		\bmbe'\bmBv \vsss \bmBv' \bmbe
		).}
\]
\end{cor}

\begin{lem}
\label{lem:anderson-weight-deri}
Let $\bmSi$ be a symmetric positive definite matrix.
Define $\bma(\bmSi) \equiv \frac{\bmSi^{-1}\oneq}{\oneq'\bmSi^{-1}\oneq}$.
Then, $
	\frac{\text{d}\bma(\bmSi)}{\text{d}\vecc[\bmSi]}
	=
	\frac{ - (\oneq'\bmSi^{-1}) \otimes \bmSi^{-1} (\oneq'\bmSi^{-1}\oneq)
	+ (\bmSi^{-1}\oneq) [ (\oneq'\bmSi^{-1}) \otimes (\oneq'\bmSi^{-1}) ]}{
	(\oneq'\bmSi^{-1}\oneq)^2}$.
\end{lem}

\begin{cor}
\label{prop:asym-tau-anderson}
Let Assumption~\ref{assu:limit-dist} hold.
Let $\wivm \equiv \bma(\bmOm)$, $\hwivm \equiv \bma(\hbmOn)$, and
$\bmG \equiv \frac{\text{d}\bma(\bmOm)}{\text{d}\vecc[\bmOm]} \msbD$.
Write $\htivm \equiv \hwivm'\hbmbn$ and $\tivm \equiv \wivm'\bmbe$.
Then,
\[
	\smash{\sqrt{n}(\htivm  - \tivm) \Dt
	\scN(0,  \wivm'\bmSi\wivm
	+
	2 \wivm' \vssb' \bmG'\bmbe
	+
	\bmbe' \bmG \vsss \bmG' \bmbe).}
\]
\end{cor}

An alternative to Assumption \ref{assu:limit-dist}, which imposes a high-level assumption on $\hbmbn$ and $\hbmOn$, is to impose a high-level assumption on $\hbmbn$ and $\hbmwn$.
A similar result can be obtained.

\begin{prop}
\label{cor:w-if}
Let $\{\bmZ_i\}^{n}_{i=1}$ be i.i.d., where $\bmZ_i \equiv (D_i, \bmX_i', \tbmY_i')' \in \bR^{1 + \dx + \dy}$.
Assume that
$\sqrt{n}
\left( \begin{smallmatrix} \hbmwn - \bmw_0 \\ \hbmbn - \bmbe \end{smallmatrix} \right)
	=
	\frac{1}{\sqrt{n}} \sum^{n}_{i=1} \bmvp(\bmZ_i) + o_{\bP}(1)$,
where $\bmvp(\bmZ_i) = (\bmvpw(\bmZ_i)', \bmvpbe(\bmZ_i)')'$ is the corresponding influence function, $\bE[\bmvp(\bmZ_i)] = \zero_{2\dy}$, and  $\bE[\bmvp(\bmZ_i)\bmvp(\bmZ_i)']$ is finite.
Write
$\si_{\tau, \bmbe}^2 \equiv  \bmw_0'\var[\bmvpbe(\bmZ_i)]\bmw_0$,
$\si_{\tau, \bmbe, \bmw} \equiv \bmw_0'\cov[\bmvpbe(\bmZ_i), \bmvpw(\bmZ_i)]\bmbe$,
$\si_{\tau, \bmw}^2 \equiv \bmbe'\var[\bmvpw(\bmZ_i)]\bmbe$,
$\htan \equiv \hbmwn'\hbmbn$, and
$\tau \equiv \bmw_0'\bmbe$.
Then, $\sqrt{n}(\htan - \tau)
	\Dt
	\scN(0, \si_{\tau, \bmbe}^2 + 2\si_{\tau, \bmbe, \bmw} + \si_{\tau, \bmw}^2)$.
\end{prop}

\subsection{Negative PCA weights in a consumer optimization problem} \label{app:consumer-optim-2}
This subsection formalizes Example \ref{eg:consumer-optim-2} as a consumer optimization problem.
In this setup, the researcher creates a summary index by applying PCA to the two goods.
I show that negative weights arise here due to a negative correlation between the two outcomes.

\begin{eg} \label{eg:subs:numeric}
Consider the following problem with substitutes (e.g., cows and goats):
\begin{align}
	\label{eq:consumer-optim-2-1}
	\begin{split}
	(Y^\star_{i,1}, Y^\star_{i,2})
	= \argmax_{y_1, y_2 \geq 0}
	\quad & y_1^{A_i} y_2^{1 - A_i}, 	\\
	 \text{ s.t. } \quad
	&
	y_1 + p_2 y_2 \leq \text{Income}_i(D_i),
	\end{split}
\end{align}
where the price of good 1 is 1,
$p_2$ is the price of good 2,
$\text{Income}_i(D_i) \geq 0$ is the income of individual $i$ depending on $D_i \in \{0, 1\}$, and $A_i \in (0, 1)$ is a random utility parameter.

The optimal solution to \eqref{eq:consumer-optim-2-1} is
$Y^\star_{i,1} = A_i \text{Income}_i(D_i)$ and
$Y^\star_{i,2} = \frac{(1 - A_i)  \text{Income}_i(D_i)}{p_2}$.
For concreteness, let $p_2 = 2$, $A_i \in \{0.2, 0.8\}$ with equal probabilities, $D_i \in \{0, 1\}$ with equal probabilities, and $\text{Income}_i(D_i) = 10 + 5D_i + V_{i}$, where $V_i \sim \text{Uniform}[0, 5]$ and $(V_i, D_i, A_i)$ are mutually independent.
Then, by direct computation, $\corr[Y^\star_{i,1}, Y^\star_{i,2}]  \approx -0.82 < 0$.
\end{eg}

The negative correlation between $Y^\star_{i,1}$ and $Y^\star_{i,2}$ arises because, for a given income, a higher $A_i$ shifts spending from good 2 to good 1.
As a result, running PCA on these two outcomes would lead to an index with a negative weight.
However, $D_i$ has a positive effect on both outcomes.
Hence, using PCA to construct an index to represent wealth or utility here is inappropriate.

\begin{eg}[Perfect substitutes]
\label{eg:app:claim-pca-2good}
Consider the same setup and notation as in Example~\ref{eg:subs:numeric}, except that goods 1 and 2 are perfect substitutes, so the objective of \eqref{eq:consumer-optim-2-1} becomes $y_1 + A_i y_2$ (see, e.g., \citealp[Chapter 5]{varian2014bk}).
Let $A_i$ be a utility parameter that follows a normal distribution, with mean $\mu$ and variance 1, truncated to $[\uA, \oA]$.
Assume $A_i \indep \text{Income}_i(D_i)$ and $0 < \uA < p_2 < \oA$.
The optimal solution is
\begin{align}
	\label{eg:run-eg-5}
	(Y_{i,1}^\star, Y_{i,2}^\star)
	=
	\begin{cases}
		(0, \frac{\text{Income}_i(D_i)}{p_2}) & \frac{A_i}{p_2} > 1,	\\
		\{(t, \frac{\text{Income}_i(D_i) - t}{p_2}) : t \in [0, \text{Income}_i(D_i)]\} & \frac{A_i}{p_2} = 1,	\\
		(\text{Income}_i(D_i), 0)& \frac{A_i}{p_2} < 1.
	\end{cases}
\end{align}

Using~\eqref{eg:run-eg-5},
$\cov[Y_{i,1}^\star, Y_{i,2}^\star]
	=
	- \frac{\bE[\text{Income}_i(D_i)]^2 \bP[A_i < p_2] \bP[A_i > p_2]}{p_2}
	< 0$.
Hence, the two outcomes are negatively correlated, so PCA assigns a negative weight to one of the goods, which is again inappropriate for representing wealth or utility.
\end{eg}

\subsection{Other methods and discussion} \label{sec:other-methods}
Other methods have been developed for aggregation.
They typically require additional assumptions, extra data, or tuning parameters.
I briefly review some of them below.

\citet{huetal2024wp} create a linear index for the latent variable.
They require identification assumptions related to measurement errors \citep{huschennach2008ecta}.
\citet{fugreen2025wp} use the treatment or other measurements as an instrument to identify the treatment effect on the latent outcome under exclusion and independence assumptions.
\citet{stoetzeretal2025ajps} propose a hierarchical item response approach to estimate the treatment effect on the latent outcome using observed indicators and require parametric assumptions.
Item response theory has also been used in economics and psychometrics.
\citet{douglas2001psyc} shows that point identification of the latent variable would require a large number of binary proxies; see also \citet{williams2019qe}.
\citet{lubotskywittenberg2006restat} study the case in which several proxies are available for a latent independent variable.
Their strategy depends on whether the error components in the proxies are independent.
My approach does not require a specific number of outcomes or such identification assumptions.

\citet{andersonmagruder2023wp} use machine learning to choose the weights to maximize power for aggregating outcomes in pre-analysis plans.
They penalize the objective using a tuning parameter on the Herfindahl--Hirschman index to avoid placing all weight on a single outcome.
They require cross-validation and a selection among possible outcomes.
Their approach is different from mine, which uses decision theory and the MSE.

When cost and benefit data are available, researchers report measures in cost-benefit terms.
For instance, \citet{heckmanetal2010jpube} estimate the benefit-cost ratio and the rate of return for the Perry Preschool Program, and \citet{bhattetal2023qje} construct an index based on social costs.
I do not assume that researchers have access to such data.

Finally, researchers sometimes motivate aggregation by concerns about multiple testing.
Whether multiple hypothesis testing should be used depends on the decision the researcher wants to make.
See \citet{romanoetal2010are} and \citet{vivianoetal2025wp}.

\subsection{Additional results}

\subsubsection{Calibrated simulations on the PCA nonstandard asymptotic distribution} \label{app:weak-sim}
This subsection examines the impact of $\bmc'\wpca$ being close to zero (i.e., when the sign constraint becomes close to binding) using simulated data calibrated to \citet{bau2022jpe}.
The DGP is as follows.
I assume that there are $\dy = 25$ outcomes that follow the linear model \eqref{eq:sep-reg-2}, where $D_i \in \{0, 1\}$, $\bP[D_i = 1] = 0.5$, $\bmU_i \sim \scN(\zeroq, \bmSU)$, and $\bmU_i \indep D_i$.
I calibrate $\bmSU$ using the data from \citet{bau2022jpe}, so that the variance matrix of $\bmY_i$ matches the correlation matrix for the asset variables in that paper.

Assume PCA is used to aggregate the outcomes, and $\bmc = \bme_1$ (i.e., the unit vector whose first entry equals 1) is used in the PCA sign identification constraint.
I set $\beta_j = 0.5$ for $j = 16, \ldots, 20$ and $\beta_j = 0$ otherwise.
I consider four DGPs that replace $\corr[Y_{i,1}, Y_{i,2}]$ with $\corr[Y_{i,1}, Y_{i,2}] + \omega$ for $\omega \in \{-0.518, -0.100, 0.042, 0.127\}$.
The corresponding first entries $w_{\operatorname{pca},1}$ of the PC1 of $\bmSY$ are $0.070$, $0.025$, $0.010$, and $0.001$, respectively.

\begin{figure}[!t]
    \centering
    \caption{Distribution of $\htpca$ under different DGPs for Appendix~\ref{app:weak-sim}.}
    \label{fig:weak-1}
		\includegraphics[scale=1]{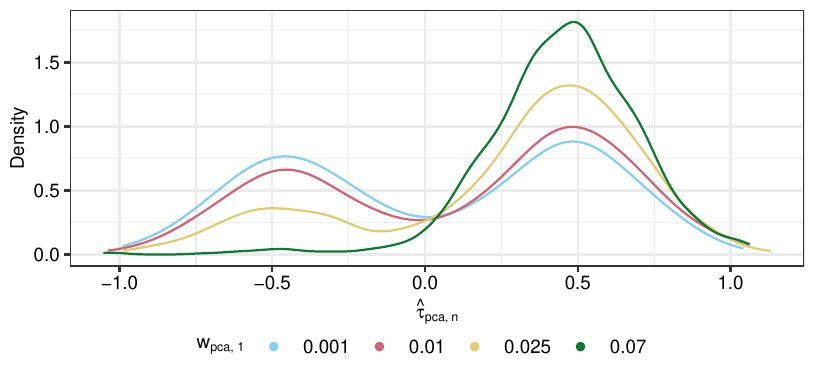}
	\mynote{
		The figure shows the sampling distribution of $\htpca$ for different values of $w_{\operatorname{pca},1}$.
		Each color represents a specific value of $w_{\operatorname{pca},1} \in \{0.070, 0.025, 0.010, 0.001\}$, obtained by adding $\omega \in \{-0.518, -0.100, 0.042, 0.127\}$, respectively, to $\corr[Y_{i,1}, Y_{i,2}]$.}
\end{figure}

Figure \ref{fig:weak-1} shows the distribution of $\htpca \equiv \hwpca'\hbmbn$ for each DGP using 1,000 replications of $n = 500$ observations.
As $w_{\operatorname{pca},1} \longrightarrow 0$, the distribution of $\htpca$ becomes bimodal.
This suggests that using ``standard'' inference methods can fail to control size, as in Proposition~\ref{prop:pca-main-weak}.

\subsubsection{Sufficient condition for IVM weights to be positive}
\label{app:ivm-pos}

\begin{prop}
\label{prop:ivm-pos-1}
Let Assumption \ref{assu:outcome} hold and $\bmSY$ be positive definite.
Let $\si_{Y, ij}$ be the $(i,j)$-th entry of $\bmSY$ for $i, j = 1, \ldots, \dy$.
Define $\rho_{Y, ij} \equiv \frac{\si_{Y,ij}}{\sqrt{\si_{Y,ii} \si_{Y,jj}}}$ for $i, j = 1, \ldots, \dy$,
$\underline\si \equiv \min_{j = 1, \ldots, \dy}  \sqrt{\si_{Y,jj}}$, and
$\overline\si \equiv \max_{j = 1, \ldots, \dy} \sqrt{\si_{Y,jj}}$.
The components of $\wivm$ in \eqref{eq:anderson-w} are strictly positive when
$\max_{i = 1, \ldots, \dy} \ \sum_{j \neq i} | \rho_{Y, ij} |  < \frac{\underline\si}{\overline \si + \underline \si}$.
\end{prop}

\subsubsection{Additional example that shows SA can lead to overcounting} \label{app:equal-v-eg}
Using a stylized example with duplicated outcomes, I show that SA can lead to overcounting.
The example models the limiting case in which some outcomes are highly correlated.
It does not suggest that duplicated outcomes are used in practice.

\begin{eg}[Duplicated outcomes and SA]
\label{eg:dup-out-1}
Consider the setting in Example \ref{eg:binary-te-1} with $\dy \geq 3$ outcomes, where 	$Y_{i,2} = \cdots = Y_{i,\dy}$
are identical copies, and $\var[Y_{i,j}] = 1$ for $j = 1, \ldots, \dy$.
Assume homogeneous treatment effects $\be_1 = \cdots = \be_{\dy} = \obe$, homoskedastic errors, and a known standardization.
Let $\vs^2_{1,\hbmbe}$ and $\vs^2_{2,\hbmbe}$ be the $(1,1)$ and $(2,2)$ entries of $\bmSb$, respectively, and let $\vs_{12,\hbmbe}$ be its $(1,2)$ entry.
Assume that $\vs_{12,\hbmbe} < \min\{\vs^2_{1,\hbmbe}, \vs^2_{2,\hbmbe}\}$.
Equal weights $\tfrac{1}{\dy}$ overcount the duplicated outcome as they put
$\tfrac{\dy - 1}{\dy}$ on $\be_2$ but only $\tfrac{1}{\dy}$ on $\be_1$, even though
both effects equal $\obe$ and both outcomes have unit variance. By contrast, minimizing $\bmw'\bmSb\bmw$
over $\bmw \in \cWcvx$ yields $w_1^\star = \tfrac{1}{2}$ and
$\sum^{\dy}_{j=2} w_j^\star = \tfrac{1}{2}$, with $w_2^\star, \ldots, w_{\dy}^\star \geq 0$ but otherwise unrestricted.
The solution is not unique.
This is because $Y_{i,2}, \ldots, Y_{i,\dy}$ are identical and only their total weight matters.
But every solution puts the weight evenly between outcome 1 and the duplicated block, so weighting by $\bmSb$ avoids the overcounting. See Appendix \ref{app:proof:app:a} for a proof.
\end{eg}

\subsubsection{Auxiliary results on the linear model} \label{supp:results-weight}
\begin{lem}
\label{lem:linear-out}
Consider $\{(D_i, \bmX_i', \bmY_i')\}^{n}_{i=1}$,
where $D_i \in \bR$,
$\bmX_i \in \bR^{\dx}$, $\bmY_i \equiv (Y_{i,1}, \ldots, Y_{i,\dy})' \in \bR^{\dy}$, and the intercept is contained in $\bmX_i$.
Consider the following linear models.
\begin{enumerate}[label=(\alph*)]
\item For each outcome $j = 1, \ldots, \dy$, consider the linear model
\begin{align}
	\label{eq:lem:linear-out-1}
	Y_{i,j} =  \beta_j D_i + \bmze_j'\bmX_i + U_{i,j},
\end{align}
where $\be_j \in \bR$, $\bmze_j \in \bR^{\dx}$, $U_{i,j} \in \bR$, and $i = 1, \ldots, n$.
\item Let $\bmw_n \in \bR^{\dy}$ be a vector of weights that is potentially data-dependent, and consider
\begin{align}
	\label{eq:lem:linear-out-2}
	\bmw_n'\bmY_i =  \tau D_i + \obmze'\bmX_i + \oU_i,
\end{align}
where $\tau \in \bR$, $\obmze \in \bR^{\dx}$, and $\oU_i \in \bR$ for $i = 1, \ldots, n$.
\end{enumerate}
Let $\hbmbn \equiv (\hbe_{n,1}, \ldots, \hbe_{n,\dy})' \in \bR^{\dy}$, where $\hbe_{n,j}$ is the ordinary least squares (OLS) estimator of $\be_j$ in~\eqref{eq:lem:linear-out-1}.
Let $\htan$ be the OLS estimator of $\tau$ in \eqref{eq:lem:linear-out-2}.
Write $\bmD \equiv (D_1, \ldots, D_n)'$, $\bmX \equiv (\bmX_1, \ldots, \bmX_n)'$, $\bmP \equiv \bmX(\bmX'\bmX)^{-1}\bmX'$, and $\bmM \equiv \bmI_n - \bmP$, where $\bmI_n$ is the identity matrix.
Assume $\bmX'\bmX$ is invertible and $\bmD'\bmM\bmD \neq 0$.
Then, $\htan = \bmw_n'\hbmbn$.
\end{lem}

Lemma~\ref{lem:linear-out} follows from two facts: the same linear estimator is applied to each outcome and to $\bmw_n'\bmY_i$, and the outcomes are aggregated linearly.
Hence, the lemma extends to other estimators with the same structure.
One example is two-stage least squares, provided that the same instruments, covariates, and sample are used for each outcome and for $\bmw_n'\bmY_i$.
Such an estimator can be studied in the same way with $\bmD$ replaced by its fitted values from the first-stage regression on the instruments and $\bmX_i$.

The following lemma formalizes the implications of Remark \ref{re:postprocess}.
When standardizing $\bmw_n'\bmY_i$, one sets $a_n$ equal to the sample mean of $\bmw_n'\bmY_i$ and $b_n$ equal to the sample standard deviation of $\bmw_n'\bmY_i$.
In the case of rescaling $\bmw_n'\bmY_i$ to $[0, 1]$, one instead sets $a_n = \min_{i=1,\ldots,n} \bmw_n'\bmY_i$ and $b_n = \max_{i=1,\ldots,n} \bmw_n'\bmY_i - \min_{i=1,\ldots,n} \bmw_n'\bmY_i$.
\begin{lem}
\label{lem:linear-post}
Consider the same assumptions and notation as in Lemma \ref{lem:linear-out}.
Let $a_n, b_n \in \bR$ be potentially data-dependent.
Define $Z_i \equiv \frac{\bmw_n'\bmY_i - a_n}{b_n}$ for $i = 1, \ldots, n$, where $b_n \neq 0$.
Consider
\begin{align}
	\label{eq:lem:linear-out-3}
	Z_i = \tau_z D_i + \bmze_z'\bmX_i + V_i,
\end{align}
where $\tau_z \in \bR$, $\bmze_z \in \bR^{\dx}$, and $V_i \in \bR$.
Then, the OLS estimator of $\tau_z$ is $\widehat\tau_{z,n} = \frac{\htan}{b_n}$.
\end{lem}

When the post-processing step yields $b_n \neq 1$, $\widehat\tau_{z,n} \neq \obe$ even if treatment effects are  homogeneous $\hbe_{n,j} = \obe \neq 0$ for $j = 1, \ldots, \dy$ and the weights sum to one $\bmw_n'\oneq = 1$.

\subsubsection{Auxiliary lemmas}

\begin{lem}
\label{lem:supp:asym-tau-1}
Let Assumption \ref{assu:eval} hold.
Let $\bvec_1$ be any unit-length leading eigenvector of $\bmSY$ such that $\bmc'\bvec_1 \neq 0$.
Let $\wpca$ be the optimal solution to the problem \eqref{eq:pca-problem}.
Then, $\wpca =   \sign(\bmc'\bvec_1) \bvec_1$.
\end{lem}

\begin{lem}
 \label{prop:t-1}
Let Assumptions \ref{assu:eval}, \ref{assu:main-reg-ols}, and \ref{assu:pre-gen-1} hold.
Consider the linear model \eqref{eq:genout-reg-1} and suppose $\bmbe \neq \zeroq$.
Let $\hRRw$ be the sample $R$-squared of \eqref{eq:genout-reg-1}.
\begin{enumerate}[label=(\alph*)]
	\item \label{lem:tstat-r-1}
		For any $\bmw \neq \zeroq$,
		$
			\hRRw
			\Pt
			\RRw
			\equiv
			\frac{
				\var[D_i] (\bmw'\bmbe)^2
			}{
				\bmw' \bmSY \bmw
			}$.
	\item \label{prop:t-1a} The optimal solutions to
	$\max_{\bmw : \|\bmw\|_2^2 = 1} \RRw$
	are $\pm \bmwr$, where
	$ \bmwr \equiv
		\frac{\bmSY^{-1} \bmbe}{\sqrt{\bmbe'\bmSY^{-2}\bmbe}}$.
	\item \label{prop:t-1b} $\bmwr = \pm \bvec_1$ if and only if $\bvec_1$ is parallel to $\bmbe$, where $\bvec_1$ is the PC1 defined in Section \ref{sec:pca-outline}.
\end{enumerate}
\end{lem}

\section{Supplemental results for Section~\ref{sec:var-min-weight}}
\label{sec:inference-b0}

In this section, I analyze the properties of the variance-minimizing weights $\hwvmc$.
These weights solve~\eqref{eq:vbmin-w}.

\begin{assu}
\label{assu:minimax-asym} \text{ }
Assumption~\ref{assu:limit-dist} holds with $\bmOm$ replaced by $\bmSi$.
\end{assu}

\begin{prop}
\label{prop:minimax-asym-c-b0}
Let Assumption \ref{assu:minimax-asym} hold.
Then, $\hwvmc \Pt \wvmc$.
\end{prop}

\begin{thm}
\label{thm:minimax-asym-d-b0}
Let the assumptions in Proposition~\ref{prop:minimax-asym-c-b0} hold.
Define $\tbmZ \equiv 2 (\wvmc' \otimes \bmI_{\dy}) \msbD \bmZS$.
Let $\bmmu^\star \equiv (\mu^\star_{1}, \ldots, \mu^\star_{\dy})'$ be the Lagrange multipliers for the nonnegativity constraints of~\eqref{eq:vbmin-w} with $\hbmSn$ replaced by $\bmSi$, when evaluated at $\wvmc$.
Let $\cJ_{\mathrm{a}} \subseteq \{1, \ldots, \dy\}$ be the set of indices for the active inequality constraints for $\cWcvx$ at $\wvmc$, and let
$\cJ_{\mathrm{a},+}(\bmmu^\star) \equiv \{ j \in \cJ_{\mathrm{a}}: \mu^\star_{j} > 0\}$.
Let $\bmh^\star(\bmze)$ be the optimal solution to
\begin{align}
	\label{eq:thm:minimax-asym-d-b0-1}
	\begin{split}
	\min_{\bmh \in \bR^{\dy}} \quad & \bmh'\bmze + \bmh'\bmSi\bmh,	\\
	\text{s.t.} \quad & \bmh'\oneq = 0,  \qquad
	h_j = 0 \text{ for $j \in \cJ_{\mathrm{a},+}(\bmmu^\star)$,}	\qquad
	h_j \geq 0 \text{ for $j \in \cJ_{\mathrm{a}} \backslash \cJ_{\mathrm{a},+}(\bmmu^\star)$.}
	\end{split}
\end{align}
Then, $\sqrt{n}(\hwvmc - \wvmc) \Dt \bmh^\star(\tbmZ)$.
\end{thm}

The following corollary shows that the limiting distribution of $\sn(\hwvmc - \wvmc)$ is normal if all entries of $\wvmc$ are positive.

\begin{cor}
\label{cor:minimax-asym-dtau-b0}
Consider the same notation and assumptions as in Theorem \ref{thm:minimax-asym-d-b0}.
Suppose $\wvmc > \zeroq$ and define \smash{$\Bvmc \equiv -\bmSi^{-1} (
		\bmI_{\dy}
		- \oneq \frac{\oneq'\bmSi^{-1}}{\oneq'\bmSi^{-1}\oneq}
	) (\wvmc' \otimes \bmI_{\dy}) \msbD$}. Then,
$ \sqrt{n}(\hwvmc - \wvmc) \Dt \scN(\zeroq, \Bvmc\csss\Bvmc')$.
\end{cor}

\section{Supplemental results for Section \ref{sec:decision}}  \label{app:main-supp-decision}

\subsection{Asymptotic validity of the FLCI} \label{app:flci-valid}
In this subsection, I establish the asymptotic validity of the FLCI.
Let $\cP$ be the class of joint distributions $P$ of the data $\{\bmZ_i\}^n_{i=1}$, where $\bmZ_i \equiv (D_i, \bmX_i', \tbmY_i')'$, and let $\Theta$ be the parameter space for $\theta$.
For $P \in \cP$, let $\bmbe_P$ denote the vector of treatment effects and $\bmStp$ the asymptotic variance of $\sn(\hbmbn - \bmbe_P)$ under $(\theta, P)$.
In addition, let $\fSnb \equiv \{ (\theta, P) \in \Theta \times \cP : \sn (\bmbe_P - \theta\oneq) \in \cS(B)\}$ as in \citet{armstrongkolesar2021qe}.
I require the following uniformity assumption.

\begin{assu}
\label{assu:uniform-flci}
\text{ }
\begin{enumerate}[label=(\alph*)]
\item $\hbmSn$ is consistent for $\bmStp$ uniformly over $(\theta, P) \in \fSnb$.
\item For any $(\theta, P) \in \fSnb$, $\bmStp \in \cM$, where $\cM$ is a compact set of positive definite matrices with eigenvalues bounded below by $\la_{\mathrm{lb}} > 0$ and above by $\la_{\mathrm{ub}} < \infty$ where $\la_{\mathrm{ub}} \geq \la_{\mathrm{lb}}$.
\item $\sqrt{n}(\hbmbn - \bmbe_P)$ converges in distribution to $\scN(\zeroq, \bmStp)$ uniformly over $(\theta, P) \in \fSnb$.
\end{enumerate}
\end{assu}

The restriction $\sn(\bmbe_P - \theta\oneq) \in \cS(B)$ in $\fSnb$ is an asymptotic device to control misspecification.
It should not be interpreted as $\bmbe_P$ being equal to $\theta \oneq$ as $n \longrightarrow \infty$.

\begin{prop}
\label{prop:flci-asym-valid-new}
Let $\alpha \in (0, 1)$.
Let Assumptions \ref{assu:outcome}, \ref{assu:convex-m2}, and \ref{assu:uniform-flci} hold.
For a finite $B \geq 0$, let $\hbmwn$ be the weight estimated from the minimax problem using $B$ or the adaptive problem over $\cB = [\uB, \oB]$ with $0 \leq \uB < \oB < \infty$, with $\bmSi$ replaced by $\hbmSn$ and $B \in \cB$.
Let $\hcvn$ be the smallest critical value such that $\max_{\bmb \in \cS(B)} \bP[|\scN(\frac{\hbmwn'\bmb}{\sqrt{\hbmwn'\hbmSn\hbmwn}}, 1)| > \hcvn] \leq \alpha$.
Then, $\cI_n$ in~\eqref{eq:minimax-flci-2} satisfies $\liminf_{n \to \infty} \inf_{(\theta, P) \in \fSnb} \bP_P[\theta \in \cI_n] \geq 1 - \alpha$.
\end{prop}

\subsection{Details for Remark \ref{re:minimax-binf}} \label{app:re:minimax-binf}
The following lemma shows that, as $B \longrightarrow \infty$, the minimax solution converges to the minimizer of $m(\bmw)$.
The intuition is that as $B$ grows large, the cost of not considering the variance information becomes ``negligible.''
Hence, the researcher would focus on $m(\bmw)$.

\begin{lem}
\label{lem:limit-minimax}
Let Assumption \ref{assu:convex-m2} hold and $m(\bmw)$ be strictly convex and continuous in $\bmw \in \cWcvx$.
Then, as $B \longrightarrow \infty$, $\bmwsb \longrightarrow \argmin_{\bmw \in \cWcvx} \ m(\bmw)$,
where $\bmwsb$ is defined in \eqref{eq:minimax-case1-1soln}.
\end{lem}

\begin{cor}
\label{cor:limit-minimax}
Let the same assumptions as in Lemma \ref{lem:limit-minimax} hold, and let the parameter space in \eqref{eq:param-b-1} use the $\ell_p$-norm for some $p \in (1, \infty)$, as in Example \ref{eg:param-b}.
Then, $\lim_{B \to \infty} \bmwsb = \frac{1}{\dy}\oneq$.
\end{cor}

\subsection{Details for Example \ref{eg:run-1}} \label{app:eg:run-1}
Suppose the Euclidean norm is used to bound $\bmb = (b_1, b_2)'$ in $\cS(B)$ as in the main text.
Since $w_1 + w_2 = 1$, the variance and the maximum squared bias can be written in terms of $w_1 \in [0, 1]$:
\begin{align}
	\label{eq:run-1-1}
	\begin{split}
	V(w_1)
	& = (1 + \si_2^2 - 2\rho\si_2) w_1^2 + 2(\rho \si_2 - \si_2^2) w_1 + \si_2^2,	\\
	\oM(B, w_1)
	&
	= B^2 \|(w_1, 1 - w_1)\|^2_2
	= B^2 ( 2w_1^2 - 2w_1 + 1)
	\equiv B^2 m(w_1),
	\end{split}
\end{align}
where $m(w_1) \equiv 2w_1^2 - 2w_1 + 1$.
The minimax risk is
\begin{align}
	\Rmm(B)
	& =
	\Rmax(B, w_1^\star(B))
	=
	\begin{cases}
		\si_2^2 + B^2
		& \text{ if $\si_2^2 - \rho \si_2 \leq - B^2$,}	\\
		1 + B^2
		& \text{ if $1 - \rho \si_2 \leq - B^2$,}	\\
		\frac{(1 - \rho^2)\si_2^2 + (1 + \si_2^2)B^2 + B^4}{1 + \si_2^2 - 2\rho\si_2 + 2B^2}
		& \text{ otherwise.}
	\end{cases}
	\label{eq:run-1-3}
\end{align}

\subsubsection{Adaptive problem}  \label{sec:proof-claim-1-app}
Assume that $\cB = [0, \infty]$ and $\rho\si_2 < 1$.
The adaptive problem is
\begin{align}
	\label{eq:run-1-4}
	\begin{split}
		\min_{t, w_1 \in \bR}
		\quad  t \qquad
		\text{s.t.} \qquad & \frac{V(w_1)}{\Rmm(0)} \leq  t,	\quad
		2(2w_1^2 - 2w_1 + 1) \leq t, \quad
		0 \leq w_1 \leq 1,
	\end{split}
\end{align}
where $V(w_1)$ is defined in \eqref{eq:run-1-1}.
The Lagrangian is
\begin{equation}
	\label{eq:run-1-5}
	\cL = t + \mu_1[\Rmm(0)^{-1}V(w_1) - t] + \mu_2(4w_1^2 - 4w_1 + 2 - t) + \mu_3(w_1 - 1) + \mu_4(-w_1),
\end{equation}
where $\{\mu_l\}^4_{l=1}$ are the Lagrange multipliers.
In Appendix \ref{app:proof:sec:c}, I show that the boundary solutions (i.e., $w_1 = 0$ or $w_1 = 1$) cannot be optimal because they violate the Karush--Kuhn--Tucker (KKT) conditions.
The optimal solution is interior and can be found by rearranging the stationarity condition $\mu_1 \Rmm(0)^{-1}[ 2(1 + \si_2^2 - 2\rho\si_2) w_1 + 2(\rho \si_2 - \si_2^2) ] + (1 - \mu_1) (8w_1 - 4) = 0$ as in~\eqref{eq:run-1-5-kkt-2}.

\subsection{Additional details for the parameter space}  \label{app:param}

\subsubsection{Shape restrictions} \label{sec:param-shape}
Consider shape restrictions of the form $\bmQ\bmb \leq \zero_l$, where $\bmQ \in \bR^{l\times \dy}$, and the following parameter spaces, the second of which restricts the magnitudes of $\bmb$ (e.g., the entrepreneurial spirit index in Example \ref{eg:param-shape}):
\begin{align}
	\cS(B) & = \{\bmb \in \bR^{\dy} : \|\bmb\|_p \leq B, \bmQ\bmb \leq \zero_l\},
	\label{eq:param-b-2} \\
	\cS_{\mathrm{abs}}(B) & = \{ \bmb \in \bR^{\dy} : \|\bmb\|_p \leq B, \bmQ|\bmb| \leq \zero_l\}. \label{eq:param-b-2-abs}
\end{align}
The following proposition simplifies the computation of the maximum squared bias.

\begin{prop}
\label{prop:abs-ineq}
Let $p \geq 1$, $B \geq 0$, $\bmQ \in \bR^{l\times \dy}$, and $\bmw \in \cWcvx$.
Let $\cS_{\mathrm{aug}}(B) \equiv \{ \bmb \in \bR^{\dy} : \|\bmb\|_p \leq B, \bmQ\bmb \leq \zero_l, \bmb \geq \zeroq\}$.
Then, $\max_{\bmb \in \cS_{\mathrm{abs}}(B)} (\bmw'\bmb)^2
	= \max_{\bmb \in \cS_{\mathrm{aug}}(B)} (\bmw'\bmb)^2$,
where $\cS_{\mathrm{abs}}(B)$ is defined in \eqref{eq:param-b-2-abs}.
\end{prop}

The above proposition shows that although the two sets $\cS_{\mathrm{abs}}(B)$ and $\cS_{\mathrm{aug}}(B) $ differ, they lead to the same maximum value.
I can also write $\oM(B, \bmw) = B^2 m(\bmw)$, where $m(\bmw) \equiv \max_{\tbmb \in \cS(1)} (\bmw'\tbmb)^2$ is convex in $\bmw$ and $\cS(B)$ is as defined in  \eqref{eq:param-b-2}.

\subsubsection{Subjective weights and a model of communication} \label{sec:param-comm}
Another concern that researchers may have is related to the reader's subjective weights on the outcomes or the communication of the weighted average of treatment effects $\hta$ to the reader \citep[see, for instance,][for recent work on scientific communication]{andrewsshapipro2021ecta, frankelkasy2022aejmi, kasyspiess2024wp}.
Readers may think that $\theta$ is a weighted average of $\bmbe$ but have a different view on what the weights should be.

The communication problem fits within the decision framework once I set $\theta = \bmga'\bmbe$, where $\bmga \in \cWcvx$ is a vector of the reader's subjective weights that can be known or unknown to the researcher.
Restricting $\bmga$ to $\cWcvx$ keeps the weights interpretable.
The choice of $\bmga$ depends on the context.
I consider the following three cases for $\bmga$:
\begin{enumerate}
	\item \label{list:amb:case1} A known $\bmga \in \cWcvx$.
	\item \label{list:amb:case2} An unknown $\bmga \in \cWcvx$.
	\item \label{list:amb:case3} A known reference weight $\bmeta \in \cWcvx$ such that $\bmga \in \cWcvx$, $\|\bmga - \bmeta\|_p \leq \Delta$, and $\Delta \geq 0$.
\end{enumerate}
Cases \ref{list:amb:case1} and \ref{list:amb:case2} are special cases of Case \ref{list:amb:case3}.
This is because setting $\Delta = 0$ in Case \ref{list:amb:case3} corresponds to Case~\ref{list:amb:case1} with $\bmeta = \bmga$.
Setting a ``large'' value of $\Delta$ in Case \ref{list:amb:case3} corresponds to allowing for any $\bmga \in \cWcvx$ as in Case \ref{list:amb:case2}.

Recall from~\eqref{eq:decision-1} and~\eqref{eq:decision-bias-1} that $\hbmbe \sim \scN(\bmbe, \bmSi)$, where $\bmSi$ is known, and that $\bmb = \bmbe - \theta \oneq$.
Substituting $\theta = \bmga'\bmbe$ gives $\bmb = \bmbe - (\bmga'\bmbe)\oneq$.
The following lemma shows that, under~\eqref{eq:decision-1} and the constraint $\|\bmb\|_p \leq B$, finding the ``worst-case $\bmb$'' does not require searching over $\bmbe \in \bR^{\dy}$.
It only requires the orthogonality condition $\bmga'\bmb = 0$.
It also shows that the communication model can be studied as a shape restriction.
\begin{lem}
\label{lem:supp:ambn-1}
Consider $B \geq 0$, $\bmga \in \cWcvx$, and the following sets:
$\cS_0(B, \bmga) \equiv \{\bmb \in \bR^{\dy} : \|\bmb\|_p \leq B, \bmb = \bmbe - (\bmga'\bmbe) \oneq \text{ for some $\bmbe \in \bR^{\dy}$}\}$ and
$\cS_1(B, \bmga) \equiv \{\bmb \in \bR^{\dy} : \|\bmb\|_p \leq B, \bmga'\bmb = 0\}$.
Then, $\cS_0(B, \bmga)  = \cS_1(B, \bmga)$.
\end{lem}

Using Lemma \ref{lem:supp:ambn-1}, I write the parameter space for Case~\ref{list:amb:case2} as
\begin{equation}
	\label{eq:app:amb:case2n-0}
	\cS_2(B) = \{ (\bmb, \bmga) \in \bR^{2\dy}: \|\bmb\|_p \leq B, \bmga'\bmb = 0, \bmga \in \cWcvx\}
\end{equation}
for a given $p \geq 1$.
The parameter space for Case~\ref{list:amb:case3} can be written as
\begin{align}
	\label{eq:app:amb:case3n-0}
	\begin{split}
	\cS_3(B, \Delta, \bmeta) =
	\{ (\bmb, \bmga) \in \bR^{2\dy} : \|\bmb\|_p \leq B, \|\bmde\|_p \leq \Delta, \bmga =  \bmeta + \bmde \in \cWcvx, \bmga '\bmb = 0
	\},
	\end{split}
\end{align}
where $B, \Delta \geq 0$ and $\bmeta \in \cWcvx$.
The following proposition bridges the analysis of the three cases.

\begin{prop}
\label{prop:app:ambiguity:case3n-1}
Let $B \geq 0$ and $\bmeta_0 \in \cWcvx$ be given.
Consider the notation and parameter spaces described in Lemma~\ref{lem:supp:ambn-1}, \eqref{eq:app:amb:case2n-0}, and \eqref{eq:app:amb:case3n-0}.
\begin{enumerate}[label=(\alph*)]
	\item \label{prop:app:ambiguity:case3n-1a}
	If $\Delta = 0$, then $\cS_3(B, 0, \bmeta_0) = \{ (\bmb, \bmga) \in \bR^{2\dy} : \bmb \in \cS_1(B, \bmeta_0), \bmga = \bmeta_0\}$.
	\item \label{prop:app:ambiguity:case3n-1b}
	If $\Delta  \geq \underline \Delta$, where $\underline \Delta \equiv \max_{\bmy \in \cWcvx} \| \bmy - \bmeta_0\|_p$, then $\cS_3(B, \Delta, \bmeta_0) = \cS_2(B)$.
\end{enumerate}
\end{prop}

\subsection{Auxiliary results} \label{app:decision:aux}

\begin{prop}
\label{prop:minimax-asym-c}
Let Assumption \ref{assu:minimax-asym} hold and consider a finite $B \geq 0$.
Let $\hbmSn$ be a consistent estimator of $\bmSi$.
Let $\hbmwn$ be the solution to the following program
\begin{equation}
	\label{eq:minimax-case1-1n}
	\hbmwn
	\equiv \argmin_{\bmw \in \cWcvx} \ \hRmaxn(B, \bmw)
	= 		 \argmin_{\bmw \in \cWcvx} \ [\bmw'\hbmSn\bmw + \oM(B, \bmw)],
\end{equation}
and $\bmwsb$ be as defined in \eqref{eq:minimax-case1-1soln}.
Then, $\hbmwn \Pt \bmwsb$.
\end{prop}

\begin{prop}
\label{prop:minimax-asym-d}
Consider the same notation and assumptions as in Proposition \ref{prop:minimax-asym-c}.
Let $\obmw(\bmze)$ be the optimal solution to the following program
\begin{align}
	\label{eq:lem:minimax-asym-a3-1}
	\min_{\bmw \in \cWcvx} \ \ [\Rmax(B, \bmw) + \bmze'\bmw],
\end{align}
where $\bmze \in \bR^{\dy}$.
Then,
$\sn ( \hbmwn - \bmwsb)
	=
	\sn \left[ \obmw\left(2(\hbmSn - \bmSi)\bmwsb\right)  - \bmwsb \right]
	+
	o_{\bP}(1)$.
\end{prop}

\begin{lem}
\label{lem:supp:areg-1}
Let Assumptions~\ref{assu:sig-pd} and \ref{assu:convex-m2} hold.
Let $\bmwsb$ be the solution to the minimax problem \eqref{eq:minimax-case1-1}.
Then, $\Rmm(B_2) \leq \Rmm(B_1) + (B_2^2 - B_1^2) m(\bmw^\star(B_1))$ for any $B_1, B_2 \geq 0$.
\end{lem}

\begin{lem}
\label{lem:supp:areg-2}
Let Assumptions~\ref{assu:sig-pd} and \ref{assu:convex-m2} hold.
Then, $m(\bmwsb)$ is nonincreasing in $B$.
\end{lem}

\begin{prop}
\label{prop:areg-new-1}
Let Assumptions \ref{assu:sig-pd} and \ref{assu:convex-m2} hold.
Consider the minimax and adaptive problems defined in \eqref{eq:minimax-case1-1} and \eqref{eq:areg-1}, respectively.
Then, for any $\bmw \in \cWcvx$, $N(B, \bmw)$ defined in the proof of Proposition \ref{prop:areg-main}\ref{prop:areg-main-a} is nondecreasing in $B \geq 0$.
\end{prop}

\begin{lem}
\label{lem:w-unif-minimax}
Let Assumption \ref{assu:convex-m2} hold.
Assume that $\cM$ is a compact set of positive definite matrices with eigenvalues bounded below by $\la_{\mathrm{lb}} > 0$ and above by $\la_{\mathrm{ub}} < \infty$ where $\la_{\mathrm{ub}} \geq \la_{\mathrm{lb}}$.
For a given finite $B \geq 0$, the optimal solution to the minimax problem in \eqref{eq:minimax-case1-1} is uniformly continuous in $\bmSi \in \cM$.
\end{lem}

\begin{lem}
\label{lem:w-unif-adapt}
Consider the same assumptions and notation as in Lemma \ref{lem:w-unif-minimax}.
Write $m(\bmw) = \max_{\bmb \in \cS(1)} (\bmw'\bmb)^2$, where $\cS(B)$ is the parameter space as in Section \ref{sec:param-space}.
For a given $\cB = [\uB, \oB]$ with $0 \leq \uB < \oB < \infty$, the optimal solution to the adaptive problem in \eqref{eq:areg-3b} is uniformly continuous in $\bmSi \in \cM$.
\end{lem}

\begin{lem}
\label{lem:minimax-asym-b-main}
Let Assumption \ref{assu:minimax-asym} hold.
Consider \eqref{eq:minimax-case1-1n}.
Let $\bmwsb$ be the optimal solution to \eqref{eq:minimax-case1-1} as defined in \eqref{eq:minimax-case1-1soln}.
Define $\msdn(\bmw) \equiv \hRmaxn(B, \bmw) - \Rmax(B, \bmw)$
for a given finite $B \geq 0$ and $\bmw \in \cWcvx$.
Then,
\begin{enumerate*}[label=(\alph*)]
	\item \label{lem:minimax-asym-b1} $\msdn(\bmw)$ is Lipschitz continuous and differentiable at $\bmwsb$.
	\item \label{lem:minimax-asym-b2} $\|\nabla \msdn(\bmw)\|_2 = O_{\bP}(\ns)$.
	\item \label{lem:minimax-asym-b3} There exists a neighborhood $\scW$ of $\bmwsb$ such that
  $\sup_{\bmw \in \scW}
	\frac{\|\nabla \msdn(\bmw) - \nabla \msdn(\bmwsb)\|_2}{\ns + \|\bmw - \bmwsb\|_2}
	= o_{\bP}(1)$.
\end{enumerate*}
\end{lem}

\begin{lem}
\label{lem:minimax-asym-a3}
Consider the same notation and assumptions as in Proposition \ref{prop:minimax-asym-c}.
Let $\obmw(\bmze)$ be the optimal solution to \eqref{eq:lem:minimax-asym-a3-1}.
There exist a neighborhood $\scW$ of $\bmwsb$ and a positive constant $\msCw$ such that for any $\bmze$ in a neighborhood of $\zeroq$, \eqref{eq:lem:minimax-asym-a3-1} has an optimal solution $\obmw(\bmze) \in \scW$ and
\begin{equation}
	\label{eq:lem:minimax-asym-a3-2}
	\Rmax(B, \bmw) + \bmze'[\bmw - \obmw(\bmze)]
	\geq
	\Rmax(B, \obmw(\bmze)) + \msCw \|\bmw - \obmw(\bmze)\|^2_2
\end{equation}
for any $\bmw \in \cWcvx \cap \scW$.
\end{lem}

\section{Details on computation} \label{app:comp}

\subsection{Minimax approach} \label{sec:comp-minimax}

\subsubsection{Minimax approach for a parameter space that bounds misspecification}
Suppose that the researcher places a bound on the misspecification as in Example \ref{eg:param-b} with a finite $B > 0$.
When the $\ell_1$-norm is used in the parameter space \eqref{eq:param-b-1}, the maximum squared bias can be written as $\oM(B, \bmw) = B^2 \|\bmw\|^2_\infty = B^2 \max_{j = 1, \ldots, \dy} w_j^2$.
Then, the optimization problem becomes the following quadratic program with linear constraints:
\begin{align*}
		\min_{\bmw \in \bR^{\dy},\ t \in \bR}\quad
		& (\bmw' \bmSi \bmw + B^2 t^2) ,	\\
		\text{s.t.} \quad & \bmw \leq t \oneq,
		\quad \bmw'\oneq = 1,
		\quad \bmw \geq \zeroq,
\end{align*}
where I introduce the auxiliary variable $t$ for $\|\bmw\|_\infty = \max_{j = 1, \ldots, \dy} w_j$ (see, for instance, \citealp[page 17]{bertsimastsitsiklis1997bk}).
Similarly, when the $\ell_2$-norm is used in~\eqref{eq:param-b-1}, I have $\oM(B, \bmw) = B^2 \|\bmw\|^2_2$, and the minimax problem becomes a quadratic program.

\subsubsection{Minimax approach for a parameter space that has shape restrictions}
\label{sec:comp-minimax-shape}
Suppose shape restrictions are imposed as in Example \ref{eg:param-shape} or Appendix \ref{sec:param-shape} (see also Proposition~\ref{prop:abs-ineq}).
The minimax problem is
\begin{align}
	\label{eq:comp-minimax-1}
	\begin{split}
	\min_{\bmw \in \cWcvx, \ t \in \bR} \quad & (\bmw'\bmSi\bmw + t^2),	\\
	\text{s.t.} \quad & \max \Big\{  \max_{\bmb \in \cS(B)} ( \bmw'\bmb ),
	 \max_{\bmb \in \cS(B)} (-\bmw'\bmb )
	\Big\} \leq t.
	\end{split}
\end{align}
The inequality constraint can be further written as two constraints
$\max_{\bmb \in \cS(B)} \bmw'\bmb \leq t$ and
$\max_{\bmb \in \cS(B)} (-\bmw'\bmb) \leq t$.
Note that $\bmQ\bmb \leq \zero_l$ is affine and, for $B > 0$, Slater's condition \citep[Chapter 5.2.3]{boydbandenberghe2004bk} is satisfied (e.g., $\bmb = \zeroq$ satisfies $\|\bmb\|_p < B$ for $B > 0$).
Under these conditions, the inner program can be written as
\begin{align}
	\max_{\bmb \in \cS(B)} \bmw'\bmb
	& =
	\min_{\bmmu \geq \zero_l} \
	\left\{ \max_{\|\bmb\|_p \leq B} \
	[ \bmw'\bmb + \bmmu'(-\bmQ\bmb) ]
	\right\}
	=
	\min_{\bmmu \geq \zero_l} \ (  B \|\bmw - \bmQ'\bmmu \|_{\ps} ),
	\label{eq:comp-minimax-2}
\end{align}
where the first equality follows from the corresponding Lagrangian dual problem (see, for instance, \citealp[Chapter 5]{boydbandenberghe2004bk}) and
the second equality follows from properties of the dual norm (see, e.g., \citealp[Appendix A.1.6]{boydbandenberghe2004bk}) with the $\ell_{\ps}$-norm being the dual norm for the $\ell_p$-norm.
Analogously to \eqref{eq:comp-minimax-2}, I have $\max_{\bmb \in \cS(B)} (-\bmw'\bmb) = \min_{\bmmu \geq \zero_l} \ ( B \|\bmw + \bmQ'\bmmu \|_{\ps} )$.
With the multipliers $\bmmu_k \in \bR^l$ for $k=1,2$, it follows that the minimax problem under the shape-constrained parameter space \eqref{eq:param-b-2} can be written as
\begin{align}
	\label{eq:comp-minimax-4}
	\begin{split}
	\min_{\bmw \in \bR^{\dy}, \ t \in \bR, \ \bmmu_1, \bmmu_2 \in \bR^l}
	\quad
	& (\bmw'\bmSi\bmw + t^2), \\
	\text{s.t.} \quad
	& \bmw'\oneq = 1,
	\quad \bmw \geq \zeroq,
	\quad \bmmu_k \geq \zero_l \ \ \text{ for $k \in \{1, 2\}$,} \\
	& B\| \bmw - \bmQ'\bmmu_1\|_{\ps} \leq t, \quad
	 B\| \bmw + \bmQ'\bmmu_2\|_{\ps} \leq t.
	\end{split}
\end{align}

\subsection{Adaptive approach} \label{sec:comp-adaptive}
As in Appendix~\ref{sec:comp-minimax}, the optimally adaptive weight in \eqref{eq:areg-3} can be computed by solving a single convex optimization problem.
This is because the worst-case adaptation regret is a maximum of two convex (or strictly convex) functions in $\bmw \in \cWcvx$.
In addition, the optimization problem for the optimally adaptive weight has the following epigraph form (see, for instance, \citealp[Chapter 4]{boydbandenberghe2004bk}):
\begin{align}
	\label{eq:areg-opt-new-1}
	\begin{split}
		\min_{\bmw \in \bR^{\dy}, \ t \in \bR}
		\quad & t ,		\\
		\text{s.t.}
		\quad &
		A(\uB, \bmw) \leq t,	\quad
		A(\oB, \bmw) \leq t,	\quad
		\bmw'\oneq = 1,	\quad
		\bmw \geq \zeroq.
	\end{split}
\end{align}

With the penalized formulation in Remark \ref{re:penalty-1},~\eqref{eq:areg-opt-new-1} can be modified by replacing $A(B, \bmw) \leq t$ with $A(B, \bmw) + \varrho \|\bmw\|^2_2 \leq t$ for $B \in \{\uB, \oB\}$ and $\varrho > 0$.

\subsubsection{Adaptive approach for a parameter space that bounds misspecification} \label{sec:comp-adapt-bound}
Suppose that the researcher places a bound on the misspecification as in Example \ref{eg:param-b}.
At a specific $B \geq 0$ and $\bmw \in \cWcvx$, the adaptation regret is $A(B, \bmw) = \frac{\bmw'\bmSi\bmw + B^2 \|\bmw\|^2_{\ps}}{\Rmm(B)}$ using \eqref{eq:decision-case1-4} and \eqref{eq:areg-1}.
Note that $\Rmm(\uB)$ and $\Rmm(\oB)$ are the minimax risks, which are the inputs to the program.
When the $\ell_2$-norm is used in \eqref{eq:param-b-1}, $\ps = 2$, so \eqref{eq:areg-opt-new-1} becomes a quadratically constrained quadratic program.

\subsubsection{Adaptive approach for a parameter space that has shape restrictions}
Under shape restrictions \eqref{eq:param-b-2}, applying the epigraph and Lagrangian-dual technique of Appendix~\ref{sec:comp-minimax-shape} (with $B_1 = \uB$, $B_2 = \oB$) turns \eqref{eq:areg-opt-new-1} into
\begin{align}
	\label{eq:areg-opt-new-4}
	\begin{split}
		\min_{
			\substack{\bmw \in \bR^{\dy}, \ t \in \bR, \ \bmmu_{i,j} \in \bR^l,\  u_{i,j} \in \bR_+}
		}
		\quad & t ,		\\
		\text{s.t.}
		 \quad
		& \bmw'\oneq = 1,			\quad
			\bmw \geq \zeroq, 	\quad
			\bmmu_{1,1}, \bmmu_{1,2}, \bmmu_{2,1}, \bmmu_{2,2} \geq \zero_l, \\
		& \bmw'\bmSi\bmw +  u_{i, j}^2 \leq t \Rmm(B_i)
			\quad \text{ for $i, j \in \{1, 2\}$}, \\
		&  B_i \|\bmw + (-1)^j \bmQ'\bmmu_{i,j}\|_{\ps} \leq u_{i,j} \quad \text{ for $i,j\in\{1,2\}$.}
	\end{split}
\end{align}

When the $\ell_2$-norm is used in \eqref{eq:param-b-2}, $\ps = 2$, so \eqref{eq:areg-opt-new-4} becomes a quadratically constrained quadratic program.

\section{Proofs for the main text}

\begin{assu} \label{assu:pca:nondegen}
$\bmc'\bmBv\vsss\bmBv'\bmc>0$, $\vsss$ does not depend on $\delta$, and $\bmBv \equiv \sum^{\dy}_{j=2} \bvec_j  \frac{\vecc[\bvec_j\bvec_1']' \msbD}{\lambda_1 - \lambda_j}$.
\end{assu}

\begin{proof}[\textbf{Proof of Proposition \ref{prop:pca-main-weak}}]
Let $\{(\bvec_j, \lambda_j)\}^{\dy}_{j=1}$ be the eigenpairs of $\bmSY$, with $\bvec_1$ being the PC1 of $\bmSY$.
By the assumptions of the proposition (with $\bmOm = \bm{\bmSi}_{Y,n}$) and $\bm{\bmSi}_{Y,n} \longrightarrow \bmSY$, I have $\hbmOn \Pt \bmSY$.
Assume without loss of generality that $\{\wpcan\}$ is a subsequence such that $\wpcan \longrightarrow \bvec_1$ with a chosen sign.
Let $\bml \in \bR^{\dy}$ be such that $\bmc \neq \bml$ and $\bml'\bvec_1 > 0$.
Let $\hbvon = \argmax_{\bmw : \bmw'\bmw = 1, \bml'\bmw \geq 0} \bmw'\hbmOn\bmw$ be an estimator of the leading eigenvector subject to the sign constraint and $\hvsn \equiv \sign(\bmc'\hbvon)$.
The estimator can be written as $\hwpca = \hvsn \hbvon$.
Next, by eigenvector perturbation as in the proof of Corollary~\ref{prop:asym-tau-standard},
\begin{align}
	\sn (\hbvon'\hbmbn - \wpcan'\bmbe)
	& =
	\hbvon' [ \sn ( \hbmbn - \bmbe ) ]
	+
	\bmbe' [ \sn (\hbvon - \wpcan) ]
	\notag \\
	& \Dt
		\begin{pmatrix}
	\bvec_1'  &  \bmbe'\bmBv
	\end{pmatrix}
	\begin{pmatrix}
	\bmZb \\ \bmZO
	\end{pmatrix}.
	\label{pf:prop:pca-main-weak-1}
\end{align}
Using the definition of $\hvsn$ and the assumption on $\bmc'\wpcan$,
\begin{align}
	\{\hvsn = 1\}
	& =
	\{ \sqrt{n}\bmc' (\hbvon - \wpcan) \geq - \delta\}.
	\label{pf:prop:pca-main-weak-2}
\end{align}
Similarly,
\begin{align}
	\sn
	\begin{pmatrix}
		\hbvon'\hbmbn - \wpcan'\bmbe	\\
		\bmc' (\hbvon - \wpcan)
	\end{pmatrix}
	& \Dt
	\begin{pmatrix}
	\bvec_1' & \bmbe' \\
	\zeroq' & \bmc'
	\end{pmatrix}
	\begin{pmatrix}
	\bmZb	\\
	\bmBv \bmZO
	\end{pmatrix}.
	\label{pf:prop:pca-main-weak-2b}
\end{align}

Then,
\begin{align}
	\sn (\htpca - \tau_{\mathrm{pca},n})
	& =
	\sn (\hvsn \hbvon'\hbmbn - \hvsn \wpcan'\bmbe + \hvsn \wpcan'\bmbe - \wpcan'\bmbe)
	\notag \\
	& =
	\hvsn \sn (\hbvon'\hbmbn - \wpcan'\bmbe)
	+
	(\hvsn - 1) \sn (\wpcan'\bmbe),
	\label{pf:prop:pca-main-weak-3}
\end{align}
where I use the definitions of $\htpca$, $\tau_{\mathrm{pca},n}$, and $\hwpca$.
Hence, for any $t \in \bR$,
\begin{align}
	& \hspace{-20pt} \bP[  \sn (\htpca - \tau_{\mathrm{pca},n}) \leq t]
	\notag \\
	& =
	\bP[  \sn (\htpca - \tau_{\mathrm{pca},n}) \leq t, \hvsn = 1] +
	\bP[  \sn (\htpca - \tau_{\mathrm{pca},n}) \leq t, \hvsn = -1]
	\notag \\
	& =
	\bP[ \sn (\hbvon'\hbmbn - \wpcan'\bmbe) \leq t, \hvsn = 1]
	\notag \\
	& \qquad +
	\bP[  -\sn (\hbvon'\hbmbn - \wpcan'\bmbe) \leq t + 2 \sn (\wpcan'\bmbe), \hvsn = -1].
	\label{pf:prop:pca-main-weak-4}
\end{align}

Since $(\wpcan'\bmbe)^2 \longrightarrow (\bvec_1'\bmbe)^2 > 0$, $|2 \sn (\wpcan'\bmbe)| \longrightarrow \infty$.
By \eqref{pf:prop:pca-main-weak-1} and Prokhorov's theorem, $\sn (\hbvon'\hbmbn - \wpcan'\bmbe)$ is tight.
In addition, $\{\hvsn = -1\} = \{ \sqrt{n}\bmc' (\hbvon - \wpcan)  < -\delta\}$ by \eqref{pf:prop:pca-main-weak-2}.
Combining the above with the joint convergence in \eqref{pf:prop:pca-main-weak-2b}, I have
\begin{equation}
	\lim_{n\to\infty}
	\bP[  -\sn (\hbvon'\hbmbn - \wpcan'\bmbe) - 2\sn (\wpcan'\bmbe) \in \cC, \hvsn = -1] = 0.
	\label{pf:prop:pca-main-weak-5}
\end{equation}
Using \eqref{pf:prop:pca-main-weak-4} and \eqref{pf:prop:pca-main-weak-5}, I have
\begin{align}
	\limsup_{n\to\infty}
	\bP[  \sn (\htpca - \tau_{\mathrm{pca},n}) \in \cC ]
	& =
	\limsup_{n\to\infty}
	\bP[ \sn (\hbvon'\hbmbn - \wpcan'\bmbe) \in \cC , \hvsn = 1]
	\notag \\
	& \leq
	\limsup_{n\to\infty}
	\bP[  \hvsn = 1]
	\notag \\
	&=
	\limsup_{n\to\infty}
	\bP[ \sqrt{n}\bmc' (\hbvon - \wpcan) \geq - \delta]
	\notag \\
	& =
	\bP[\bmc'\bmBv \bmZO \geq -\delta],
\end{align}
where the third line follows from  \eqref{pf:prop:pca-main-weak-2}, and
the fourth line follows from  \eqref{pf:prop:pca-main-weak-2b} and $\bP[\bmc'\bmBv\bmZO = -\delta] = 0$, which holds by Assumption~\ref{assu:pca:nondegen}.
The proposition follows as $\delta \downarrow 0$ by Assumption \ref{assu:pca:nondegen} and the normality of $\bmc'\bmBv \bmZO$ with positive variance.
\end{proof}

\begin{proof}[\textbf{Proof of Proposition \ref{prop:kling-var-1}}]
The problem is strictly convex because $\obmSi$ is positive definite and the constraint is affine.
Let the Lagrangian of the optimization problem be
$\cL = \bmw'\obmSi\bmw + \mu (1 - \bmw'\oneq)$.
From the first-order conditions, the optimal $\bmw$ is
$\frac{\obmSi^{-1}\oneq}{\oneq'\obmSi^{-1}\oneq}$.

When $\obmSi$ is an equicorrelation matrix, $\obmSi = (1 - \orho) \bmI_{\dy} +  \orho \oneq\oneq'$ for some $\orho \in (-\frac{1}{\dy-1}, 1)$.
By direct computation, the optimal $\bmw$ is $\frac{1}{\dy}\oneq$.
\end{proof}

\begin{proof}[\textbf{Proof of Proposition \ref{prop:vmc-b0-d}}]
Since $|\bmw'(\hbmSn - \bmSi)\bmw| \leq \max_{j,k}|(\hbmSn - \bmSi)_{jk}|$ for all $\bmw \in \cWcvx$, $\sup_{\bmw \in \cWcvx}|\bmw'\hbmSn\bmw - \bmw'\bmSi\bmw| \Pt 0$.
In addition, $\wvmc$ is the unique minimizer of $\bmw'\bmSi\bmw$ over the compact set $\cWcvx$ because $\bmSi$ is positive definite.
Hence, $\hwvmc \Pt \wvmc$ by Theorem 2.12(i) of \citet{kosorok2008bk}.
Since $\be_j = \theta$ for all $j = 1, \ldots, \dy$ and $\hwvmc'\oneq = \wvmc'\oneq = 1$, $\sn(\hwvmc'\hbmbn - \wvmc'\bmbe) = \hwvmc'\sn(\hbmbn - \bmbe) \Dt \scN(0, \wvmc'\bmSi\wvmc)$ by Slutsky's theorem.
\end{proof}

\begin{proof}[\textbf{Proof of Proposition \ref{prop:areg-main}}]
{\noindent \it \underline{Part \ref{prop:areg-main-a}:}}
Let $ N(B, \bmw)
	\equiv
	\Rmm(B) \frac{\partial \Rmax(B, \bmw)}{\partial (B^2)}
	- \Rmax(B, \bmw) \frac{\partial \Rmm(B)}{\partial (B^2)}$ for any $B \geq 0$ and $\bmw \in \cWcvx$.
Here,
$\frac{\partial A(B, \bmw)}{\partial (B^2)}
	=
	\frac{\Rmm(B) \frac{\partial \Rmax(B, \bmw)}{\partial (B^2)}
	- \Rmax(B, \bmw) \frac{\partial \Rmm(B)}{\partial (B^2)}}{[\Rmm(B)]^2}
	=
	\frac{N(B, \bmw)}{[\Rmm(B)]^2}$,
for any given $\bmw \in \cWcvx$; see Proposition \ref{prop:areg-new-1}.
From Proposition \ref{prop:areg-new-1}, there are only three possibilities for the shape of $A(B, \bmw)$ against $B \in [\uB, \oB]$ for a given $\bmw \in \cWcvx$.

First, if $N(0, \bmw) \geq 0$, then $N(B, \bmw) \geq 0$ for any $B \geq 0$.
Thus, $\frac{\partial A(B, \bmw)}{\partial (B^2)} \geq 0$, i.e., $A(B, \bmw) $ is nondecreasing in $B$.
Second, if $N(0, \bmw) \leq 0$, then there are two scenarios to consider.
The first scenario is that $N(B, \bmw) \leq 0$ for any $B \geq 0$.
Hence, $\frac{\partial A(B, \bmw)}{\partial (B^2)} \leq 0$, i.e., $A(B, \bmw)$ is nonincreasing in $B$.
The remaining scenario is that there exists $B_0$ such that $N(B, \bmw) \leq 0$ for any $B \in [0, B_0)$ and $N(B, \bmw) \geq 0$ for any $B \in [B_0, \infty)$.
This means $A(B, \bmw)$ is nonincreasing in $B$ for any $B \in [0, B_0)$ and nondecreasing in $B$ for any $B \in [B_0, \infty)$.

{\noindent \it \underline{Part \ref{prop:areg-main-b}:}}
In each of the possible cases discussed in Proposition \ref{prop:areg-main}\ref{prop:areg-main-a}, the maximum of $A(B, \bmw)$ over $B \in [\uB, \oB]$, where $\oB \geq \uB \geq 0$, must be achieved at one of the endpoints.
In the case with $\oB = \infty$, I show that $\lim_{B \to \infty} A(B, \bmw)$ exists.
By the definition of $A(B, \bmw)$ in \eqref{eq:areg-1}, Assumption \ref{assu:convex-m2}, and $\bmwsb$ in \eqref{eq:minimax-case1-1soln}, I have
\begin{align*}
	\lim_{B \to \infty} A(B, \bmw)
	& = \lim_{B \to \infty}  \tfrac{\bmw'\bmSi\bmw + B^2 m(\bmw)}{\min\limits_{\tbmw \in \cWcvx} \ [\tbmw'\bmSi\tbmw + B^2 m(\tbmw)]}
	= \lim_{B \to \infty}   \tfrac{\bmw'\bmSi\bmw + B^2 m(\bmw)}{(\bmwsb)'\bmSi\bmwsb + B^2 m(\bmwsb)}
	= \tfrac{m(\bmw)}{\lim\limits_{B \to \infty} m(\bmwsb)}.
\end{align*}
The limit above exists and is finite.
To see this, note that $m(\bmwsb)$ is nonincreasing in $B$ by Lemma~\ref{lem:supp:areg-2} and lies in $[\min_{\tbmw \in \cWcvx} m(\tbmw), 1]$.
In addition, $\bmwsb \in \cWcvx$ is nonzero (it sums to 1) and every parameter space in Section~\ref{sec:param-space} imposes $\|\bmb\|_p \leq B$.
Hence, $\cS(1) \subseteq \{\|\bmb\|_p \leq 1\}$, so
$m(\bmw) \leq \max_{\bmb : \|\bmb\|_p \leq 1} (\bmw'\bmb)^2 \leq \|\bmw\|^2_{\ps} \leq 1$
for any $\bmw \in \cWcvx$ since $\|\bmw\|_{\ps} \leq 1$ for any $\ps \in [1, \infty]$.
Next, $m(\bmw) > 0$ for every $\bmw \in \cWcvx$ because $\cS(1)$ has a nonempty interior and $\bmw \neq \zeroq$; in particular, $m(\bmwsb) > 0$.
\end{proof}

\begin{proof}[\textbf{Proof of Proposition \ref{prop:same-opt-adpt}}]
Since $\uB$ and $\oB$ are held fixed in the proof, I define $\uA(\bmw) \equiv A(\uB, \bmw)$ and $\oA(\bmw) \equiv A(\oB, \bmw)$ for any $\bmw \in \cWcvx$.
I also define $\Amax(\bmw) \equiv \Amax(\cB,\bmw)$.
Using the assumptions on $A(B, \bmw)$ and the definition of $\bmwa$ in \eqref{eq:areg-3b}, I have $\bmwa = \argmin_{\bmw \in \cWcvx} \Amax(\bmw)$.
Assume to the contrary that $\uA(\bmwa) \neq \oA(\bmwa)$.
Without loss of generality, suppose that
\begin{equation}
	\label{pf:prop:same-opt-adpt-1}
	\uA(\bmwa) > \oA(\bmwa).
\end{equation}
This means $\Amax(\bmwa) = \uA(\bmwa)$.

Define $\Delta_{A} \equiv \uA(\bmwa) - \oA(\bmwa) > 0$.
Let $\bmw_1 \equiv \argmin_{\bmw \in \cWcvx} \uA(\bmw)$.
Note that $\bmw_1$ is unique since $\uA(\bmw)$ is strictly convex in $\bmw \in \cWcvx$.
In addition, $\uA(\bmw_1) = 1$ by the definition of adaptation regret and $\uA(\bmw) = \frac{\Rmax(\uB, \bmw)}{\min_{\bmu \in \cWcvx} \Rmax(\uB, \bmu)}$.
Thus, $\min_{\bmw \in \cWcvx} \uA(\bmw) = \frac{\min_{\bmw \in \cWcvx} \Rmax(\uB, \bmw)}{\min_{\bmu \in \cWcvx} \Rmax(\uB, \bmu)} = 1$.
By the same argument, $\min_{\bmw \in \cWcvx} \oA(\bmw) = 1$.

Since $\bmwa \in \cWcvx$, I have
$\uA(\bmwa) \geq \min_{\bmw \in \cWcvx} \uA(\bmw)$ and
$\oA(\bmwa) \geq \min_{\bmw \in \cWcvx} \oA(\bmw)$.
Together with $\uA(\bmwa) > \oA(\bmwa)$ in \eqref{pf:prop:same-opt-adpt-1}, it follows that
\begin{equation}
	\label{pf:prop:same-opt-adpt-2}
	\uA(\bmwa) > \oA(\bmwa) \geq \min_{\bmw \in \cWcvx} \oA(\bmw) = 1.
\end{equation}
From the above, it follows that $\bmwa \neq \bmw_1$ because $\uA(\bmw_1) = 1$ and $\uA(\bmwa) > 1$.
Since $\uA(\bmw)$ is strictly convex in $\bmw \in \cWcvx$ by assumption, for any $t \in (0, 1)$, I have
\begin{equation}
	\label{pf:prop:same-opt-adpt-3}
	\uA((1 - t)\bmwa + t \bmw_{1})
	<
	(1 - t)\uA(\bmwa) + t \uA(\bmw_{1})
	=
	(1 - t)\uA(\bmwa) + t
	\leq
	\uA(\bmwa),
\end{equation}
where the last inequality follows from \eqref{pf:prop:same-opt-adpt-2}.

Next, $\oA(\bmw)$ is continuous in $\bmw \in \cWcvx$.
In particular, it is continuous at $\bmwa$.
Thus, for any $\varepsilon > 0$, there exists $\varphi_\varepsilon > 0$ such that for any $\bmw \in \cWcvx$, $\|\bmw - \bmwa\|_2 < \varphi_\varepsilon$ implies $|\oA(\bmw) - \oA(\bmwa)| < \varepsilon$.
Let $t_\varepsilon \equiv \frac{\varphi_\varepsilon}{2\|\bmw_1 - \bmwa\|_2} > 0$.
As shown above, $\bmw_1 \neq \bmwa$, so this choice of $t_\varepsilon$ is well-defined.
If $t_\varepsilon \geq 1$, divide it by a large enough number so that $t_\varepsilon \in (0, 1)$.
This choice of $t_\varepsilon$ satisfies
$ \|[(1 - t_\varepsilon )\bmwa + t_\varepsilon \bmw_1] - \bmwa\|
	=
	\|t_\varepsilon \bmw_1 - t_\varepsilon\bmwa\|
	=
	t_\varepsilon \|\bmw_1 - \bmwa\|
	\leq \frac{\varphi_\varepsilon}{2}
	<
	\varphi_\varepsilon$.
In addition, $(1 - t_\varepsilon)\bmwa +  t_\varepsilon \bmw_1 \in \cWcvx$.
This is because $t_\varepsilon \in (0, 1)$ and $\bmwa, \bmw_1 \in \cWcvx$, which give $(1 - t_\varepsilon)\bmwa +  t_\varepsilon\bmw_1 \geq \zeroq$, and
$[(1 - t_\varepsilon)\bmwa +  t_\varepsilon\bmw_1]'\oneq =
(1 -  t_\varepsilon) +  t_\varepsilon = 1$.
Hence,
\begin{equation}
	\label{pf:prop:same-opt-adpt-4}
	\oA(\bmwa) -\varepsilon  < \oA((1 - t_\varepsilon)\bmwa +  t_\varepsilon\bmw_1)  < \oA(\bmwa) + \varepsilon.
\end{equation}
Set $\varepsilon = \frac{\Delta_A}{4} > 0$, so that
\begin{equation}
	\label{pf:prop:same-opt-adpt-5}
	\oA((1 - t_\varepsilon)\bmwa +  t_\varepsilon\bmw_1)
	< \oA(\bmwa) + \varepsilon
	= \uA(\bmwa) - \Delta_A + \varepsilon
	= \uA(\bmwa) - \frac{3}{4}\Delta_A,
\end{equation}
where the inequality follows from \eqref{pf:prop:same-opt-adpt-4},
the first equality follows from the definition of $\Delta_A$,
and the last equality follows from the choice of $\varepsilon$.

Combining the results in \eqref{pf:prop:same-opt-adpt-3} and \eqref{pf:prop:same-opt-adpt-5}, I have
$
	\Amax((1 - t_\varepsilon)\bmwa +  t_\varepsilon\bmw_1)
	= \max\{\uA((1 - t_\varepsilon)\bmwa +  t_\varepsilon\bmw_1),
	\oA((1 - t_\varepsilon)\bmwa +  t_\varepsilon\bmw_1) \}
	<
	\max\left\{\uA(\bmwa), \uA(\bmwa) - \frac{3}{4} \Delta_A\right\}
	\leq \uA(\bmwa)
$,
which shows that $\bmwa$ cannot minimize $\Amax(\bmw)$.
\end{proof}

\section{Proofs for the appendices}

\subsection{Proofs for Appendix \ref{app:main-supp-problems}} \label{app:proof:app:a}

\begin{proof}[\textbf{Proof of Proposition~\ref{prop:pre-gen-1}}]
The convergence in probability follows by the continuous mapping theorem (CMT) and \eqref{eq:pre-gen-2}.
For the limiting distribution of $\hbmbn$,
\begin{align}
	\sn (\hbmbn - \bmbe)
	& =
	 \diag[\hbSn]^{-1} \sn (\htbmbn - \tbmbe)
	 +
	 \diag[ \hbSn]^{-1} \diag[\bms]^{-1}\diag[\tbmbe]   \sqrt{n} ( \bms -  \hbSn   )
	\notag \\
	& \Dt
	\diag[\bms]^{-1} \bmZtb
	-
	\diag[\bms]^{-2}  \diag[\tbmbe] \bmZs
	\notag \\
	& =
	\diag[\bms]^{-1} \bmZtb
	-
	\bmH \bmZs
	\sim
	\scN(\zeroq,
	\bmSb
	),
	\label{eq:pre-gen-3}
\end{align}
by $(\diag[\hbSn]^{-1} - \diag[\bms]^{-1}) \tbmbe = \diag[(\tfrac{\tbe_1}{\hS_{n,1} s_1}, \ldots, \tfrac{\tbe_\dy}{\hS_{n,\dy} s_\dy})'](\bms - \hbSn)$, the CMT, and Slutsky's theorem,
with $\bmH$ and $\bmSb$ as in the proposition and \eqref{eq:pre-gen-4}.
\end{proof}

\begin{proof}[\textbf{Proof of Proposition \ref{eg:prop:gen-precision}}]
{\noindent \it \underline{Part \ref{prop:pca-eq-1}:}}
The PCA problem in \eqref{eq:pca-problem} coincides with the precision maximization problem if and only if the leading eigenvectors of $\bmSY$ and $\bmSb^{-1}$ are parallel to each other because $\argmin_{\bmw \in \cWunit} \bmw'\bmSb\bmw= \argmax_{\bmw \in \cWunit}\bmw'\bmSb^{-1}\bmw$.

{\noindent \it \underline{Part \ref{prop:anderson-eq-hetero-2}:}}
Write $ \smash{\wb \equiv \argmin_{\bmw \in \cWsto} \bmw'\bmSb\bmw = \tfrac{\bmSb^{-1}\oneq}{\oneq'\bmSb^{-1}\oneq}}$.
If $\wivm = \wb$, then I have
$ \bmSY^{-1}\oneq
	=
	({\tfrac{\oneq'\bmSY^{-1}\oneq}{\oneq'\bmSb^{-1}\oneq}} ) \bmSb^{-1}\oneq
$.
Here, $ \smash{\tfrac{\oneq'\bmSY^{-1}\oneq}{\oneq'\bmSb^{-1}\oneq}} \neq 0$ because $\bmSY$ and $\bmSb$ are positive definite.

If $\bmSY^{-1}\oneq = a \bmSb^{-1}\oneq$ for some $a \neq 0$, then
$
	 \smash{\wivm =
	\tfrac{\bmSY^{-1}\oneq}{\oneq'\bmSY^{-1}\oneq}
	=
	\tfrac{a \bmSb^{-1}\oneq}{a \oneq'\bmSb^{-1}\oneq}
	=
	\wb}
$.

{\noindent \it \underline{Part \ref{prop:sa-eq-1}:}}
If $\wsa = \wb$, then
$\oneq
	=
	 {\dy\frac{\bmSb^{-1}\oneq}{\oneq'\bmSb^{-1}\oneq}}
$.
Here, $ \smash{\frac{\dy}{\oneq'\bmSb^{-1}\oneq} }\neq 0$ because $\bmSb$ is positive definite.
If $\oneq = a \bmSb^{-1}\oneq$ for some $a \neq 0$, then
\smash{$
	\wsa =
	\frac{1}{\dy} \oneq
	=
	\frac{1}{\oneq'\oneq} \oneq
	=
	\frac{a \bmSb^{-1}\oneq}{a \oneq'\bmSb^{-1}\oneq}
	=
	\wb
$}.
\end{proof}

\begin{proof}[\textbf{Proof of (\ref{eq:pre-gen-7b})}]
From \eqref{eq:pre-gen-5}, $\htbe_{n,j} = \tbe_j + \frac{1}{n_1}\sum_{i=1}^n D_i\tU_{i,j} - \frac{1}{n_0}\sum_{i=1}^n (1-D_i)\tU_{i,j}$ for $j = 1, \ldots, \dy$, where $n_1 \equiv \sum_{i=1}^n D_i$ and $n_0 \equiv \sum_{i=1}^n (1 - D_i)$.
By Lemma \ref{lem:linear-out},
$
  \sqrt{n}(\bmw'\htbmbn - \bmw'\tbmbe)
  = \tfrac{n}{n_1}\tfrac{1}{\sqrt n}\textstyle\sum_{i=1}^n D_i(\bmw'\tbmU_i)
   - \tfrac{n}{n_0}\tfrac{1}{\sqrt n}\sum_{i=1}^n (1-D_i)(\bmw'\tbmU_i)
$.
Note that $\tfrac{n}{n_1}\Pt \tfrac{1}{p_D}$,
$\tfrac{n}{n_0}\Pt\tfrac{1}{1-p_D}$, and
$D_i(1-D_i)=0$.
The central limit theorem with Assumption \ref{assu:main-reg-ols} and
  Slutsky's theorem give the asymptotic variance $\frac{\bmw'\var[D_i\tbmU_i]\bmw}{p_D^2} + \frac{\bmw'\var[(1-D_i)\tbmU_i]\bmw}{(1-p_D)^2}$. As $D_i$ is binary, $\var[D_i\tbmU_i] = p_D\bmStUi$ and
  $\var[(1-D_i)\tbmU_i]=(1-p_D)\bmStUo$, giving \eqref{eq:pre-gen-7b}.
\end{proof}

\begin{proof}[\textbf{Proof of Proposition \ref{prop:t-final-all}}]
{\noindent \it \underline{Part \ref{prop:t-final}:}}
For any $\bmw, \bmw_0 \in \bR^{\dy}$ such that $\|\bmw\|_2 = \|\bmw_0\|_2 = 1$ and $R^2(\bmw_0) \in (0, 1)$, I have
	$\frac{[\Tnw]^2}{[\widehat{T}_n(\bmw_0)]^2}
	\Pt
	\cUpsilon(\bmw; \bmw_0)$
by  Lemma \ref{prop:t-1}\ref{lem:tstat-r-1}, the CMT, and the definition of $[\Tnw]^2$ in \eqref{eq:r2-t-1}.

For a fixed $\bmw_0$ as in the proposition, $\cUpsilon(\bmw; \bmw_0)$ is increasing in $\RRw$ for $\RRw \in [0, 1)$.
This is because $\frac{\text{d}(\frac{x}{1-x})}{\text{d}x} = \frac{1}{(1-x)^2} > 0$ for any $x \neq 1$.
Hence, $\cUpsilon(\bmw; \bmw_0)$ is maximized when $\RRw$ is maximized.
Thus, the result follows from Lemma \ref{prop:t-1}\ref{prop:t-1a} and \ref{prop:t-1b}.

{\noindent \it \underline{Part \ref{prop:t-ivm}:}}
As in the proof of Lemma \ref{prop:t-1}, any $\bmw$ that maximizes $\RRw$ is parallel to $\bmSY^{-1}\bmbe$.
If $\oneq'\bmSY^{-1}\bmbe = 0$, no $\bmw \in \cWsto$ is parallel to $\bmSY^{-1}\bmbe$, so the maximum over $\cWsto$ is not attained.
Otherwise, since $\RRw$ is homogeneous of degree 0, the optimal solution is \smash{$\bmw_{\text{I}} \equiv \frac{\bmSY^{-1}\bmbe}{\oneq'\bmSY^{-1}\bmbe}$}.
Suppose $\bmbe$ is parallel to $\oneq$, i.e., $\bmbe = t \oneq$ for some $t \neq 0$.
Then, $\bmw_{\text{I}} = \frac{t \bmSY^{-1} \oneq}{t \oneq' \bmSY^{-1} \oneq} = \wivm$.
Next, suppose that $\bmw_{\text{I}}  = \wivm$.
This means
$\bmbe = \smash{\frac{\oneq'\bmSY^{-1}\bmbe}{\oneq'\bmSY^{-1}\oneq}\oneq}$ because $\bmSY$ is invertible and $\oneq'\bmSY^{-1}\bmbe \neq 0$, i.e., $\bmbe$ is parallel to $\oneq$.

{\noindent \it \underline{Part \ref{prop:t-sa}:}}
If $\wsa$ is parallel to $\bmSY^{-1}\bmbe$, then $t \bmSY^{-1}\bmbe = \wsa = \frac{1}{\dy}\oneq$ for some $t \neq 0$.
Thus,
$
\bmw_{\text{I}}
= \smash{\frac{\bmSY^{-1}\bmbe}{\oneq'\bmSY^{-1}\bmbe}}
= \wsa
$.
If $\bmw_{\text{I}}  = \wsa$, then
$
	\smash{\frac{\bmSY^{-1}\bmbe}{\oneq'\bmSY^{-1}\bmbe} = \frac{1}{\dy} \oneq = \wsa}
$, i.e., $\wsa$ is parallel to $\bmSY^{-1}\bmbe$.
\end{proof}

\begin{proof}[\textbf{Proof of Proposition \ref{prop:asym-tau-general}}]
The decomposition $\htan - \tau = \hbmwn'(\hbmbn - \bmbe) + (\hbmwn - \bmw_0)'\bmbe$, $\hbmwn \Pt \bmw_0$, the CMT, and Slutsky's theorem give
\[
	\sn(\htan - \tau)
	= \begin{pmatrix} \bmw_0' & \bmbe'\bmM \end{pmatrix}
	\begin{pmatrix} \sn(\hbmbn - \bmbe) \\ \sn\vech[\hbmOn - \bmOm] \end{pmatrix} + \opo
	\Dt \begin{pmatrix} \bmw_0' & \bmbe'\bmM \end{pmatrix}
	\begin{pmatrix} \bmZb \\ \bmZO \end{pmatrix}.
\]
Evaluating the variance of the limit with Assumption \ref{assu:limit-dist}\ref{assu:limit-dist-2} gives the result.
\end{proof}

\begin{proof}[\textbf{Proof of Lemma \ref{lem:asym-tau-standard-c}}]
Note that $\hwpca'\hbmOn\hwpca = \sup_{\bmw \in \cWunit} \bmw'\hbmOn\bmw$ and the leading eigenvalue is unique.
Moreover, $-\wpca \notin \cWunit$ because $\bmc'\wpca > 0$.
For any $\bmw \in \cWunit$, $|\bmw'\hbmOn\bmw - \bmw'\bmOm\bmw| \leq \|\hbmOn - \bmOm\|_\infty \|\bmw\|^2_2$.
Hence, $\sup_{\bmw \in \cWunit} |\bmw'\hbmOn \bmw - \bmw' \bmOm \bmw | \Pt 0$ by Assumption \ref{assu:limit-dist}.
Therefore, $\hwpca \Pt \wpca$ by Theorem 2.12(i) of \citet{kosorok2008bk}.
\end{proof}

\begin{proof}[\textbf{Proof of Corollary \ref{prop:asym-tau-standard}}]
By Lemma \ref{lem:asym-tau-standard-c}, $\hwpca \Pt \wpca$. Since $\bmOm$ is symmetric and $\wpca = \bvec_1$, eigenvector perturbation \citep[Theorem 3.11, Chapter 1]{stewart2001bk} gives $\sn(\hwpca - \wpca)
= \sn \sum^{\dy}_{j=2}  \frac{\vecc[\bvec_j\bvec_1']' \vecc[\hbmOn - \bmOm]}{\lambda_1 - \lambda_j} \bvec_j + \opo
= \bmBv\, \sn\vech[\hbmOn - \bmOm] + \opo$. The result follows from Proposition \ref{prop:asym-tau-general} with $\bmM = \bmBv$.
\end{proof}

\begin{proof}[\textbf{Proof of Lemma \ref{lem:anderson-weight-deri}}]
By matrix derivative equalities and properties of the Kronecker product (Example 18.8a and Theorem 18.5 of~\citealp{magnusneudecker2019bk}),
I have
$\text{d}(\oneq'\bmSi^{-1}\oneq)
= -[ (\oneq'\bmSi^{-1}) \otimes (\oneq'\bmSi^{-1}) ] \vecc[\text{d}\bmSi]$
and
$\text{d}(\bmSi^{-1}\oneq)
=
-[ (\oneq'\bmSi^{-1}) \otimes \bmSi^{-1} ] \vecc[\text{d}\bmSi]
$.
It follows that
$\frac{\text{d}(\bmSi^{-1}\oneq)}{\text{d}\vecc[\bmSi]}
	=
	- (\oneq'\bmSi^{-1}) \otimes \bmSi^{-1}$ and
$\frac{\text{d}(\oneq'\bmSi^{-1}\oneq)}{\text{d}\vecc[\bmSi]}
	=
	- (\oneq'\bmSi^{-1}) \otimes (\oneq'\bmSi^{-1})$.
Hence, the derivative of $\bma(\bmSi)$ with respect to $\vecc[\bmSi]$ can be obtained by the quotient rule.
Since $\bmSi$ is positive definite, $\oneq'\bmSi^{-1}\oneq > 0$, so the denominator of the derivative is nonzero.
\end{proof}

\begin{proof}[\textbf{Proof of Corollary \ref{prop:asym-tau-anderson}}]
By Assumption~\ref{assu:limit-dist}, $\msbD$ satisfies $\vecc[\hbmOn - \bmOm] = \msbD \vech[\hbmOn - \bmOm]$.
By Lemma~\ref{lem:anderson-weight-deri} and the delta method, $\sqrt{n}[\bma(\hbmOn) - \bma(\bmOm)] = \bmG \sqrt{n} \vech[ \hbmOn - \bmOm] + \opo$.
In addition,  $\hwivm = \bma(\hbmOn) \Pt \bma(\bmOm) = \wivm$. The result follows from Proposition \ref{prop:asym-tau-general} with $\bmM = \bmG$.
\end{proof}

\begin{proof}[\textbf{Proof of Proposition~\ref{cor:w-if}}]
As in the proof of Proposition~\ref{prop:asym-tau-general},
$\sn(\htan - \tau) = \bmw_0'\sn(\hbmbn - \bmbe) + \bmbe'\sn(\hbmwn - \bmw_0) + \opo
= \frac{1}{\sqrt n}\sum^{n}_{i=1} [\bmw_0'\bmvpbe(\bmZ_i) + \bmbe'\bmvpw(\bmZ_i)] + \opo$
by the given assumption.
The central limit theorem gives the stated variance.
\end{proof}

\begin{proof}[\textbf{Proof of Proposition \ref{prop:ivm-pos-1}}]
Write $\bmSY = \bmD \bmC \bmD$, where
$\bmD \equiv \diag[\bmSY]^{1/2}$ and
$\bmC$ is the corresponding correlation matrix.
Hence, $\bmSY^{-1} = \bmD^{-1} \bmC^{-1} \bmD^{-1}$.
By \eqref{eq:anderson-w}, the sign of each component of $\wivm$ is determined by the corresponding component of $\bmSY^{-1} \oneq$ since $\oneq'\bmSY^{-1}\oneq > 0$.

Since $\bmC$ is a correlation matrix, it can be decomposed as $\bmC = \bmI_{\dy} + \bmGa$, where $\bmGa$ is the matrix of off-diagonal entries of $\bmC$ and $\bmI_{\dy}$ is the $\dy \times \dy$ identity matrix.
Then, $\bmC^{-1}$ can be written as a Neumann series, which converges if $\| \bmGa \|_{\infty} < 1$ (see, for instance, \citealp[pages 497 and 618]{meyer2023bk}).
This condition holds because $\max_{i = 1, \ldots, \dy} \
	\sum_{j \neq i} | \rho_{Y, ij} |  < \frac{\underline\si}{\overline \si + \underline \si}$ is assumed, $\underline\si, \overline\si > 0$, and $\frac{\underline\si}{\overline \si + \underline \si} < 1$.
Thus, $ \bmC^{-1}
	= (\bmI_{\dy} + \bmGa)^{-1}
	= \sum^\infty_{k=0} (-\bmGa)^{k}$ and
	$ \bmSY^{-1} \oneq
	=
	 \bmD^{-1}  \sum^\infty_{k=0} (-1)^k \bmGa^{k} \bmD^{-1} \oneq
	 =
	 \bmD^{-1} [
	 	\bmD^{-1} \oneq + \sum^\infty_{k=1} (-1)^k \bmGa^k \bmD^{-1} \oneq ]$.
For $i = 1, \ldots, \dy$, the $i$-th component of $\bmSY^{-1} \oneq$ is
\begin{align}
	(\bmSY^{-1} \oneq)_i
	& =
	\frac{1}{\si_{Y, ii}^{1/2}}
	\left[ \frac{1}{\si_{Y, ii}^{1/2}}
	+
	\sum^\infty_{k=1} (-1)^k (\bmGa^k \bmD^{-1} \oneq)_i
	\right]
	\geq
	\frac{1}{\si_{Y,ii}^{1/2}}
	\left[
		\frac{1}{\si_{Y, ii}^{1/2}}
		- \sum^\infty_{k=1} \|\bmGa\|^k_{\infty}  \|\bmD^{-1}\oneq\|_\infty
	\right]
	\notag \\
	& =
	\frac{\|\bmD^{-1}\oneq\|_\infty}{\si_{Y,ii}^{1/2}}
	\left[
		\frac{1}{\si_{Y,ii}^{1/2} \|\bmD^{-1} \oneq\|_\infty}
		- \sum^\infty_{k=1} \|\bmGa\|^k_{\infty}
	\right]
	\geq
	\frac{\|\bmD^{-1}\oneq\|_\infty}{\si_{Y,ii}^{1/2}}
	\left[
		\frac{\underline\si}{\overline \si}
		- \frac{\|\bmGa\|_\infty}{1 - \|\bmGa\|_\infty}
	\right],
	\notag
\end{align}
where the first inequality uses $a \geq -|a|$, $|\bmGa^k \bmD^{-1}\oneq|_i \leq \|\bmGa\|_\infty^k \|\bmD^{-1}\oneq\|_\infty$, sums the geometric series (convergent under the given condition), and applies $\si_{Y,ii}^{1/2} \leq \overline\si$ and $\|\bmD^{-1}\oneq\|_\infty^{-1} = \underline\si$.
Since $\|\bmGa\|_\infty < 1$, each entry is positive if $\frac{\underline\si}{\overline \si}  - \frac{\|\bmGa\|_\infty}{1 - \|\bmGa\|_\infty} > 0$, or equivalently,
$\|\bmGa\|_\infty < \frac{\frac{\underline\si}{\overline \si}}{1 + \frac{\underline\si}{\overline \si}}
	=
	\frac{\underline\si}{\overline \si + \underline \si}$.
By the definition of $\bmGa$, $\max_{i=1,\ldots,\dy}\sum_{j \neq i} | \rho_{Y, ij} | = \|\bmGa\|_\infty$.
Hence, the conclusion follows.
\end{proof}

\begin{proof}[\textbf{Derivation of the optimal weights in Example \ref{eg:dup-out-1}}]
As the example assumes, the standardization is known, so that $\bmSb = \diag[\bms]^{-1}\psbb\diag[\bms]^{-1}$.
Using the notation in Example \ref{eg:binary-te-1},
let $\si_{1,j}^2 \equiv \var[D_i U_{i,j}]$,
$\si_{0,j}^2 \equiv \var[(1 - D_i) U_{i,j}]$, and
$\vs_{j, \hbmbe}^2 \equiv \frac{\si_{1,j}^2}{p_D^2} + \frac{\si_{0,j}^2}{(1 - p_D)^2}$ for $j = 1,2$.
Then,
	\smash{$\bmw'\bmSb\bmw
	= \vs_{1, \hbmbe}^2 w_1^2
	+ 2 \vs_{12, \hbmbe}w_1 ( 1 - w_1)
	+ \vs_{2, \hbmbe}^2 (1 - w_1)^2$},
where the equality follows since $\bmw'\oneq = 1$ and $\bmw \in \cWcvx$.
The first-order condition gives \smash{$w_1^\star = \frac{ \vs_{2, \hbmbe}^2 - \vs_{12, \hbmbe} }{ \vs_{1, \hbmbe}^2 +  \vs_{2, \hbmbe}^2  - 2\vs_{12, \hbmbe} }$}.

In this example, $\be_j = \obe$ for all $j = 1, \ldots, \dy$, and the errors are homoskedastic.
Since $Y_{i,j} = \xi_j + \obe D_i + U_{i,j}$ and $\var[Y_{i,j}] = 1$ for all $j$, $\var[U_{i,j}] = 1 - \obe^2 p_D(1 -p_D)$ for all $j$.
Hence,
	$ \si_{1,j}^2
	=
	p_D [  1 - \obe^2 p_D(1 -p_D) ]$
and
	$ \si_{0,j}^2
	=
	(1 - p_D) [  1 - \obe^2 p_D(1 -p_D) ]$
for all $j$.
As a result, $ \vs_{1, \hbmbe}^2 =  \vs_{2, \hbmbe}^2 $.
Whenever $\vs_{12,\hbmbe} < \min\{\vs^2_{1, \hbmbe}, \vs^2_{2, \hbmbe}\}$, $w_1^\star = \frac{1}{2}$ and $\sum^{\dy}_{j=2} w_j^\star = \frac{1}{2}$.
\end{proof}

\begin{proof}[\textbf{Proof of Lemma \ref{lem:linear-out}}]
Let $\tbmY_j \equiv (Y_{1,j}, \ldots, Y_{n,j})'$ for $j = 1, \ldots, \dy$ be the vector that stacks all observations on outcome $j$, and $\bmY \equiv (\bmY_1, \ldots, \bmY_{\dy})$.
Then, from \eqref{eq:lem:linear-out-1} and the FWL theorem, I have $\widehat \be_{n,j} = (\bmD'\bmM\bmD)^{-1} (\bmD'\bmM \tbmY_j)$ for $j = 1,\ldots, \dy$.
Hence, for \eqref{eq:lem:linear-out-2}, I have $\htan
	=
	(\bmD'\bmM\bmD)^{-1} [\bmD'\bmM (\tbmY\bmw_n)]
	=
	\bmw_n'\hbmbn$.
\end{proof}

\begin{proof}[\textbf{Proof of Lemma \ref{lem:linear-post}}]
Using the definition of $Z_i$,
define
$ \bmZ \equiv (Z_1, \ldots, Z_n)'
	= \frac{1}{b_n}(\bmw_n'\bmY_1 - a_n, \ldots, \bmw_n'\bmY_n - a_n)'
	= \frac{1}{b_n}(\tbmY\bmw_n - a_n \bm{1}_n)$.
Using the FWL theorem on \eqref{eq:lem:linear-out-3}, I have
$ \widehat\tau_{z,n}
	=
	\frac{1}{b_n} (\bmD'\bmM\bmD)^{-1} (\bmD'\bmM \tbmY\bmw_n )
	- \frac{a_n}{b_n} (\bmD'\bmM\bmD)^{-1} (\bmD'\bmM \bm{1}_n )
	=
	\frac{1}{b_n}\htan$
by Lemma \ref{lem:linear-out}.
\end{proof}

\begin{proof}[\textbf{Proof of Lemma \ref{lem:supp:asym-tau-1}}]
Denote $\tbv_1 \equiv \sign(\bmc'\bvec_1) \bvec_1$.
Assume to the contrary that $\wpca \neq \tbv_1$.
Note that $\tbv_1$ is an optimal solution to \eqref{eq:pca-problem} because $\bmc'\tbv_1 = \sign(\bmc'\bvec_1) \bmc'\bvec_1 > 0$ as $\bmc'\bvec_1 \neq 0$, and $\tbv_1'\tbv_1 = \bvec_1'\bvec_1 = 1$ since $\bvec_1$ has unit length, and
$ \tbv_1'\bmSY\tbv_1
	= \sign(\bmc'\bvec_1)^2 \bvec_1'\bmSY\bvec_1
	= \bvec_1'\bmSY\bvec_1
	= \lambda_1$,
because $(\bvec_1, \lambda_1)$ is an eigenpair of $\bmSY$.

Since the leading eigenvalue is unique by Assumption \ref{assu:eval}, $\wpca$ and $\tbv_1$ are parallel to each other.
As $\wpca \neq \tbv_1$ and both vectors have unit length, it can only be that $\wpca = -\tbv_1$.
But this implies
$ \bmc'\wpca
	=
	-\bmc'\tbv_1
	=
	-\sign(\bmc'\tbv_1) \bmc'\tbv_1
	< 0$,
where the inequality follows because $\sign(\bmc'\tbv_1) \bmc'\tbv_1 \geq 0$ and  $\bmc'\tbv_1 \neq 0$ by the hypothesis.
This contradicts the fact that $\wpca$ solves \eqref{eq:pca-problem} because it satisfies $\bmc'\wpca \geq 0$.
\end{proof}

\begin{proof}[\textbf{Proof of Lemma \ref{prop:t-1}}]
{\noindent \it \underline{Part \ref{lem:tstat-r-1}:}}
Let $\oyn \equiv \frac{1}{n} \sum^n_{i=1} \bmw'\bmY_i$,
$\htssw \equiv \sum^n_{i=1} (\bmw'\bmY_i - \oyn)^2$,
$\oDn \equiv \frac{1}{n}\sum^n_{i=1} D_i$, and
$\hessw \equiv \sum^n_{i=1} ( \hxin + \htan D_i - \oyn)^2$ where $\hxin$ is the estimator of $\overline\xi$.
In \eqref{eq:genout-reg-1},
$\hxin + \htan D_i
= \oyn + \htan(D_i - \oDn)$ and
$\htan = \frac{\sum^n_{i=1} (D_i - \oDn)(\bmw'\bmY_i - \oyn)}{ \sum^n_{i=1}  (D_i-\oDn)^2}$.
Thus, $\hessw = \htan^2 \sum^n_{i=1}  (D_i-\oDn)^2$ and
$\hRRw = \frac{\hessw}{\htssw} = \frac{
		[\sum^n_{i=1} (D_i - \oDn)(\bmw'\bmY_i - \oyn)]^2
	}{
		[ \sum^n_{i=1}  (D_i-\oDn)^2 ]
		\htssw
	}$.

Write
$\otYn \equiv \frac{1}{n} \sum^n_{i=1} \tbmY_i$,
$\oyn = \bmw'\diag[\hbSn]^{-1} \otYn$,
$\htssw	=  \bmw'\diag[\hbSn]^{-1} [ \sum^n_{i=1}(\tbmY_i-\otYn)(\tbmY_i-\otYn)' ] \diag[\hbSn]^{-1} \bmw$,
and $\sum^n_{i=1} (D_i - \oDn)(\bmw'\bmY_i - \oyn) = \bmw'\diag[\hbSn]^{-1} \sum^n_{i=1} (D_i - \oDn)(\tbmY_i - \otYn)$.
Under the given assumptions,
$\diag[\hbSn] \Pt \diag[\bms]$,
$\frac{1}{n}  \sum^n_{i=1} (D_i - \oDn)(\tbmY_i - \otYn) \Pt \cov[D_i, \tbmY_i] = \tbmbe \var[D_i]$,
$\frac{1}{n} \sum^n_{i=1} (D_i - \oDn)^2 \Pt \var[D_i]$, and
$\frac{1}{n} \sum^n_{i=1}(\tbmY_i-\otYn)(\tbmY_i-\otYn)' \Pt \var[\tbmY_i]$.
By \eqref{eq:pre-gen-2} and the CMT, $\hRRw
	\Pt
	\frac{
		\var[D_i] (\bmw'\bmbe)^2
	}{
		 \bmw' \bmSY \bmw
	}$.
Here, $\bmw'\bmSY\bmw > 0$ for any $\bmw \neq \zeroq$ (since $\bmSY$ is positive definite) and $\var[D_i]>0$ by Assumption \ref{assu:main-reg-ols}.

{\noindent \it \underline{Part \ref{prop:t-1a}:}}
Write $\si_D^2 \equiv \var[D_i]$.
By Lemma \ref{prop:t-1}\ref{lem:tstat-r-1},
$\RRw = \si_D^2 \frac{\bmw'(\bmbe \bmbe')\bmw}{\bmw'\bmSY\bmw}$.
This is a generalized Rayleigh quotient and is homogeneous of degree 0.
Since $p_D \in (0, 1)$ from Assumption \ref{assu:main-reg-ols}, it follows that $\si_D^2 = p_D(1 - p_D) > 0$.
It is equivalent to a generalized eigenvalue problem, and the optimal solution satisfies \citep[Section 4.3]{ghojoghetal2019wp}
\begin{equation}
	\label{eq:pf:prop:t-1-5}
	[\bmSY^{-1} (\bmbe \bmbe')]\bmw
	=
	\lambda \bmw,
\end{equation}
because $\bmSY$ is invertible and $\bmbe \neq \zeroq$.
If $\lambda = 0$, then $\bmw'\bmbe = 0$ because $\bmbe \neq \zeroq$ and $\bmSY^{-1}$ has full rank (so $\bmSY^{-1}\bmbe \neq \zeroq$).
But such $\lambda =0$ does not correspond to the maximum.
If $\lambda \neq 0$, $\bmw$ is parallel to $\bmSY^{-1}\bmbe$ because $\bmbe'\bmw$ is a scalar.
Let $\bmw = t \bmSY^{-1}\bmbe$ for some nonzero $t \in \bR$.
Since $\bmw$ has unit length, $t = \pm (\bmbe'\bmSY^{-2}\bmbe)^{-\frac{1}{2}}$.
Note that $t$ is well-defined because $\bmbe'\bmSY^{-2}\bmbe = \|\bmSY^{-1}\bmbe\|^2_2 > 0$ by positive definiteness of $\bmSY$ and $\bmSY^{-1}\bmbe \neq \zeroq$.
Recall that the generalized Rayleigh quotient is homogeneous of degree 0.
Therefore, $\bmw^\star = \pm \smash{\frac{\bmSY^{-1}\bmbe}{\sqrt{\bmbe'\bmSY^{-2}\bmbe}}}$.

Next, multiplying $\bmw'$ on both sides of \eqref{eq:pf:prop:t-1-5} gives $\lambda = \bmw'[\bmSY^{-1} (\bmbe \bmbe')]\bmw$.
Evaluating this at $\bmw = \bmw^\star$ gives the corresponding eigenvalue $\lambda^\star \equiv \bmbe'\bmSY^{-1}\bmbe > 0$ because $\bmbe \neq \zeroq$ and $\bmSY$ is positive definite.
Since $\bmbe\bmbe'$ has rank $1$ and $\bmSY^{-1}$ is invertible, $\bmSY^{-1}(\bmbe\bmbe')$ has rank $1$ and hence a single nonzero
eigenvalue.
As $\lambda^\star > 0$, it is the largest, so $\bmw^\star$ maximizes $\RRw$.

{\noindent \it \underline{Part \ref{prop:t-1b}:}}
First, assume that $\bmwr = \pm \bvec_1$.
Using Lemma \ref{prop:t-1}\ref{prop:t-1a}, I have $\bmbe = \pm \sqrt{\bmbe'\bmSY^{-2}\bmbe} \bmSY\bvec_1$.
This holds because $\bmbe'\bmSY^{-2}\bmbe > 0$ as discussed in the proof of Lemma \ref{prop:t-1}\ref{prop:t-1a} and $\bmSY$ is a full-rank matrix.
Moreover, $\bmSY \bvec_1 = \lambda_1 \bvec_1$ because $(\bvec_1, \lambda_1)$ is an eigenpair of $\bmSY$.
It follows that $\bmbe
	= \pm \lambda_1  \sqrt{\bmbe'\bmSY^{-2}\bmbe} \bvec_1$,
which means $\bmbe$ is parallel to $\bvec_1$ because $\lambda_1  \sqrt{\bmbe'\bmSY^{-2}\bmbe}$ is a scalar.

Conversely, assume that $\bmbe$ is parallel to $\bvec_1$, i.e., $\bmbe = t \bvec_1$ for some $t \in \bR \backslash \{0\}$.
Then,
	$\bmwr
	=
	\frac{\bmSY^{-1} \bmbe}{\sqrt{\bmbe'\bmSY^{-2}\bmbe}}
	=
	\frac{\bmSY^{-1}(t \bvec_1)}{\sqrt{(t \bvec_1)'\bmSY^{-2}(t \bvec_1)}}
	=
	\frac{t \lambda_1^{-1}\bvec_1}{\sqrt{t^2 \lambda_1^{-2} \bvec_1'\bvec_1}}
	=
	\frac{t \lambda_1^{-1}\bvec_1}{\sqrt{t^2 \lambda_1^{-2}}}
	=
	\pm \bvec_1$
by  Lemma \ref{prop:t-1}\ref{prop:t-1a} and the facts that $(\bvec_1, \lambda_1)$ is an eigenpair of $\bmSY$, that $\bvec_1$ has unit length, and that $\la_1 > 0$.
\end{proof}

\subsection{Proofs for Appendix \ref{sec:inference-b0}}

\begin{proof}[\textbf{Proof of Proposition \ref{prop:minimax-asym-c-b0}}]
This follows from Proposition \ref{prop:minimax-asym-c} with $\oM(B, \bmw) = 0$.
\end{proof}

\begin{proof}[\textbf{Proof of Theorem \ref{thm:minimax-asym-d-b0}}]
The problem \eqref{eq:vbmin-w} with $\hbmSn$ replaced by $\bmSi$ is the same as setting $\oM(B, \bmw) = 0$ in \eqref{eq:minimax-case1-1} (or setting $B = 0$).
Thus, I have $\wvmc = \bmws(0) = \obmw(\zeroq)$, where $\obmw(\cdot)$ is defined as a solution to \eqref{eq:lem:minimax-asym-a3-1}.
Applying Proposition \ref{prop:minimax-asym-d} with $\oM(B, \bmw) = 0$ gives
	$\sn(\hwvmc - \wvmc)
	= \sn [ \obmw(\nabla\msdn(\wvmc)) - \obmw(\zeroq) ] + o_{\bP}(1)
	= \bmD\obmw_0(\sn\nabla\msdn(\wvmc)) + o_{\bP}(1)$,
where $\bmD\obmw_0 $ is the directional derivative of $\obmw(\cdot)$ at $\zeroq$,
and $\msdn(\cdot)$ is given in \eqref{pf:lem:minimax-asym-b1-1}.

At least one component of any $\bmw \in \cWcvx$ is nonzero.
Hence, the gradients of the active inequality constraints (i.e., those such that $w_j = 0$) and the gradient $\oneq$ of the equality constraint in $\cWcvx$ are linearly independent.
Thus, the linear independence constraint qualification is satisfied at the optimal solution (see, e.g., \citealp[Definitions 12.1 and 12.4]{nocedalwright2006bk}) and the Lagrange multipliers are unique by \citet{wachsmuth2013orl}.
Then,
\begin{align*}
	\sqrt{n} \nabla \msdn(\wvmc)
	& = 2 (\wvmc' \otimes \bmI) \msbD\vech[ \sqrt{n}(\hbmSn - \bmSi ) ]
	\Dt
	2 (\wvmc' \otimes \bmI) \msbD \bmZS
	\equiv \tbmZ,
\end{align*}
where the above uses the properties of the Kronecker product and vectorization (Theorem 18.5 of \citealp{magnusneudecker2019bk}), the CMT, and the definition of $\tbmZ$.
The directional derivative $\bmD\obmw_0(\bmze)$ is given by $\bmh^\star(\bmze)$ in \eqref{eq:thm:minimax-asym-d-b0-1}.
It is also unique because $\bmSi$ is positive definite.
Then, the convergence in distribution follows from page 163 of \citet{shapiroetal2021bk}.
\end{proof}

\begin{proof}[\textbf{Proof of Corollary \ref{cor:minimax-asym-dtau-b0}}]
Here, \eqref{eq:thm:minimax-asym-d-b0-1} reduces to
$\min_{\bmh \in \bR^{\dy} : \bmh'\oneq = 0} \ \ \bmh'\bmze + \bmh'\bmSi\bmh$.
The optimal solution is
$\smash{
	\bmh^\star(\bmze)
	= - \frac{1}{2}\bmSi^{-1}
	(\bmze
		- \frac{\oneq'\bmSi^{-1}\bmze}{\oneq'\bmSi^{-1}\oneq} \oneq
	)
	=
	-\frac{1}{2}\bmSi^{-1} (
		\bmI_{\dy}
		- \oneq \frac{\oneq'\bmSi^{-1}}{\oneq'\bmSi^{-1}\oneq}
	)
	\bmze}
$.
Recall from Theorem \ref{thm:minimax-asym-d-b0} that $\tbmZ \equiv 2(\wvmc'\otimes \bmI)\msbD\bmZS$.
Then,
$\sqrt{n}(\hwvmc - \wvmc)
 \Dt \bmh^\star(\tbmZ)
	= -\frac{1}{2}
	\bmSi^{-1} (
		\bmI_{\dy}
		- \oneq \frac{\oneq'\bmSi^{-1}}{\oneq'\bmSi^{-1}\oneq}) \tbmZ$.
The result follows from defining $\Bvmc$ as in the statement.
\end{proof}

\subsection{Proofs for Appendix \ref{app:main-supp-decision}} \label{app:proof:sec:c}

\begin{proof}[\textbf{Proof of Proposition \ref{prop:flci-asym-valid-new}}]
Let $\bmws(B, \bmSi)$ be the solution to the minimax problem and $\bmwa(\bmSi)$ be the solution to the adaptive problem over $\cB$.
In what follows, $\hbmwn$ is either $\bmws(B, \hbmSn)$ or $\bmwa(\hbmSn)$, where $\hbmSn$ is as defined in the statement of the proposition.
Similarly, for any $(\theta, P) \in \fSnb$, let $\bmw_{\theta, P}$ be $\bmws(B, \bmStp)$ or $\bmwa(\bmStp)$, respectively.

The argument adapts the proof of Theorem C.4 of \citet{armstrongkolesar2021qe}.
Let $\widehat s_n \equiv (\hbmwn'\hbmSn\hbmwn)^{1/2}$,
$s_{\theta, P} \equiv (\bmw_{\theta, P}'\bmStp\bmw_{\theta, P})^{1/2}$, and
$\widehat t_n \equiv \max_{\bmb \in \cS(B)} \frac{|\hbmwn'\bmb|}{ \widehat s_n}$.
For $t \geq 0$, let $\cv_\alpha(t)$ be the $(1 - \alpha)$ quantile of
$|\scN(t, 1)|$, which is continuous in $t$, and note that $\hcvn = \cv_\alpha(\widehat t_n)$.
By Assumption \ref{assu:uniform-flci}(a)--(b), with probability approaching 1 uniformly over $\fSnb$, $\hbmSn$ lies in a compact neighborhood of $\cM$, on which Lemmas \ref{lem:w-unif-minimax} and
\ref{lem:w-unif-adapt} apply.
Hence, $\hbmwn - \bmw_{\theta, P} \Pt \zeroq$ and
$\widehat s_n - s_{\theta, P} \Pt 0$ uniformly over $\fSnb$.

Note that $\frac{\la_{\mathrm{lb}}}{\dy} \leq s^2_{\theta, P} \leq \la_{\mathrm{ub}}$
because $\frac{1}{\dy} \leq \|\bmw\|_2^2 \leq 1$ for every $\bmw \in \cWcvx$.
Assumption~\ref{assu:uniform-flci}(c) and the boundedness of $\frac{\|\bmw_{\theta, P}\|_2}{s_{\theta, P}}$ give $\frac{\bmw_{\theta, P}'\sn(\hbmbn - \bmbe_P)}{s_{\theta, P}} \Dt \scN(0, 1)$
uniformly over $\fSnb$.
Let $T_n \equiv \frac{\hbmwn'\sn(\hbmbn - \bmbe_P)}{\widehat s_n}$.
By the above and the uniform tightness of $\sn(\hbmbn - \bmbe_P)$, $\sup_{(\theta, P) \in \fSnb} \sup_{x \in \bR} |\bP[T_n \leq x] - \Phi(x)| \longrightarrow 0$ holds.
Observe that
$\sqrt{n}(\htan - \theta)
= \sqrt{n}(\hbmwn'\hbmbn - \theta)
= \hbmwn' \sqrt{n}(\hbmbn - \bmbe_P) + \hbmwn'  \sqrt{n}(\bmbe_P - \theta\oneq)
= \hbmwn' \sqrt{n}(\hbmbn - \bmbe_P) + \hbmwn'\tbmb$, where $\tbmb \equiv \sn(\bmbe_P - \theta \oneq)$.
By the definition of $\cI_n$, $\theta \in \cI_n$ if and only if $|T_n + \widehat t_n'| \leq \cv_\alpha(\widehat t_n)$, where $\widehat t_n' \equiv \frac{\hbmwn'\tbmb}{\widehat s_n}$.
Let $t_{\theta, P} \equiv \max_{\bmb \in \cS(B)} \frac{ |\bmw_{\theta, P}'\bmb|}{s_{\theta, P}}$ and $t_{\theta, P}' \equiv \frac{\bmw_{\theta, P}'\tbmb}{s_{\theta, P}}$.
Since $\tbmb \in \cS(B)$, $|t_{\theta, P}'| \leq t_{\theta, P}$.
Moreover, $t_{\theta, P}$ is bounded uniformly over $\fSnb$ as $\cS(B)$ is bounded
and $s_{\theta, P}$ is bounded away from zero.
Hence, $\widehat t_n - t_{\theta, P} \Pt 0$ and
$\widehat t_n' - t_{\theta, P}' \Pt 0$ uniformly over $\fSnb$.

By the uniform continuity of $\cv_\alpha(\cdot)$ on bounded sets, $\cv_\alpha(\widehat t_n) - \cv_\alpha(t_{\theta, P}) \Pt 0$ uniformly over $\fSnb$.
Hence, with $Z \sim \scN(0, 1)$, uniformly over $(\theta, P) \in \fSnb$, I have
$\bP[\theta \notin \cI_n]
= \bP[|Z + t_{\theta, P}'| > \cv_\alpha(t_{\theta, P})] + o(1)
\leq \alpha + o(1)$,
where the inequality holds because $|t_{\theta, P}'| \leq t_{\theta, P}$ and
$\bP[|\scN(t', 1)| > c]$ is increasing in $|t'|$.
\end{proof}

\begin{proof}[\textbf{Proof of Lemma \ref{lem:limit-minimax}}]
The set $\cWcvx$ is compact and $V(\bmw)$ is continuous in $\bmw$.
Moreover, $\oM(B, \bmw) = \max_{\bmb \in \cS(B)} (\bmw'\bmb)^2$ is continuous by Berge's maximum theorem \citep[Theorem 17.31]{aliprantisborder2006bk}, and $m(\bmw)$ is continuous by assumption.
It follows that $V(\bmw)$ and $m(\bmw)$ are bounded on $\cWcvx$.
In addition, let $\bmwm \equiv \argmin_{\bmw \in \cWcvx} m (\bmw)$.
Define $\msC_V \equiv  \max_{\bmw \in \cWcvx} | V(\bmw) - V(\bmwm) | \geq 0$.
Hence, for any $\bmw \in \cWcvx$, it must be that $|V(\bmw) - V(\bmwm)| \leq \msC_V$.
Thus, $V(\bmw) - V(\bmwm) \geq -\msC_V$.
For any $\de > 0$, let $\cW_{m}(\de) \equiv \{ \bmw \in \cWcvx : \| \bmw - \bmwm \|_2 \geq \de\}$.
By the continuity and strict convexity of $m(\bmw)$, there exists $\varepsilon_{\de} > 0$ such that $m(\bmw) \geq m(\bmwm) + \varepsilon_{\de}$ for any $\bmw \in \cW_{m}(\de)$.
Therefore, the following holds for any $\bmw \in \cW_m(\de)$:
\begin{align}
	\Rmax(B, \bmw) - \Rmax(B, \bmwm)
		\geq -\msC_V + B^2 \varepsilon_\de.
	\label{pf:lem:limit-minimax-3}
\end{align}
Set $B_0 = \sqrt{\frac{\msC_V}{\varepsilon_\de}} + 1 > 0$, so that $-\msC_V + B_0^2 \varepsilon_\de > 0$.
By \eqref{pf:lem:limit-minimax-3}, for any $B \geq B_0$, I have
\begin{align}
	\Rmax(B, \bmw) - \Rmax(B, \bmwm)
	\geq -\msC_V + B^2 \varepsilon_\de
	\geq -\msC_V + B_0^2 \varepsilon_\de
	> 0
	\label{pf:lem:limit-minimax-4}
\end{align}
for any $\bmw \in \cW_m(\de)$.
Recall that $\bmwsb$ is the minimax solution in \eqref{eq:minimax-case1-1soln}.
This implies that $\Rmax(B, \bmwsb) \leq \Rmax(B, \bmw)$ for any $\bmw \in \cWcvx$.
In particular,
\begin{align}
	\Rmax(B, \bmwsb) \leq \Rmax(B, \bmwm).
	\label{pf:lem:limit-minimax-5}
\end{align}
Then, for any $B \geq B_0$, \eqref{pf:lem:limit-minimax-4} implies that $\bmwsb \in \cWcvx \backslash \cW_m(\de)$.
This is because if $\bmwsb \in \cW_m(\de)$, then $\Rmax(B, \bmwsb) > \Rmax(B, \bmwm)$, contradicting \eqref{pf:lem:limit-minimax-5}.

The above shows that for any $\de > 0$, there exists $B_0$ such that for any $B \geq B_0$, $\bmwsb \in \cWcvx \backslash \cW_m(\de)$, or equivalently, $\|\bmwsb - \bmwm\|_2 < \de$.
\end{proof}

\begin{proof}[\textbf{Proof of Corollary \ref{cor:limit-minimax}}]
Let the $\ell_{\ps}$-norm be the dual norm of the $\ell_p$-norm.
Then, $m(\bmw) = \|\bmw\|^2_{\ps} = [\sum^{\dy}_{j=1} h(w_j)]^{\frac{2}{\ps}}$ where $\ps = \frac{p}{p-1}$ and $h(w_j) \equiv |w_j|^{\ps}$ from Example \ref{eg:param-b}.

Let $a_j = 1$ for $j = 1, \ldots, \dy$.
Then, $\smash{\frac{1}{\dy} \sum^{\dy}_{j=1} h(w_j)
	= \frac{\sum^{\dy}_{j=1} a_j h(w_j)}{\sum^{\dy}_{j=1} a_j}
	\geq h ( \frac{\sum^{\dy}_{j=1} a_j w_j}{\sum^{\dy}_{j=1} a_j} )}
 	=  h(\dy^{-1})
	= \dy^{-\ps}$,
by Jensen's inequality as $h$ is convex (see, e.g., Proposition 3.8 of \citealp{aragonetal2019bk}) and $\sum^{\dy}_{j=1} w_j = 1$ for any $\bmw \in \cWcvx$.
Since $\ps \in (1, \infty)$, equality holds if and only if the $w_j$ are the same for all $j = 1, \ldots, \dy$. It follows that $\argmin_{\bmw \in \cWcvx} m(\bmw) = \frac{1}{\dy}\oneq$.
\end{proof}

\noindent\emph{\textbf{KKT conditions for the adaptive problem in Example \ref{eg:run-1}.}}
Beyond primal feasibility and dual feasibility
($\mu_1,\ldots,\mu_4 \geq 0$), the stationarity and complementary slackness conditions are
\begin{align}
	& 1 - \mu_1 - \mu_2 = 0, \label{eq:run-1-5-kkt-1} \\
	& \tfrac{\mu_1 [ 2(1 + \si_2^2 - 2\rho\si_2) w_1 + 2(\rho \si_2 - \si_2^2) ]}{\Rmm(0)}
					+ \mu_2 (8w_1 - 4) + \mu_3 - \mu_4 = 0, \label{eq:run-1-5-kkt-2} \\
	& \mu_1[\Rmm(0)^{-1}V(w_1) - t] = 0, \   \mu_2[(2w_1-1)^2 + 1 - t] = 0, \
	\mu_3(w_1 - 1) = 0, \  \mu_4 w_1 = 0. \label{eq:run-1-5-kkt-cs}
\end{align}

\begin{proof}[\textbf{Proof that (\ref{eq:run-1-4}) has no corner solution}]
  At $w_1^\star = 0$, the constraint $-w_1 \leq 0$ is active and $w_1 - 1 \leq 0$ is slack, so
  $\mu_3 = 0$ and $\mu_4 \geq 0$. With $V(0) = \si_2^2$ and $m(0) = 1$, complementary
  slackness gives $\mu_1[\si_2^2 - \Rmm(0)t] = 0$, primal feasibility gives $t \geq 2$, and
  stationarity (with $\mu_2 = 1 - \mu_1$) gives
  $\mu_4 = 2\mu_1\Rmm(0)^{-1}(\rho\si_2 - \si_2^2) + 4(\mu_1 - 1)$.
  If $\mu_1 = 0$, then $\mu_4 = -4 < 0$.
	If $\mu_1 > 0$, complementary slackness forces $t = \frac{\si_2^2}{\Rmm(0)}$.
	There are two subcases:
	(a) when $\si_2^2 \leq \rho\si_2$, $\Rmm(0) = \si_2^2$ by \eqref{eq:run-1-3}, so $t = 1$, contradicting $t \geq 2$;
	(b) when $\si_2^2 > \rho\si_2$, both terms of $\mu_4$ are nonpositive with the first strictly negative, so $\mu_4 < 0$. In every case, the KKT conditions fail, so $w_1^\star = 0$ is not optimal.
	The $w_1^\star = 1$ case can be analyzed similarly.
  \end{proof}

\begin{proof}[\textbf{Derivation of the interior solution in (\ref{eq:run-1-adapt-optim})}]
Suppose that $w_1^\star \in (0, 1)$.
Then, \eqref{eq:run-1-5-kkt-cs} requires $\mu_1, \mu_2 \geq 0$ and $\mu_3 = \mu_4 = 0$.
Moreover, \eqref{eq:run-1-5-kkt-2} becomes
		$\tfrac{\mu_1 [ 2(1 + \si_2^2 - 2\rho\si_2) w_1 + 2(\rho \si_2 - \si_2^2) ]}{\Rmm(0)} + (1 - \mu_1) (8w_1 - 4) = 0$.
The result follows by rearranging this condition.
\end{proof}

\begin{proof}[\textbf{Proof of Proposition \ref{prop:abs-ineq}}]
Let $m_{\mathrm{abs}} \equiv \max_{\bmb \in \cS_{\mathrm{abs}}(B)} (\bmw'\bmb)^2$ and
$m_{\mathrm{aug}} \equiv \max_{\bmb \in \cS_{\mathrm{aug}}(B)} (\bmw'\bmb)^2$.
For any $\bmb \in  \cS_{\mathrm{aug}}(B)$, $|\bmb| = \bmb$ as $\bmb \geq \zeroq$, so $\bmQ |\bmb| \leq \zero_l$.
Thus, $\bmb \in \cS_{\mathrm{abs}}(B)$ and $m_{\mathrm{aug}} \leq m_{\mathrm{abs}}$.

Next, for any $\bmb \in  \cS_{\mathrm{abs}}(B)$, $\bmQ|\bmb|\leq \zero_l$ and $|\bmb|\geq \zeroq$.
Since $\||\bmb|\|_p = \|\bmb\|_p \leq B$, it follows that $|\bmb| \in \cS_{\mathrm{aug}}(B)$.
The objective does not decrease when $\bmb$ is replaced by $|\bmb|$ because
	$(\bmw'\bmb)^2
	= |\bmw'\bmb|^2
	\leq (|\bmw|'|\bmb|)^2
	= (\bmw'|\bmb|)^2$,
using the triangle inequality and $\bmw \in \cWcvx$, so $\bmw \geq \zeroq$.
As a result, for each $\bmb  \in  \cS_{\mathrm{abs}}(B)$, there exists $|\bmb| \in \cS_{\mathrm{aug}}(B)$ such that the objective is weakly larger.
Hence, $m_{\mathrm{abs}} \leq m_{\mathrm{aug}}$.
Therefore, $m_{\mathrm{abs}} = m_{\mathrm{aug}}$.
\end{proof}

\begin{proof}[\textbf{Proof of Lemma \ref{lem:supp:ambn-1}}]
For any $\bmx \in \cS_0(B, \bmga)$, there exists $\bmbe \in \bR^{\dy}$ such that $\bmx = \bmbe - (\bmga'\bmbe)\oneq$.
Hence,
$\bmga'\bmx
	=
	\bmga'\bmbe
	-
	\bmga'\bmbe(\bmga'\oneq)
	=0$
because $\bmga \in \cWcvx$.
In addition, $\|\bmx\|_p \leq B$ since $\bmx \in \cS_0(B, \bmga)$.
It follows that $\bmx \in \cS_1(B, \bmga)$.
This means $\cS_0(B, \bmga) \subseteq \cS_1(B, \bmga)$.

Next, for any $\bmx \in \cS_1(B, \bmga)$ and $t \in \bR$, let $\bmy \equiv \bmx + t \oneq$.
Then,
$ \bmy - (\bmga'\bmy)\oneq
	=
	\bmx + t\oneq
	-
	(\bmga'\bmx + t \bmga'\oneq)\oneq
	=
	\bmx + t \oneq - t\oneq
	=
	\bmx$,
where the second equality uses $\bmga'\bmx = 0$ as $\bmx \in \cS_1(B, \bmga)$.
Since $\|\bmx\|_p \leq B$, it follows that $\bmx \in \cS_0(B, \bmga)$.
Thus, $\cS_1(B, \bmga) \subseteq \cS_0(B, \bmga)$.
\end{proof}

\begin{proof}[\textbf{Proof of Proposition \ref{prop:app:ambiguity:case3n-1}}]
{\it \noindent \underline{Part \ref{prop:app:ambiguity:case3n-1a}:}}
Suppose that $\Delta = 0$.
This implies that $\bmde = \zeroq$.
Hence, $\bmga = \bmeta_0 + \bmde = \bmeta_0$.
This means $\cS_3(B, \Delta, \bmeta_0)$ in \eqref{eq:app:amb:case3n-0} becomes
	$\cS_3(B, 0, \bmeta_0)
	=
	\{ (\bmb, \bmga) \in \bR^{2\dy} : \|\bmb\|_p \leq B, \|\bmde\|_p \leq 0, \bmga = \bmeta_0 + \bmde \in \cWcvx, \bmga'\bmb = 0\}
	=
	\{ (\bmb, \bmga) \in \bR^{2\dy} : \bmb \in \cS_1(B, \bmeta_0), \bmga = \bmeta_0\}$,
as desired, where $\cS_1(B, \bmeta_0)$ is given in Lemma~\ref{lem:supp:ambn-1}.

{\it \noindent \underline{Part \ref{prop:app:ambiguity:case3n-1b}:}}
Any $(\bmb, \bmga) \in \cS_3(B, \Delta, \bmeta_0)$ satisfies $\bmga'\bmb = 0$, $\bmga \in \cWcvx$, and $\|\bmb\|_p \leq B$ by construction.
Hence, $(\bmb, \bmga) \in \cS_2(B)$, so $\cS_3(B, \Delta, \bmeta_0) \subseteq \cS_2(B)$.

Next, for any $(\bmb, \bmga) \in \cS_2(B)$, write $\bmde_{\bmga} \equiv \bmga - \bmeta_0$.
By definition, $\| \bmde_{\bmga} \|_p  = \|\bmga - \bmeta_0\|_p \leq \max_{\bmy \in \cWcvx} \|\bmy - \bmeta_0\|_p \leq \underline \Delta \leq \Delta$.
In addition, $\|\bmb\|_p \leq B$, $\bmga \in \cWcvx$, and $\bmga'\bmb = 0$ by the definition of $\cS_2(B)$.
Hence, $(\bmb, \bmga) \in \cS_3(B, \Delta, \bmeta_0)$.
This gives $\cS_2(B) \subseteq \cS_3(B, \Delta, \bmeta_0)$.
\end{proof}

\begin{proof}[\textbf{Proof of Proposition \ref{prop:minimax-asym-c}}]
Write $M_n(\bmw) \equiv -\hRmaxn(B, \bmw)$
and $M(\bmw) \equiv -\Rmax(B, \bmw)$
for any $\bmw \in \cWcvx$ and finite $B \geq 0$.
I also write $\bmwo \equiv \bmwsb$.
I show that the proposition holds by verifying the assumptions in Theorem 2.12(i) of \citet{kosorok2008bk}.

First, suppose that for some sequence $\{\bmw_n\}$ in $\cWcvx$, $\liminf_{n \to \infty} M(\bmw_n) \geq M(\bmwo)$ but $\|\bmw_n - \bmwo\|_2 \longrightarrow 0$ is false, i.e.,
there exist $\varepsilon > 0$ and a subsequence $\{\bmw_{n_k}\}$ such that
\begin{equation}
	\label{pf:prop:minimax-asym-c-1}
	\|\bmw_{n_k} - \bmwo\|_2 \geq \varepsilon.
\end{equation}
Note that $\Rmax(B, \bmw)$ is strongly convex in $\bmw \in \cWcvx$.
As a result, using an argument similar to that of \eqref{pf:lem:w-unif-minimax-5} and \eqref{pf:lem:w-unif-minimax-10}, and noting that $\bmwo$ is the optimal solution to the minimax problem, it follows that for any $\bmw \in \cWcvx$, $\Rmax(B, \bmw) \geq \Rmax(B, \bmwo) + \kappa \|\bmw - \bmwo\|^2_2$ for some $\kappa > 0$.
Such $\kappa$ can be taken to be $\frac{1}{2}\lambda_{\min}(\bmSi)$ where $\lambda_{\min}(\bmSi)$ is the smallest eigenvalue of $\bmSi$.
Here, $\kappa > 0$ because $\bmSi$ is positive definite.
This means
\begin{equation}
	\label{pf:prop:minimax-asym-c-2}
	-M(\bmw_{n_k}) \geq -M(\bmwo) + \kappa  \|\bmw_{n_k} - \bmwo\|^2_2
	\geq -M(\bmwo) + \kappa \varepsilon^2,
\end{equation}
where the first inequality follows from the earlier discussion on strong convexity and the second inequality uses \eqref{pf:prop:minimax-asym-c-1}.
Rearranging \eqref{pf:prop:minimax-asym-c-2} gives $M(\bmw_0) - \kappa \varepsilon^2 \geq M(\bmw_{n_k})$.
This implies $M(\bmw_0) - \kappa \varepsilon^2 \geq \liminf_{k \to \infty} M(\bmw_{n_k}) \geq \liminf_{n \to \infty} M(\bmw_n)$.
This contradicts $\liminf_{n \to \infty} M(\bmw_n) \geq M(\bmwo)$, so the hypothesis in Theorem 2.12 of \citet{kosorok2008bk} holds.

Next, $M_n(\hbmw_n) = \sup_{\bmw \in \cWcvx} M_n(\bmw)$ by the property of the minimax problem.
In addition, for any $\bmw \in \cWcvx$,
$|M_n(\bmw) - M(\bmw)| = |\bmw'(\hbmSn - \bmSi)\bmw| \leq \|\hbmSn- \bmSi\|_F \|\bmw\|_2^2 \leq \|\hbmSn- \bmSi\|_F$.
By Assumption~\ref{assu:minimax-asym}, $\|\hbmSn- \bmSi\|_F = o_{\bP}(1)$.
As a result, $\sup_{\bmw \in \cWcvx} |M_n(\bmw) - M(\bmw)|  \Pt 0$.
The assumptions in  Theorem 2.12(i) of \citet{kosorok2008bk} also hold.
Hence, the consistency result of this proposition follows.
\end{proof}

\begin{proof}[\textbf{Proof of Proposition \ref{prop:minimax-asym-d}}]
Following the discussion in Section \ref{sec:minimax}, \eqref{eq:minimax-case1-1} has a unique optimal solution.
Hence, Assumption A1 of  \citet{shapiro1993mor} holds.
Next, Assumptions A3 and B1--B3 of  \citet{shapiro1993mor} also follow from Lemmas \ref{lem:minimax-asym-b-main} and \ref{lem:minimax-asym-a3}.
$\hbmwn$ is also consistent for $\bmwsb$ by Proposition \ref{prop:minimax-asym-c}.
It follows from Theorem 2.1 of  \citet{shapiro1993mor} that $\hbmwn = \obmw(\nabla\msdn(\bmwsb)) + o_{\bP}(\ns)$.
Since $\bmwsb = \obmw(\zeroq)$,
$\sn ( \hbmwn - \bmwsb)
	=
	\sn \left[ \obmw(\nabla \msdn(\bmwsb))  - \obmw(\zeroq) \right]
	+
	o_{\bP}(1)$,
where $\nabla \msdn(\bmwsb)$ is given in \eqref{pf:lem:minimax-asym-b2-1}.
\end{proof}

\begin{proof}[\textbf{Proof of Lemma \ref{lem:supp:areg-1}}]
Recall that the minimax problem \eqref{eq:minimax-case1-1} has a unique solution for each $B \geq 0$.
Then, for any $B_1, B_2 \geq 0$, I have
$\Rmm(B_2)
= V(\bmw^\star(B_2)) + B^2_2 m(\bmw^\star(B_2))
\leq V(\bmw^\star(B_1)) + B^2_2 m(\bmw^\star(B_1))
= \Rmm(B_1) +  (B^2_2 - B^2_1)m(\bmw^\star(B_1))$,
where the first equality uses the definition of the minimax risk from \eqref{eq:minimax-case1-1},
the inequality follows because $\bmw^\star(B_1)$ is not necessarily an optimal solution to the minimax problem for $B = B_2$,
and the last equality follows from the definition of the minimax risk from \eqref{eq:minimax-case1-1} again.
\end{proof}

\begin{proof}[\textbf{Proof of Lemma \ref{lem:supp:areg-2}}]
For any $B_1 > B_2 \geq 0$ (I consider strict inequality to ensure the division below is well-defined), I have
$\Rmm(B_1) \leq \Rmm(B_2) + (B_1^2 - B_2^2) m(\bmw^\star(B_2))$ and
$\Rmm(B_2) \leq \Rmm(B_1) + (B_2^2 - B_1^2) m(\bmw^\star(B_1))$ from Lemma \ref{lem:supp:areg-1}.
Since $B_1^2 - B_2^2 > 0$, the inequalities imply that
$m(\bmw^\star(B_2))
	\geq
	\frac{\Rmm(B_1) - \Rmm(B_2)}{B_1^2 - B_2^2}
	\geq
	m(\bmw^\star(B_1))$.
\end{proof}

\begin{proof}[\textbf{Proof of Proposition \ref{prop:areg-new-1}}]
To begin with, note that
$\frac{\partial \Rmm(B)}{\partial (B^2)} = m(\bmwsb)$
by Assumption \ref{assu:convex-m2} and Danskin's theorem (see, for instance, \citealp{bertsekas2009bk}) because for each given $\bmw \in \cWcvx$, $\Rmax(B, \bmw) = V(\bmw) + (B^2) m(\bmw)$ is affine in $(B^2)$ and continuous, $\cWcvx$ is compact, and $\bmwsb$ is unique.

Since $\oM(B, \bmu) = B^2 m (\bmu)$ and $\bmu \in \cWcvx$, $\frac{\partial \oM(B, \bmu)}{\partial (B^2)} = m(\bmu)$ and $\left.\frac{\partial \oM(B, \bmu)}{\partial (B^2)}\right|_{\bmu = \bmwsb} = m(\bmwsb)$.
Hence, $N(B, \bmu)$ in the proof of Proposition \ref{prop:areg-main}\ref{prop:areg-main-a} can be written as
\begin{equation}
	\label{eq:pf:prop:areg-1-1}
	N(B, \bmu) = \Rmm(B) m(\bmu) - \Rmax(B, \bmu) m(\bmwsb).
\end{equation}
For any $B_1 > B_2 \geq 0$, I have $B_1^2 m(\bmw^\star(B_1)) - B_2^2 m(\bmw^\star(B_2))
= B_2^2 [ m(\bmw^\star(B_1)) - m(\bmw^\star(B_2))] + (B_1^2 - B_2^2) m(\bmw^\star(B_1))$.
For any $\bmu \in \cWcvx$ and $B_1 \geq B_2 \geq 0$, I have
\begin{align*}
	N(B_1, \bmu) - N(B_2, \bmu)
	& = m(\bmu) [\Rmm(B_1) - \Rmm(B_2) + (B_2^2 - B_1^2) m(\bmw^\star(B_1))]
	\notag 	\\
	& \qquad - \Rmax(B_2, \bmu) [ m(\bmw^\star(B_1)) - m(\bmw^\star(B_2))]
	\geq 0,
\end{align*}
using \eqref{eq:minimax-case1-1}, \eqref{eq:pf:prop:areg-1-1}, $m(\bmu) \geq 0$, $\Rmax(B_2, \bmu) \geq 0$, and Lemmas \ref{lem:supp:areg-1} and \ref{lem:supp:areg-2}.
\end{proof}

\begin{proof}[\textbf{Proof of Lemma \ref{lem:w-unif-minimax}}]
Under Assumption \ref{assu:convex-m2}, the maximum risk function can be written as $\Rmax(B, \bmw, \bmSi)
	\equiv
	V(\bmw, \bmSi) + B^2 m(\bmw)
	=
	\bmw'\bmSi\bmw + B^2 m(\bmw)$.
In the above, $V(\bmw, \bmSi)$ is strongly convex in $\bmw \in \cWcvx$ for a given $\bmSi \in \cM$.
The function $m(\bmw)$ is convex in $\bmw \in \cWcvx$.
By the definitions of strong convexity and convexity (see \citealp{aragonetal2019bk}), it follows that $\Rmax(B, \bmw, \bmSi)$ is strongly convex in $\bmw \in \cWcvx$.
Therefore, for any $\bmw_1, \bmw_2 \in \cWcvx$ and $\bmSi \in \cM$,
\begin{align}
	\label{pf:lem:w-unif-minimax-5}
	\Rmax(B, \bmw_1, \bmSi)
	\geq
	\Rmax(B, \bmw_2, \bmSi) + \bmg'(\bmw_1 - \bmw_2)
	+ \frac{\msC_R}{2}\|\bmw_1 - \bmw_2\|^2_2
\end{align}
for any $\bmg \in \partial \Rmax(B, \bmw_2, \bmSi) $ (where $\bmg$ is a subgradient of the maximum risk $\Rmax(B, \bmw_2, \bmSi)$ and $\partial \Rmax(B, \bmw_2, \bmSi)$ is the set of all subgradients) and for some $\msC_R > 0$.

Next, observe that $\Rmax(B, \bmw, \bmSi_1) - \Rmax(B, \bmw, \bmSi_2) =  \bmw'(\bmSi_1 - \bmSi_2)\bmw$
for any $\bmSi_1, \bmSi_2 \in \cM$ and $\bmw \in \cWcvx$.
Hence, by the Cauchy--Schwarz inequality and noting that $\bmw \in \cWcvx$,
\begin{align}
	|\Rmax(B, \bmw, \bmSi_1) - \Rmax(B, \bmw, \bmSi_2)|
	& \leq
	\|\bmSi_1 - \bmSi_2 \|_2 \|\bmw\|^2_2
	\leq
	\|\bmSi_1 - \bmSi_2 \|_2.
	\label{pf:lem:w-unif-minimax-7}
\end{align}

Since the maximum risk is strictly convex in $\bmw \in \cWcvx$ for any $\bmSi \in \cM$, the optimal solution is unique.
For any $\bmSi_1, \bmSi_2 \in \cM$, let $\bmws(B, \bmSi_k)$ be the optimal solutions to the minimax problem with matrix $\bmSi_k$ for $k \in \{1, 2\}$.
By the optimality of the solutions,
\begin{align}
	\Rmax(B, \bmws(B, \bmSi_1), \bmSi_1)
	& \leq \Rmax(B, \bmws(B, \bmSi_2), \bmSi_1).
	\label{pf:lem:w-unif-minimax-8}
\end{align}

Note that $\cWcvx$ is a polyhedral set because it is defined by a finite collection of inequalities.
For any $\bmw \in \cWcvx$, $\Rmax(B, \bmw, \bmSi_2) < \infty$ for any finite $B \geq 0$.
Let $\textrm{ri}$ be the relative interior and $\textrm{dom}$ be the effective domain (see \citealp{bertsekas2009bk}, for definitions).
In addition, $\textrm{ri}(\textrm{dom}(\Rmax)) \cap \cWcvx \neq \emptyset$.
Recall that $\bmws(B, \bmSi_2)$ minimizes the maximum risk $\Rmax(B, \bmw, \bmSi_2)$ over $\bmw \in \cWcvx$.
By Proposition 5.4.7 of \citet{bertsekas2009bk}, there exists $\bmg_2 \in \partial \Rmax(B, \bmws(B, \bmSi_2), \bmSi_2) $ such that the following holds for any $\bmw \in \cWcvx$:
\begin{align}
	\bmg_2'(\bmw - \bmws(B, \bmSi_2)) \geq 0.
	\label{pf:lem:w-unif-minimax-10}
\end{align}

Applying \eqref{pf:lem:w-unif-minimax-5} with $\bmw_1 = \bmws(B, \bmSi_1)$, $\bmw_2 = \bmws(B, \bmSi_2)$, $\bmg = \bmg_2$, and $\bmSi = \bmSi_2$, I have
\begin{align}
	\frac{\msC_R}{2} \|\bmws(B, \bmSi_1) - \bmws(B, \bmSi_2)\|^2_2 & \leq
	\Rmax(B, \bmws(B, \bmSi_1), \bmSi_2)
	- \Rmax(B, \bmws(B, \bmSi_2), \bmSi_2)
	\notag \\
	& \qquad
	- \bmg_2'[\bmws(B, \bmSi_1) - \bmws(B, \bmSi_2)]
	\notag \\
	& \leq
	\Rmax(B, \bmws(B, \bmSi_1), \bmSi_2)
	- \Rmax(B, \bmws(B, \bmSi_2), \bmSi_2)
	\notag \\
	& \leq
	[\Rmax(B, \bmws(B, \bmSi_1), \bmSi_2)  - \Rmax(B, \bmws(B, \bmSi_1), \bmSi_1) ]
	\notag \\
	& \qquad +
	[\Rmax(B, \bmws(B, \bmSi_2), \bmSi_1)
	- \Rmax(B, \bmws(B, \bmSi_2), \bmSi_2)]
	\notag \\
	& \leq 2 \|\bmSi_1 - \bmSi_2 \|_2,
	\label{pf:lem:w-unif-minimax-11}
\end{align}
where the second, third, and last inequalities follow from \eqref{pf:lem:w-unif-minimax-10}, \eqref{pf:lem:w-unif-minimax-8}, and \eqref{pf:lem:w-unif-minimax-7}, respectively.
Thus, the above shows that for any $\varepsilon > 0$, there exists $\delta = \frac{\msC_R}{4}\varepsilon^2 > 0$ such that for any $\bmSi_1,\bmSi_2 \in \cM$ with $ \|\bmSi_1 - \bmSi_2 \|_2 < \delta$, $ \|\bmws(B, \bmSi_1) - \bmws(B, \bmSi_2)\|_2 < \varepsilon$ holds.
\end{proof}

\begin{proof}[\textbf{Proof of Lemma \ref{lem:w-unif-adapt}}]
As in the proof of Lemma \ref{lem:w-unif-minimax}, I write the maximum risk and the adaptation regret as functions of $\bmSi$
(e.g.,
$A( B, \bmw, \bmSi) \equiv \frac{ \Rmax(B, \bmw, \bmSi) }{\Rmm(B, \bmSi)}$,
where
$\Rmm(B, \bmSi) \equiv \min_{\bmw \in \cWcvx} \Rmax(B, \bmw, \bmSi)$).
Let $\cB = [\uB, \oB]$.
As the solution to the adaptive problem is unique and $\cB$ is fixed, I define
$ \bmwa(\bmSi) \equiv \argmin_{\bmw \in \cWcvx} \Amax(\bmw, \bmSi)$
and $\Amax(\bmw, \bmSi) \equiv \max\{A(\uB, \bmw, \bmSi), A( \oB, \bmw, \bmSi) \}$.

In the proof of Lemma \ref{lem:w-unif-minimax}, I show that $\Rmax(B, \bmw, \bmSi)$ is uniformly continuous in $\bmSi \in \cM$.
For any $\bmSi_1, \bmSi_2 \in \cM$, let $\bmws(B, \bmSi_1)$ and $\bmws(B, \bmSi_2)$ be the optimal solutions to \eqref{eq:minimax-case1-1}, as in Lemma \ref{lem:w-unif-minimax}.
Then, $\Rmm(B, \bmSi_k) = \Rmax(B, \bmws(B, \bmSi_k), \bmSi_k)$ for $k \in \{1, 2\}$.
Thus,
\begin{align}
	\Rmm(B, \bmSi_1)  - \Rmm(B, \bmSi_2)
	&
	= \Rmax(B, \bmws(B, \bmSi_1), \bmSi_1) - \Rmax(B, \bmws(B, \bmSi_2), \bmSi_2)
	\notag \\
	&
	\leq
	\Rmax(B, \bmws(B, \bmSi_2), \bmSi_1) - \Rmax(B, \bmws(B, \bmSi_2), \bmSi_2)
	\notag \\
	& \leq
	|\bmws(B, \bmSi_2)' (\bmSi_1 - \bmSi_2) \bmws(B, \bmSi_2) |
	\notag \\
	& \leq
	\|\bmSi_1 - \bmSi_2\|_2,
	\label{pf:lem:w-unif-adapt-2}
\end{align}
where the second line follows from $ \Rmax( B, \bmws(B, \bmSi_1), \bmSi_1)  \leq  \Rmax(B, \bmw, \bmSi_1) $ for any $\bmw \in \cWcvx$,
and the last line uses that $\|\bmws(B, \bmSi_2)\|^2_2 \leq 1$ because $\bmws(B, \bmSi_2) \in \cWcvx$.
By similar reasoning, I have $\Rmm(B, \bmSi_2)  - \Rmm(B, \bmSi_1) \leq \|\bmSi_1 - \bmSi_2\|_2$.
Therefore, I obtain
\begin{align}
	 | \Rmm(B, \bmSi_1)  - \Rmm(B, \bmSi_2) |
	 \leq
	\|\bmSi_1 - \bmSi_2\|_2.
	\label{pf:lem:w-unif-adapt-4}
\end{align}

Next, I bound $\Rmm(B, \bmSi)$.
For any $\bmSi \in \cM$ and finite $B \geq 0$, I have
\begin{align}
	\Rmm(B, \bmSi)
	& =
	\min_{\bmw \in \cWcvx} \ [ \bmw'\bmSi\bmw + B^2 m(\bmw) ]
	\geq \min_{\bmw \in \cWcvx} \ \bmw'\bmSi\bmw
	\geq \frac{\la_{\mathrm{lb}}}{\dy},
	\label{pf:lem:w-unif-adapt-5}
\end{align}
because $m(\bmw) \geq 0$, $\bmSi - \la_{\mathrm{lb}} \bmI_{\dy}$ is positive semidefinite, and $\|\bmw\|^2_2 \geq \frac{1}{\dy}$ by the Cauchy--Schwarz inequality.
For the upper bound, $m(\bmw) = \max_{\bmb \in \cS(1)} (\bmw'\bmb)^2 \leq 1$.
Hence,
\begin{align}
	\Rmm(B, \bmSi)
	=
	\min_{\bmw \in \cWcvx} \ [\bmw'\bmSi\bmw + B^2 m(\bmw)]
	\leq
	\min_{\bmw \in \cWcvx} \ [\la_{\mathrm{ub}} \|\bmw\|^2_2 + B^2]
	\leq
	\la_{\mathrm{ub}}  + B^2
	\label{pf:lem:w-unif-adapt-6}
\end{align}
because $\la_{\mathrm{ub}}\bmI_{\dy} - \bmSi$ is positive semidefinite.
By \eqref{pf:lem:w-unif-adapt-5}, $\Rmm(B, \bmSi) > 0$ for any $\bmSi \in \cM$, so $\frac{1}{\Rmm(B, \bmSi)} > 0$ holds as well.
Moreover,
\begin{align}
	\left|
		\frac{1}{\Rmm(B, \bmSi_1)} - \frac{1}{\Rmm(B, \bmSi_2)}
	\right|
	&
	=
	\left|
		\frac{\Rmm(B, \bmSi_1) - \Rmm(B, \bmSi_2)}{\Rmm(B, \bmSi_1) \Rmm(B, \bmSi_2)}
	\right|
	\leq
	\frac{\dy^2 \|\bmSi_1 - \bmSi_2\|_2}{\la_{\mathrm{lb}}^2},
	\label{pf:lem:w-unif-adapt-7}
\end{align}
using \eqref{pf:lem:w-unif-adapt-4} and \eqref{pf:lem:w-unif-adapt-5}.
The function $\Rmm(B, \bmSi)$ is uniformly continuous in $\bmSi \in \cM$ by \eqref{pf:lem:w-unif-adapt-4}.
For any $\bmSi_1, \bmSi_2 \in \cM$,
\begin{align}
	\left| A( B, \bmw, \bmSi_1) - A( B, \bmw, \bmSi_2) \right|
	& =
	\left|
	\frac{ \Rmax(B, \bmw, \bmSi_1) }{\Rmm(B, \bmSi_1)} - \frac{ \Rmax(B, \bmw, \bmSi_2) }{\Rmm(B, \bmSi_2)} \right|
	\notag \\
	& \leq
	\frac{\dy \|\bmSi_1 - \bmSi_2\|_2}{\la_{\mathrm{lb}}}
	+ \frac{\dy^2(\la_{\mathrm{ub}} + B^2) \|\bmSi_1 - \bmSi_2\|_2}{\la_{\mathrm{lb}}^2}
	\notag 	\\
	& \equiv
	\msK_1(B)  \|\bmSi_1 - \bmSi_2\|_2,
	\label{pf:lem:w-unif-adapt-8}
\end{align}
where the second-to-last line uses \eqref{pf:lem:w-unif-minimax-7}, \eqref{pf:lem:w-unif-adapt-5}, \eqref{pf:lem:w-unif-adapt-6}, \eqref{pf:lem:w-unif-adapt-7}, and $\Rmax(B, \bmw, \bmSi_2) > 0$,
and the last line defines $\msK_1(B) \equiv \frac{\dy}{\la_{\mathrm{lb}}} + \frac{\dy^2(\la_{\mathrm{ub}} + B^2)}{\la_{\mathrm{lb}}^2} \geq 0$.

Next, since $|\max\{a_1, a_2\} - \max\{b_1, b_2\}| \leq \max\{|a_1 - b_1|, |a_2 -b_2|\}$ and $\msK_1(B) \geq 0$ in \eqref{pf:lem:w-unif-adapt-8}, I have
\begin{align}
	& \hspace{-20pt} \left| \Amax(\bmw, \bmSi_1) - \Amax(\bmw, \bmSi_2) \right| \notag \\
	& \leq
	 \max\{ | A(\uB, \bmw, \bmSi_1) - A(\uB, \bmw, \bmSi_2) |  ,
		   | A(\oB, \bmw, \bmSi_1) - A(\oB, \bmw, \bmSi_2) | \}
	\notag \\
	& \leq
	\max \{  \msK_1(\uB)  \|\bmSi_1 - \bmSi_2\|_2, \msK_1(\oB)  \|\bmSi_1 - \bmSi_2\|_2 \}
	\notag \\
	& \leq \max \{  \msK_1(\uB) , \msK_1(\oB) \} \|\bmSi_1 - \bmSi_2\|_2
	\equiv \msK_2 \|\bmSi_1 - \bmSi_2\|_2.
	\label{pf:lem:w-unif-adapt-9}
\end{align}

Now, for any $\bmSi_1, \bmSi_2 \in \cM$, consider $\bmwa(\bmSi_1)$ and $\bmwa(\bmSi_2)$, which are optimal solutions to the adaptive problem with matrices $\bmSi_1$ and $\bmSi_2$, respectively.
By the definition of the adaptation regret,
\begin{align}
	\Amax(\bmwa(\bmSi_2), \bmSi_1)
	- \Amax(\bmwa(\bmSi_1), \bmSi_1)
	& \geq {\msK_3} \| \bmwa(\bmSi_2) - \bmwa(\bmSi_1)\|_2^2,
	\label{pf:lem:w-unif-adapt-12}
\end{align}
where $\msK_3 > 0$ and the inequality follows because $A( B, \bmw, \bmSi)$ is strongly convex in $\bmw \in \cWcvx$
(the proof follows from the proof of Lemma~\ref{lem:w-unif-minimax}, particularly  equations \eqref{pf:lem:w-unif-minimax-5},  \eqref{pf:lem:w-unif-minimax-10}, and \eqref{pf:lem:w-unif-minimax-11}, because $A( B, \bmw, \bmSi)$ is a (positive) scalar multiple of $\Rmax(B, \bmw, \bmSi)$ when $B$ and $\bmSi$ are held fixed) and because $\Amax$ is the maximum of two strongly convex functions.
It follows from \eqref{pf:lem:w-unif-adapt-9} and \eqref{pf:lem:w-unif-adapt-12} that
${\msK_3}  \| \bmwa(\bmSi_2) - \bmwa(\bmSi_1)\|^2_2
	\leq \Amax(\bmwa(\bmSi_2), \bmSi_1) - \Amax(\bmwa(\bmSi_1), \bmSi_1)
	\leq 2  \msK_2 \|\bmSi_1 - \bmSi_2\|_2$.
Therefore, for any $\varepsilon > 0$, there exists $\delta = \frac{\msK_3 }{2\msK_2}\varepsilon^2 > 0$ such that for any $\bmSi_1,\bmSi_2 \in \cM$ with $ \|\bmSi_1 - \bmSi_2 \|_2 < \delta$, $ \|\bmwa(\bmSi_1) - \bmwa(\bmSi_2)\|_2 < \varepsilon$ holds.
\end{proof}

\begin{proof}[\textbf{Proof of Lemma \ref{lem:minimax-asym-b-main}}]
{\it \underline{Part \ref{lem:minimax-asym-b1}:}}
Note that $\msdn(\bmw)$ can be written as
\begin{equation}
	\label{pf:lem:minimax-asym-b1-1}
	\msdn(\bmw)
	= \hRmaxn(B, \bmw) - \Rmax(B, \bmw)
	= \bmw'(\hbmSn - \bmSi)\bmw.
\end{equation}
The function $\msdn$ is differentiable on $\cWcvx$ (and at $\bmwsb$).
For Lipschitz continuity, note that for any $\bmw_1, \bmw_2 \in \cWcvx$,
$| \msdn(\bmw_1) - \msdn(\bmw_2)|
= | (\bmw_1 + \bmw_2)'(\hbmSn - \bmSi)(\bmw_1 - \bmw_2) |
\leq 2\|\hbmSn - \bmSi\|_F \|\bmw_1 - \bmw_2\|_2$
using the triangle inequality, the Cauchy--Schwarz inequality, the symmetry of $\hbmSn - \bmSi$, and $\max_{\bmw \in \cWcvx} \|\bmw\|_2 \leq 1$.
Hence, $\msdn$ is Lipschitz continuous.

{\it \noindent \underline{Part \ref{lem:minimax-asym-b2}:}}
From \eqref{pf:lem:minimax-asym-b1-1},
\begin{equation}
	\label{pf:lem:minimax-asym-b2-1}
	\nabla\msdn(\bmw)
	= 2(\hbmSn - \bmSi)\bmw.
\end{equation}
Hence, $\| \nabla\msdn(\bmw)  \|_2
	\leq
	2\| \hbmSn - \bmSi \|_F$
using the Cauchy--Schwarz inequality and $\|\bmw\|_2 \leq 1$.
Here, $\hbmSn - \bmSi = O_{\bP}(\ns)$ by Assumption \ref{assu:minimax-asym}.
Thus, the result follows.

{\it \noindent \underline{Part \ref{lem:minimax-asym-b3}:}}
To begin with, note that
$\|\nabla \msdn(\bmw) - \nabla \msdn(\bmwsb)\|_2 \leq
	2 \|\hbmSn- \bmSi\|_F \| \bmw - \bmwsb \|_2$
using \eqref{pf:lem:minimax-asym-b2-1} and the Cauchy--Schwarz inequality.
Hence,
\begin{align*}
	\tfrac{\|\nabla \msdn(\bmw) - \nabla \msdn(\bmwsb)\|_2}{\ns + \|\bmw - \bmwsb\|_2}
	& \leq
	2\tfrac{ \|\hbmSn- \bmSi\|_F \| \bmw - \bmwsb \|_2}{\ns + \|\bmw - \bmwsb\|_2}
	 \leq
	 2\|\hbmSn- \bmSi\|_F ,
\end{align*}
using $\frac{ \| \bmw - \bmwsb \|_2}{\ns +  \| \bmw - \bmwsb \|_2} \leq 1$.
The result follows because $\hbmSn \Pt \bmSi$ under Assumption \ref{assu:minimax-asym}.
The neighborhood can be taken to be a small ball around $\bmwsb$.
\end{proof}

\begin{proof}[\textbf{Proof of Lemma \ref{lem:minimax-asym-a3}}]
The objective of \eqref{eq:lem:minimax-asym-a3-1} is strictly convex in $\bmw \in \cWcvx$.
Thus, the optimal solution $\obmw(\bmze)$ is unique.
The minimum eigenvalue of $\bmSi$ is positive as $\bmSi$ is positive definite by Assumption \ref{assu:minimax-asym}.
Here, $\Rmax(B, \bmw) = f_1(\bmw) + f_2(\bmw)$ where $f_1(\bmw) = \oM(B, \bmw)$ is convex in $\bmw$ and $f_2(\bmw) = \bmw'\bmSi\bmw$ is twice continuously differentiable, with $\nabla^2f_2(\bmw) = 2\bmSi$.
The lemma follows from pages 832--833 of \citet{shapiro1993mor}.
\end{proof}

\end{document}